\documentclass[useAMS]{mn2e}
\usepackage{epsfig}
\usepackage{longtable}
\usepackage{mathrsfs}
\usepackage{graphicx}
\usepackage{amssymb}

\usepackage[totalwidth=500pt, totalheight=640pt]{geometry}

\usepackage[varg]{txfonts}
\usepackage{astrojournals}
\usepackage{graphicx}
\usepackage{microtype}
\usepackage{xcolor}
\usepackage{fixltx2e}
\usepackage{siunitx}

\title[The protoplanetary disc HST~10 in Orion]{Chemical abundances in Orion protoplanetary discs: integral field spectroscopy and photoevaporation models of HST~10\thanks{Based on observations made with ESO telescopes at the Paranal Observatory under
programme 078.C-0247(A).}}
\author[Y. G. Tsamis et al.]{Y. G. Tsamis$^{1}$\thanks{E-mail:
ygtsamis@gmail.com}, N. Flores-Fajardo$^{2}$, W. J. Henney$^{2}$, J. R. Walsh$^{1}$ and A. Mesa-Delgado$^{3}$ \\
$^{1}$European Southern Observatory, Karl-Schwarzschild-Str. 2, D-85748 Garching 
bei M$\ddot{u}$nchen, Germany\\
$^{2}$Centro de Radioastronom\'ia y Astrof\'isica, Universidad Nacional Aut\'onoma de M\'exico, Campus Morelia, Apartado Postal 3-72, 58090 Morelia, Michoac\'an, Mexico\\
$^{3}$Departamento de Astronom\'ia y Astrof\'isica, Facultad de F\'isica, Pontificia Universidad Cat\'olica de Chile, Av. Vicu\~na Mackenna 4860,782-0436 Macul, Santiago, Chile \\
}

\newcommand{\hst}{{\it HST\/}}
\newcommand{\eld}{$N_{\rm e}$}

\newcommand{\elt}{$T_{\rm e}$}

\newcommand{\cmt}{cm$^{-3}$}

\newcommand{\cp}{C$^+$}
\newcommand{\cpp}{C$^{2+}$}

\newcommand{\op}{O$^+$}

\newcommand{\opp}{O$^{2+}$}

\newcommand{\np}{N$^+$}
\newcommand{\sulp}{S$^+$}
\newcommand{\sulpp}{S$^{2+}$}

\newcommand{\nepp}{Ne$^{2+}$}

\newcommand{\clpp}{Cl$^{2+}$}
\newcommand{\arpp}{Ar$^{2+}$}
\newcommand{\arppp}{Ar$^{3+}$}
\newcommand{\fepp}{Fe$^{2+}$}
\newcommand{\foiii}{[O~{\sc iii}]}

\newcommand{\foi}{[O~{\sc i}]}
\newcommand{\foii}{[O~{\sc ii}]}
\newcommand{\fsii}{[S~{\sc ii}]}
\newcommand{\fsiii}{[S~{\sc iii}]}

\newcommand{\fni}{[N~{\sc i}]}
\newcommand{\fnii}{[N~{\sc ii}]}

\newcommand{\fariv}{[Ar~{\sc iv}]}
\newcommand{\fcliii}{[Cl~{\sc iii}]}

\newcommand{\fneii}{[Ne~{\sc ii}]}
\newcommand{\fneiii}{[Ne~{\sc iii}]}

\newcommand{\ffeiii}{[Fe~{\sc iii}]}

\newcommand{\oii}{O~{\sc ii}}
\newcommand{\cii}{C~{\sc ii}}

\newcommand{\fciii}{C~{\sc iii}]}

\newcommand{\fariii}{[Ar~{\sc iii}]}

\newcommand{\hi}{H\,{\sc i}}
\newcommand{\hii}{H~{\sc ii}}
\newcommand{\hei}{He~{\sc i}}

\newcommand{\hp}{H$^+$}
\newcommand{\hep}{He$^+$}
\newcommand{\hepp}{He$^{2+}$}
\newcommand{\ha}{H$\alpha$}
\newcommand{\hb}{H$\beta$}
\newcommand{\hg}{H$\gamma$}
\newcommand{\hd}{H$\delta$}
\newcommand{\oric}{$\theta^1$\,Ori~C}
\newcommand{\lam}{$\lambda$}
\newcommand{\kms}{km\,s$^{-1}$}

\newcommand\texttheta{\ensuremath{\theta}}
\newcommand\thC{\texttheta\textsuperscript{1}\,Ori~C}

\begin{document}

\date{Accepted ... Received ...}

\pagerange{\pageref{firstpage}--\pageref{lastpage}} \pubyear{2002}

\maketitle

\label{firstpage}

\begin{abstract}

Photoevaporating protoplanetary discs (proplyds) in the vicinity of hot massive stars, such as those found in Orion, are important objects of study for the fields of star formation, early disc evolution, planetary formation, and \hii\ region astrophysics. Their element abundances are largely unknown, unlike those of the main-sequence stars or the host Orion nebula. We present a spectroscopic analysis of the Orion proplyd HST~10, based on integral field observations with the Very Large Telescope/FLAMES fibre array with 0.31$''$ $\times$ 0.31$''$ spatial pixels. The proplyd and its vicinity are imaged in a variety of emission lines across a 6.8$''$ $\times$ 4.3$''$ area. The reddening, electron density and temperature are mapped out from various line diagnostics. The abundances of helium, and eight heavy elements are measured relative to hydrogen using the direct method based on the \foiii\ electron temperature. The abundance ratios of O/H and S/H are derived without resort to ionization correction factors. We construct dynamic photoevaporation models of HST~10 with the Cloudy microphysics code that validate the oxygen and sulfur abundances. With the exception of \foi\ \lam6300 and  \fsii\ \lam4069, the model fit is satisfactory for all spectral lines arising from the proplyd. The models show that the classic ionization correction factor for neon significantly underestimates (0.4 dex) this element's abundance in the low ionization conditions of HST~10. Apart from iron, whose gas-phase abundance is $\sim$0.3 dex lower than in the local Orion nebula, most other elements in the proplyd do not show substantially different gas-phase abundances from the nebula. The abundances of carbon, oxygen and neon in HST~10 are practically the same as those in B-type stars in Orion.



\end{abstract}


\section{Introduction}

HST~10 (e.g. Johnstone et al. 1998) is an Orion proplyd of impressive appearance that lies farther away from the Trapezium than the compact group of bright `LV' proplyds (Laques \& Vidal 1979). It is thus subjected to a much smaller incident ionizing flux from the Trapezium stars. Fig.\,~1 shows a composite image of HST~10 based on the VLT dataset presented in this paper that highlights its `teardrop' morphology and the presence of both neutral and ionized gas emission. The proplyd is designated as 182--413 according to the nomenclature of Orion sources by O'Dell \& Wenn (1994).    
It is larger and fainter than the proplyds found close to the Trapezium, with a less elongated and less symmetric tail. This is in line with the general trends seen in these sources 
(Bally et al. 1998; O'Dell 1998), which can be understood in terms of a model whereby protostellar discs around the young low-mass stars in the nebula are evaporated by the ultraviolet radiation from the high-mass Trapezium stars (Johnstone et al. 1998; Henney \& Arthur 1998; St\"orzer \& Hollenbach 1999; Vasconcelos et al. 2011).  

In this study, we present optical integral field spectroscopy (IFS) of HST~10 and its immediate Orion environment, constructing photoevaporation models of the proplyd. Our aim is to enlarge the number of sources with element abundance determinations following the studies of LV\,2 (Tsamis et al. 2011; Tsamis \& Walsh 2012) and HST~1 (Mesa-Delgado et al. 2012) proplyds.

\begin{figure*}
  \setkeys{Gin}{width=\linewidth}
  \centering
  \includegraphics{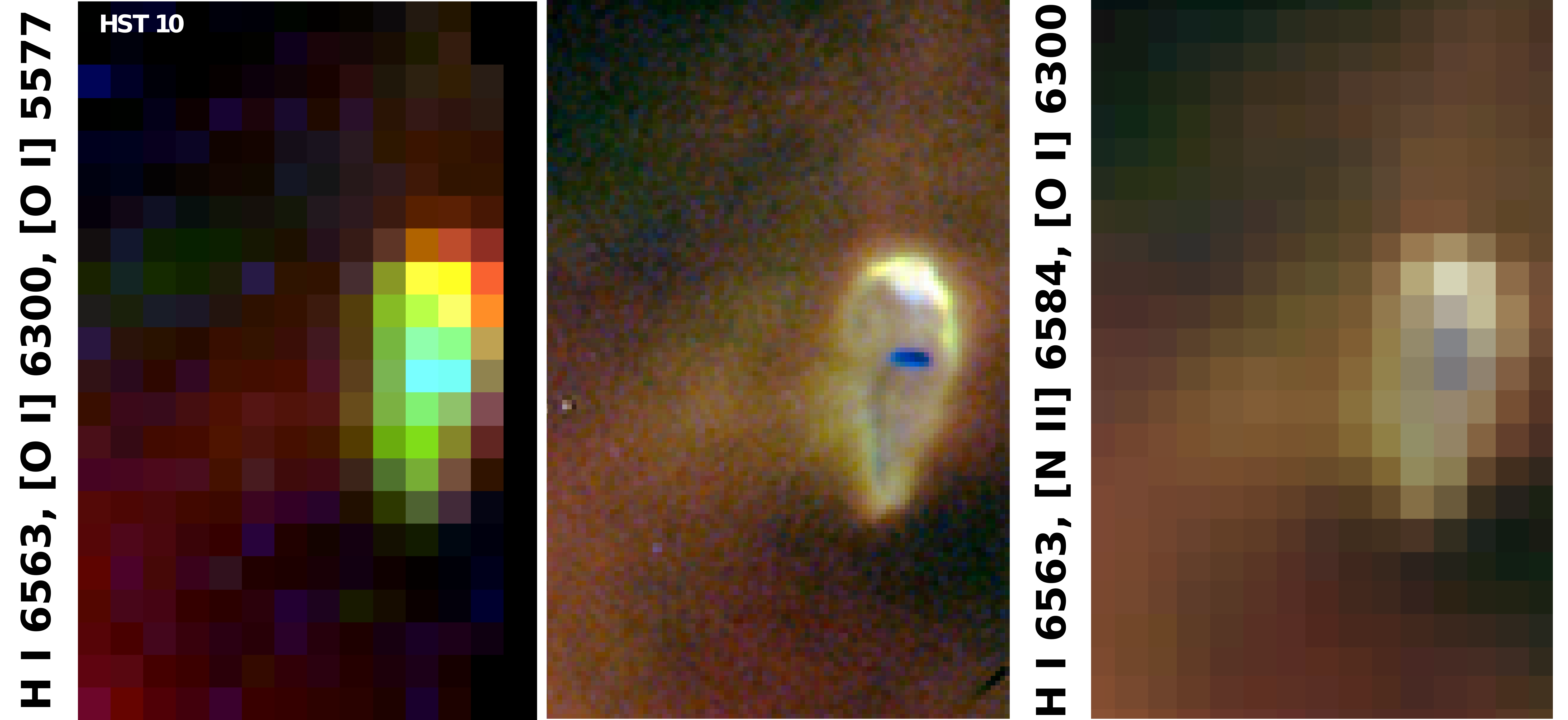}
\caption{Comparison of VLT versus {\it HST} images of the HST 10 proplyd: \emph{(Left)}
composite of VLT Argus monochromatic channel maps with \ha\ $\lambda$6563 (red), \foi\ $\lambda$6300 (green), \foi\ $\lambda$5577 (blue). The embedded circumstellar disc dominates the emission of \foi\ $\lambda$5577. The field of view is 6.8$''$ $\times$ 4.3$''$; \emph{(Centre)}: {\it HST} composite using filters F656N (red), F658N (green), F631N (red); \emph{(Right)} The {\it HST} image
convolved with a Gaussian filter of 0.8$''$ FWHM (approximately equal to the
seeing during the VLT observations) and rebinned to the VLT pixel scale. North is up and east is to the left-hand side.}
\end{figure*}

\begin{figure*}
  \setkeys{Gin}{width=\linewidth}
  \centering
  \includegraphics{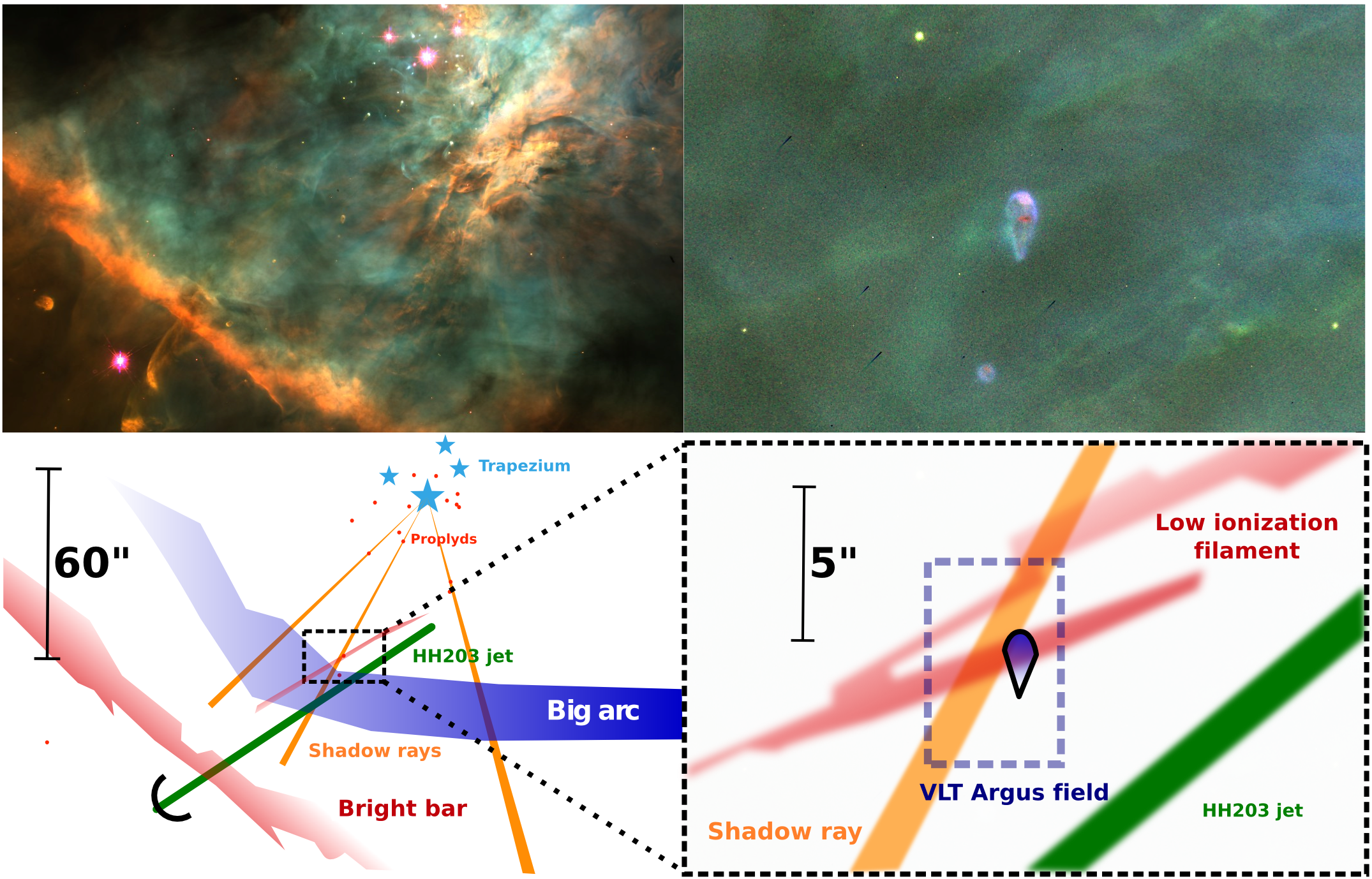}
  \caption{HST~10 in the context of the Orion Nebula (to be viewed anti-clockwise). {\it Left upper panel}: Large scale, three-color image of the Orion nebula in the filters of \fnii\ (red), \ha\ (green), and \foiii\ (blue),
    based on a mosaic of \textit{HST} WFC observations 
    described in O'Dell \& Wong (1996). {\it Left lower panel}: Schematic view of various components in Orion.  
     {\it Right lower panel}: Sketch of the HST~10 field (dashed cyan box) and its immediate environs. {\it Right upper panel}: Three-color image centred on HST~10 in the filters of \foi\ (red), \fsii\ (green), and \fnii\ (blue), based on \textit{HST} PC observations 
    described in O'Dell (1998).  
    The labelled objects are described in the text.}
  \label{fig:wfpc2-images}
\end{figure*}

\subsection{HST~10 in the Orion nebula context}

The location on the sky of HST~10 is halfway between 
the inner Trapezium cluster and the Bright Bar region (see Fig.~\ref{fig:wfpc2-images}), 
at an angular separation of about \(1'\) to the south-east from \thC{} (O7V spectral type; e.g. Sim{\'o}n-D{\'{\i}}az et al. 2006), the dominant illuminating star  of the Orion nebula.  
Kinematic studies of the emission from the proplyd (Henney \& O'Dell 1999) 
suggest that it is situated in the foreground of the nebula, 
with a true separation from \thC{} of 0.2--0.3~pc.  Unlike in many of the smaller, brighter proplyds, 
the circumstellar disc of mass $\sim$0.0054\,M$_\odot$ (see Section 4.3) is clearly visible in HST~10.  
In most emission lines, the disc is seen in absorption, 
but it is seen in emission in the H\(_2\)~2.12\(\mu\)m line (Chen et al. 1998), 
the \foi\ \lam6300 line (Johnstone et al. 1998), and the \foi\ \lam5577 line (our Fig.\,~1). 
The \lam6300 line emission is shown in red in the right panel of Fig.~\ref{fig:wfpc2-images}.
There is also a faint high-velocity microjet detected in \foi\ \lam6300 (Henney \& O'Dell 1999),
which extends perpendicular to the disc (Bally et al. 2000). 

Although the proplyds close to the Trapezium tend to be very symmetric 
about the line that joins them to \thC{},
this is not so true of HST~10, which seems to be governed by two different axes. 
The rotational axis of the disc and jet is oriented north-south, 
while the direction to \thC{} is at roughly \(30^\circ\) to this,
at a position angle of \(330^\circ\).  
This produces considerable distortion in the shape of the proplyd,
with the tail, in particular, seemingly more governed by the jet axis 
than the ionizing radiation field.

The Orion Nebula is highly structured at all scales 
and many previously studied nebular features pass through or near 
the immediate vicinity of HST~10, as illustrated in Fig.\,~\ref{fig:wfpc2-images}.
The shadow rays ({O'Dell} 2000), are low-ionization linear features, which are seen outside the positions of some of the brighter proplyds,
and are exactly aligned with the outer segment of the line joining \thC{} and the proplyd
(see fig.~2 of O'Dell et al. 2009). 
Three such rays are indicated in Fig.\,~\ref{fig:wfpc2-images}: the two most prominent, 
which are cast by 177--341 (HST~1) and 159--350 (HST~3), 
plus a much fainter one, which is cast by 170--337 
and which passes within \(2''\) of HST~10 in projection. 
Unlike HST~10, 170--337 shows evidence of being located behind the Trapezium stars (Henney \& O'Dell 1999), and therefore its shadow must be also, therefore it is very unlikely to be physically close to HST~10. 

The bar features are another type of linear structure that is very common in the nebula of which the Bright Bar is the most prominent example (e.g. {O'Dell} \& {Yusef-Zadeh} 2000; Mesa-Delgado et al. 2011; Rubin et al. 2011). 
These are regions where the line of sight is tangential to a local ionization front, 
but the exact geometry is unclear in many cases. 
{Garc{\'{\i}}a-D{\'{\i}}az} \& {Henney} (2007) found that some faint compact bars are associated with dark lanes
that are seen as linear extinction features in red-shifted velocity channels. 
Such is the case for the jagged low ionization filament that crosses the field of HST~10, 
as shown in the right panel of Fig.\,~\ref{fig:wfpc2-images}. A section of this filament has been detected by VLT Argus and its electron temperature and density have been determined from our spectra. 
This type of emission feature may represent the ionized skin of a dense molecular filament 
that is protruding into the \hii{} region, in which case it is likely to be close to the principal ionization front in the background of the nebula. Therefore again, the line of sight position of the feature is probably far from HST~10.

A different type of linear structure is the high velocity collimated jets 
that drive the numerous Herbig-Haro bow shocks seen in the nebula. 
The driving jet of the HH~203 bow shock (Doi et al. 2004) passes within \(5''\) of HST~10, 
although it is just outside the field of view of our VLT observations. A much larger scale kinematic feature is the so-called Big Arc (Doi et al. 2004; {Garc{\'{\i}}a-D{\'{\i}}az} \& {Henney} 2007), 
which is a blue-shifted high-ionization structure 
that extends over several arcminutes and whose origin is unclear. 
Although HST~10 is very close to the northern boundary of this feature, 
it probably does not affect our observations.

\section{The VLT observations and data reduction}

\setcounter{table}{0}
\begin{table}
\centering
\begin{minipage}{75mm}
\caption{Journal of VLT/FLAMES Argus observations.}
\begin{tabular}{@{}lcccc@{}}
\hline
              Date                   &$\lambda$-range &Grating   &$\lambda$/$ \delta
\lambda$   &Exp. time   \\
               (UT)                  &(\AA)           &          &   &(s)               
\\
\hline

2006/12/15      &3620--4081    &LR1      &12\,800    &2 $\times$ 150 \\ 
2006/12/15      &3964--4567    &LR2      &10\,200    &3 $\times$ 390 \\ 
2006/12/15		 &4501--5078	 &LR3		&12\,000	  &4 $\times$ 206 \\
2006/10/03		 &5015--5831	 &LR4		&9\,600		  &3 $\times$ 254 \\
2006/10/03		 &5741--6524 	 &LR5		&11\,800	&3 $\times$ 255 \\
2006/10/03		 &6438--7184 	 &LR6		&13\,700	&4 $\times$ 179 \\

\hline
\end{tabular}
\end{minipage}
\end{table}

Integral field spectroscopy of HST~10 ($\alpha_{\rm J2000}$ $=$ 5$^{\rm h}$ 35$^{\rm m}$ 18.204$^{\rm s}$, $\delta_{\rm J2000}$ $=$ $-$5$^{\circ}$ 24$'$ 13.38$''$) was performed on the 8.2-m VLT/UT2 Kueyen at the ESO observatory on Paranal during two nights in 2006 October and December with the fibre-fed FLAMES Giraffe
spectrograph (Pasquini et al. 2002) and the Argus microlens array in high magnification mode. A field of view of
6.8$''$ $\times$ 4.3$''$ was observed and 297 positional spectra were recorded in the
optical range (3620--7184\,\AA) using a 2K $\times$ 4K EEV CCD (with pixels of 15$\mu$m size). Six low
resolution (LR) grating settings were employed
(Table~1). The size of the angular resolution element was 0.31$''$ $\times$ 0.31$''$, corresponding to a spatial scale of 127 $\times$ 127 AU$^{2}$ at the distance to M42 (410 pc; Reid et al 2009). The measured seeing during the observations with the various gratings was 0.7$''$ (LR1), 0.7--0.9$''$ (LR2), 0.9$''$ (LR3), 0.8--0.9$''$ (LR4), 0.7--0.8$''$ (LR5), and 0.7$''$ FWHM (LR6). 

Due to a slight misalignment of the guide/fiducial star fibres used for the acquisition of the Argus field of view, the proplyd is not exactly centred in the array and is shifted towards the western edge of the field (Fig.\,~1, left panel). In Fig.\,~1 (right panel) we show an {\it HST} image of the proplyd that has been degraded to the resolution of the Argus spaxels, taking into account the seeing measured during our VLT observations. Comparison of these two accurately aligned images shows that no appreciable loss of flux has resulted from the proplyd based on its position within the Argus field of view. The resolution achieved allows us to clearly distinguish three regions in HST~10: cusp, core/disk area, and tail. We cannot however very well resolve the structure within these.

The 2D data arrays were cosmic ray cleaned, flat fielded, wavelength calibrated, and extracted utilizing the girBLDRS pipeline developed by the Geneva Observatory (Blecha \& Simond 2004). The flux calibration was done within {\sc iraf} using contemporaneous exposures of various
spectrophotometric standards for the grating settings: EG~21 (LR1), Feige~110 (LR2, LR3), LTT~7987 (LR4, 5, 6). The reduction products consisting of separate signal and error arrays were converted into ($X$, $Y$, $\lambda$) data cubes using {\sc iraf} routines and the Argus fibre positioning table.

The data cubes were corrected for the effects of differential atmospheric refraction (DAR) using the algorithm of Walsh and Roy (1990) as in Tsamis et al. (2011). In this way, the various emission line or continuum maps were registered on an identical spatial grid across the entire spectral range of Table~1, allowing us to form high fidelity ratio maps of spectral features widely separated in wavelength. The shifts involved in the correction for DAR were extremely small as the airmasses during the observations were low ($<$1.3). There was a detectable shift in the positioning of Argus between the two nights and as a result the LR4, 5, 6 spatial grid was shifted by one spaxel due east to match the LR1, 2, 3 cubes. The necessary rebinning has contributed to the edge effects apparent on the left-hand side of some of the maps discussed in Section~3.1. The LR5 and LR6 spectra were scaled by a factor of 0.813 via the flux of the \fnii\ \lam5755 line which falls on both LR4 and LR5 gratings in order to match the bluer spectral coverage. The LR1 spectrum was scaled to the LR2 spectrum using the \hi\ \lam3970 line common to both gratings.

We selected emission lines of interest in the various grating set-ups and fitted them with Gaussian profiles in all $\sim$300 individual positional spectra. An automated procedure was used based on {\sc midas} tasks utilizing the Newton-Raphson least-squares method. The local continuum bracketing each line was fitted using a first order polynomial or a cubic spline. The resulting arrays were further processed within {\sc midas} and {\sc iraf} to form maps of the lines or continua and their signal-to-noise (S/N) ratios.

\section{Mapping the physical diagnostics}

\subsection{Emission line maps}

\begin{figure*}



  \setkeys{Gin}{width=.32\linewidth}
  \setlength\tabcolsep{0pt}
  \centering
    \includegraphics[]{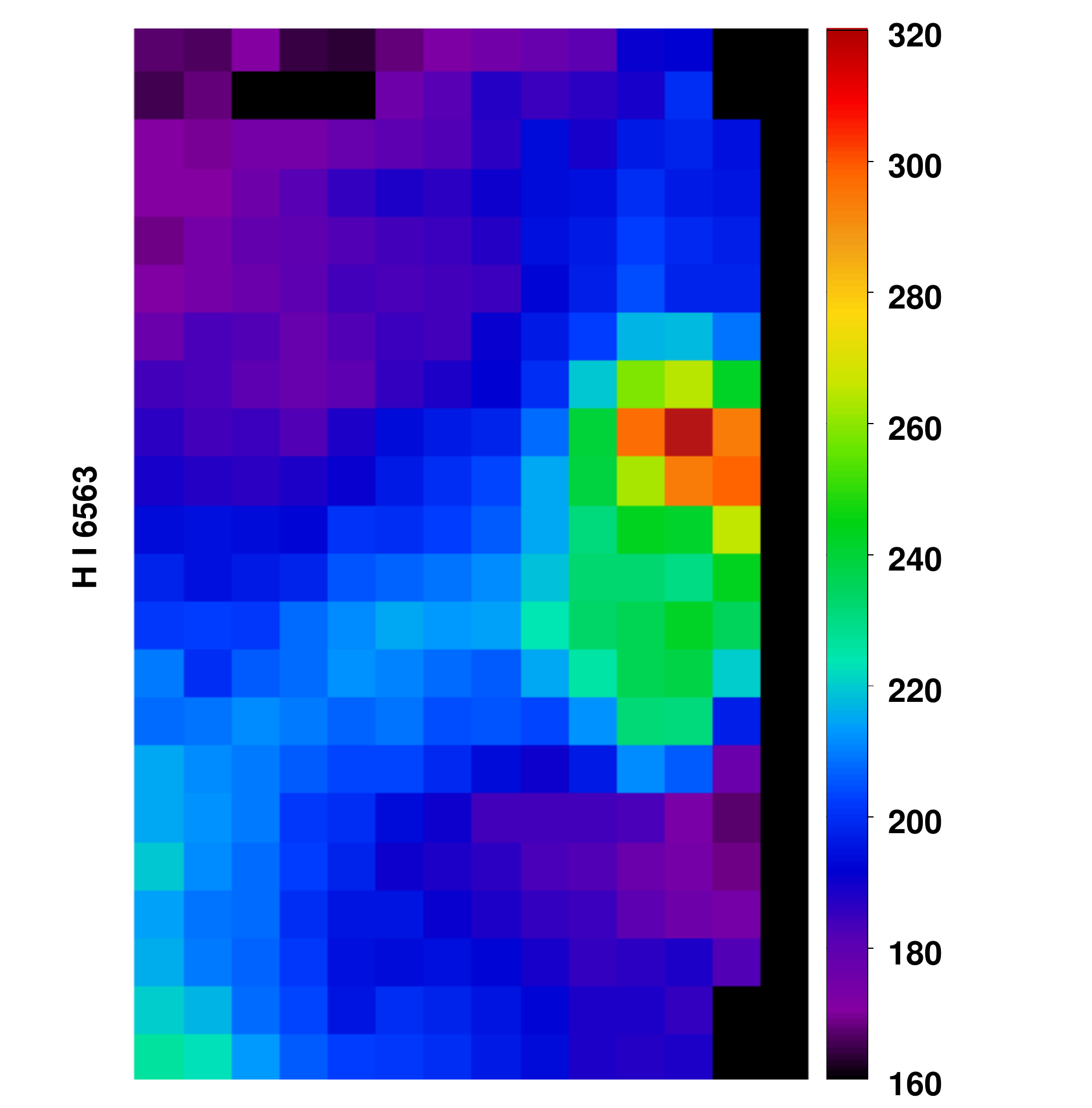}
    \includegraphics[]{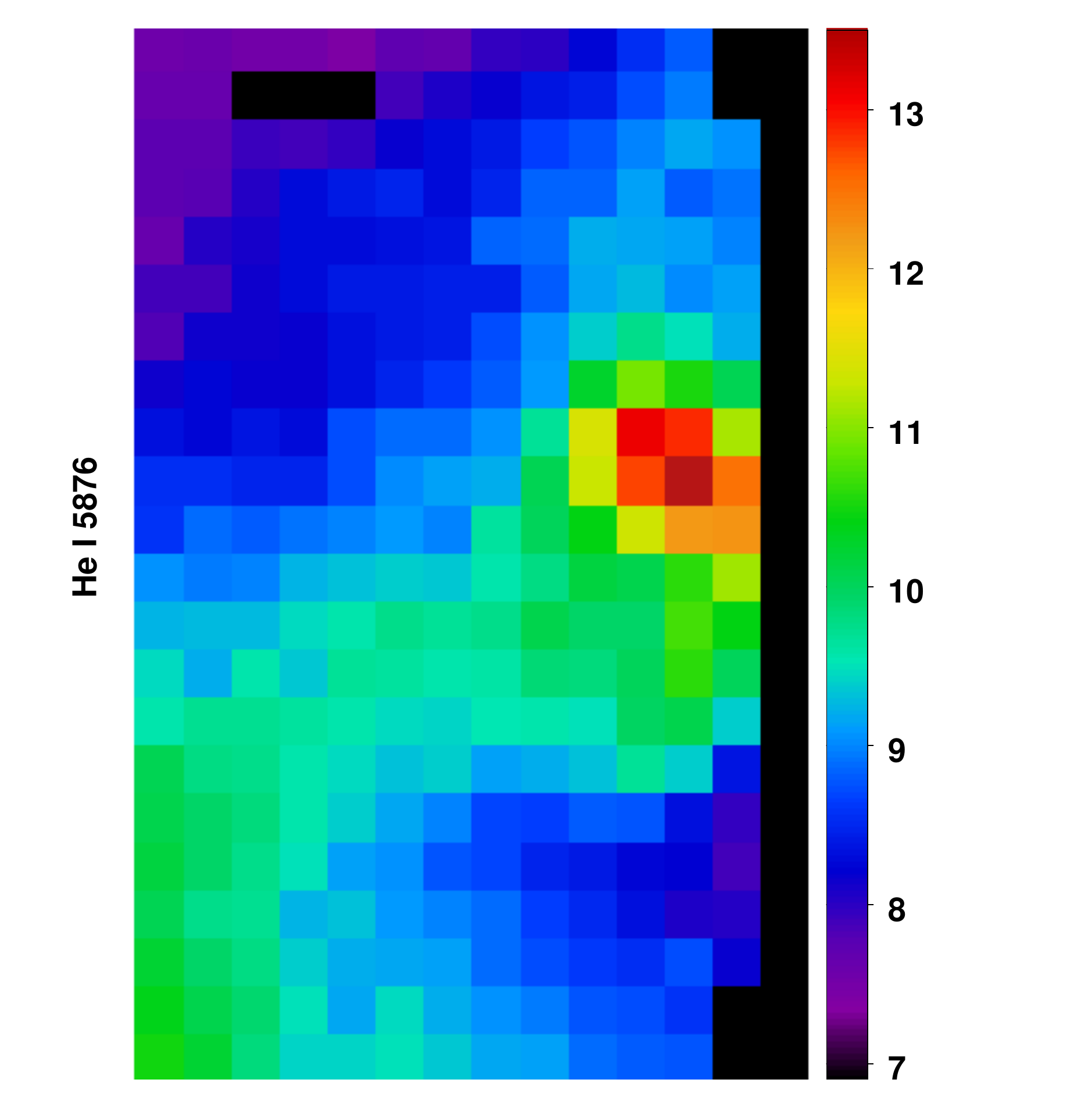}
    \includegraphics[]{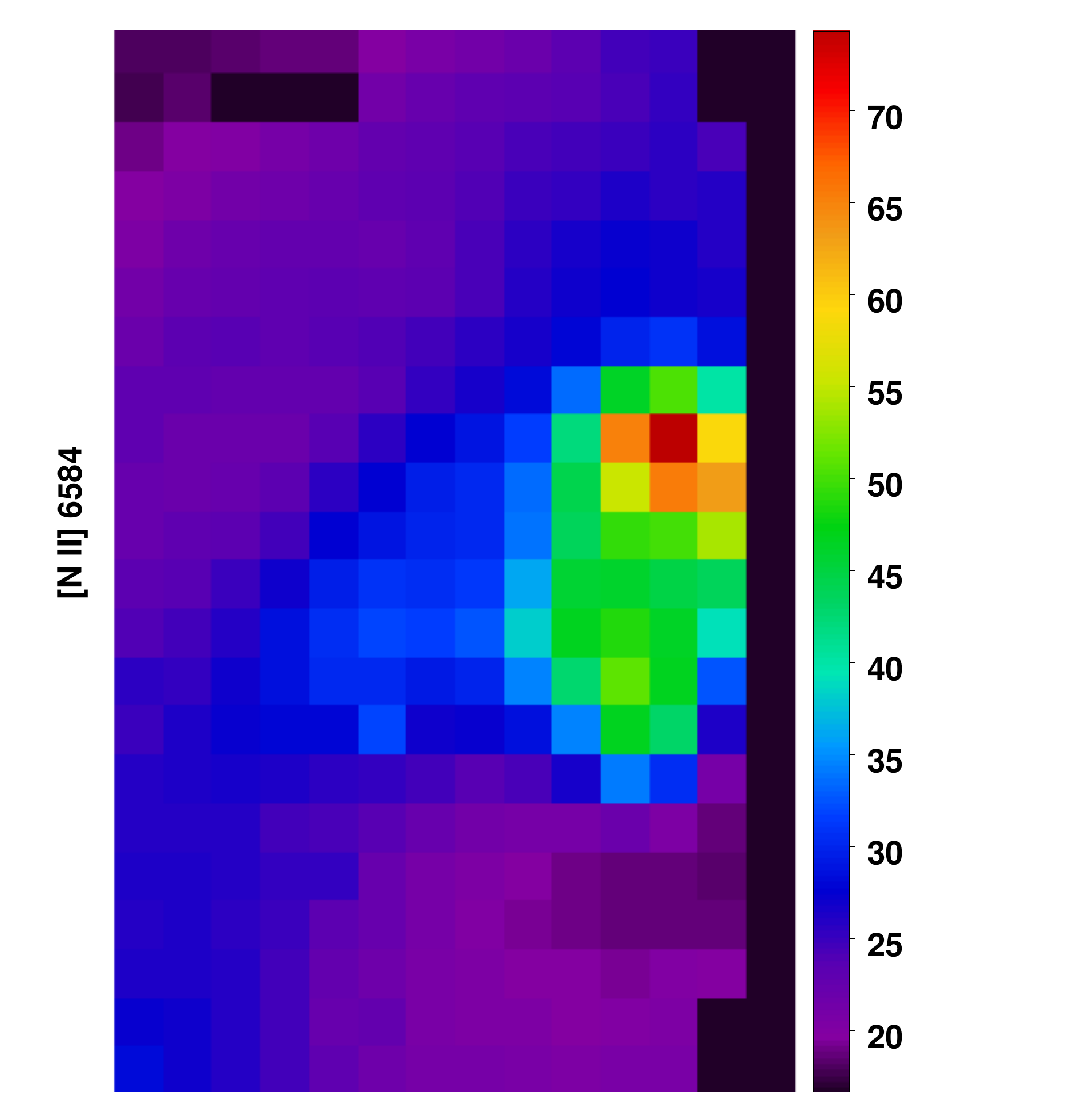}
    \includegraphics[]{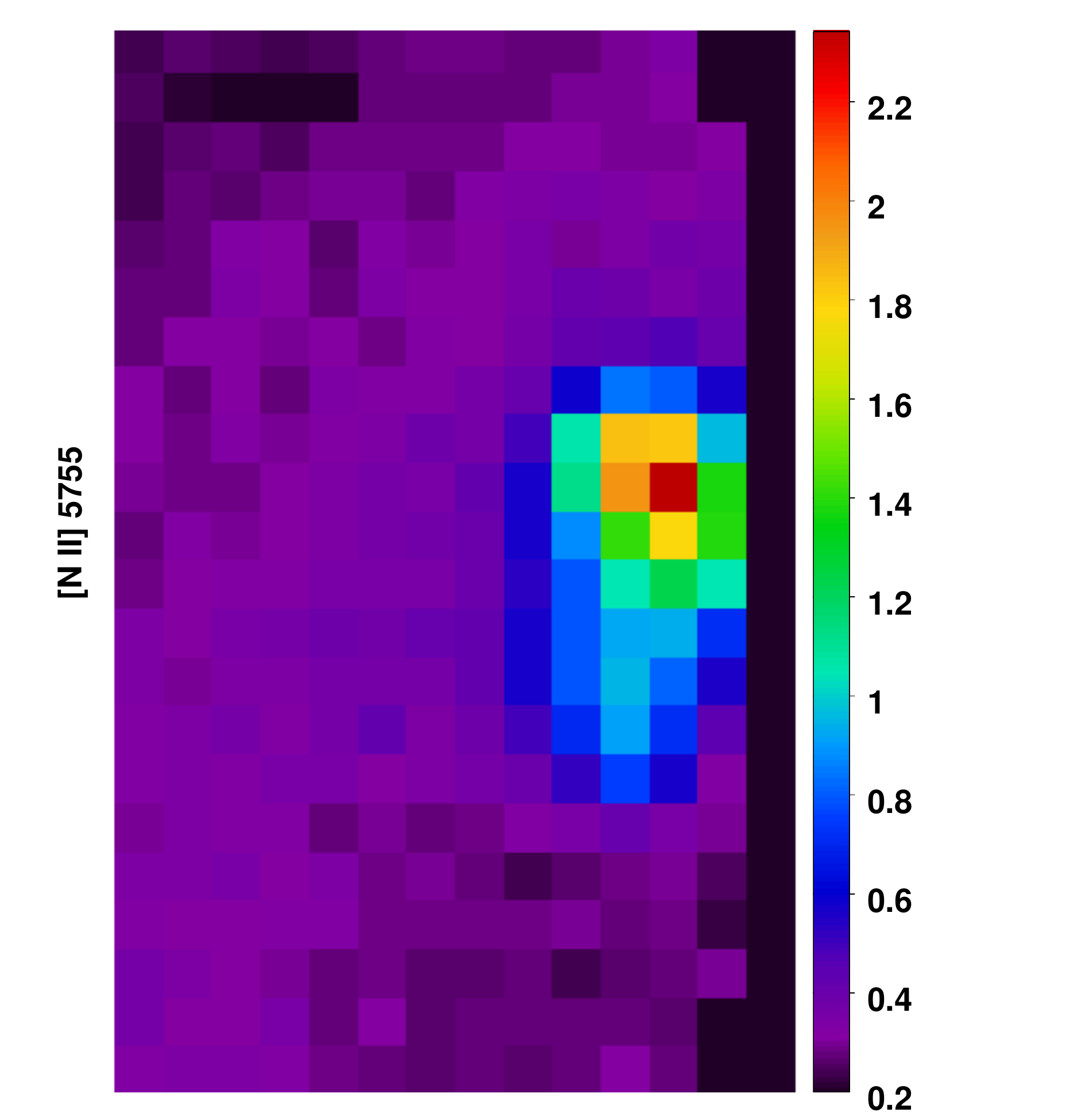}
    \includegraphics[]{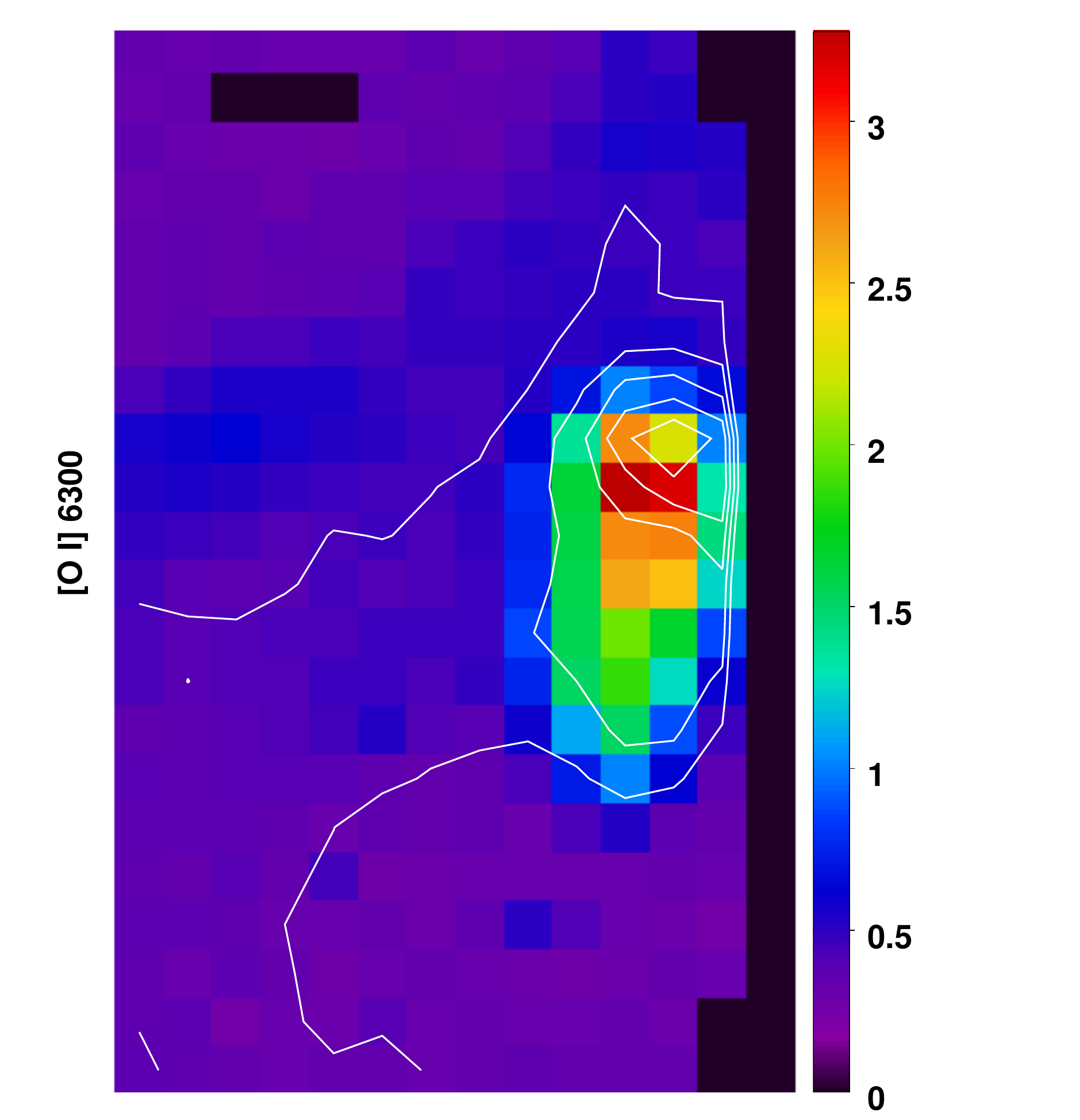}
    \includegraphics[]{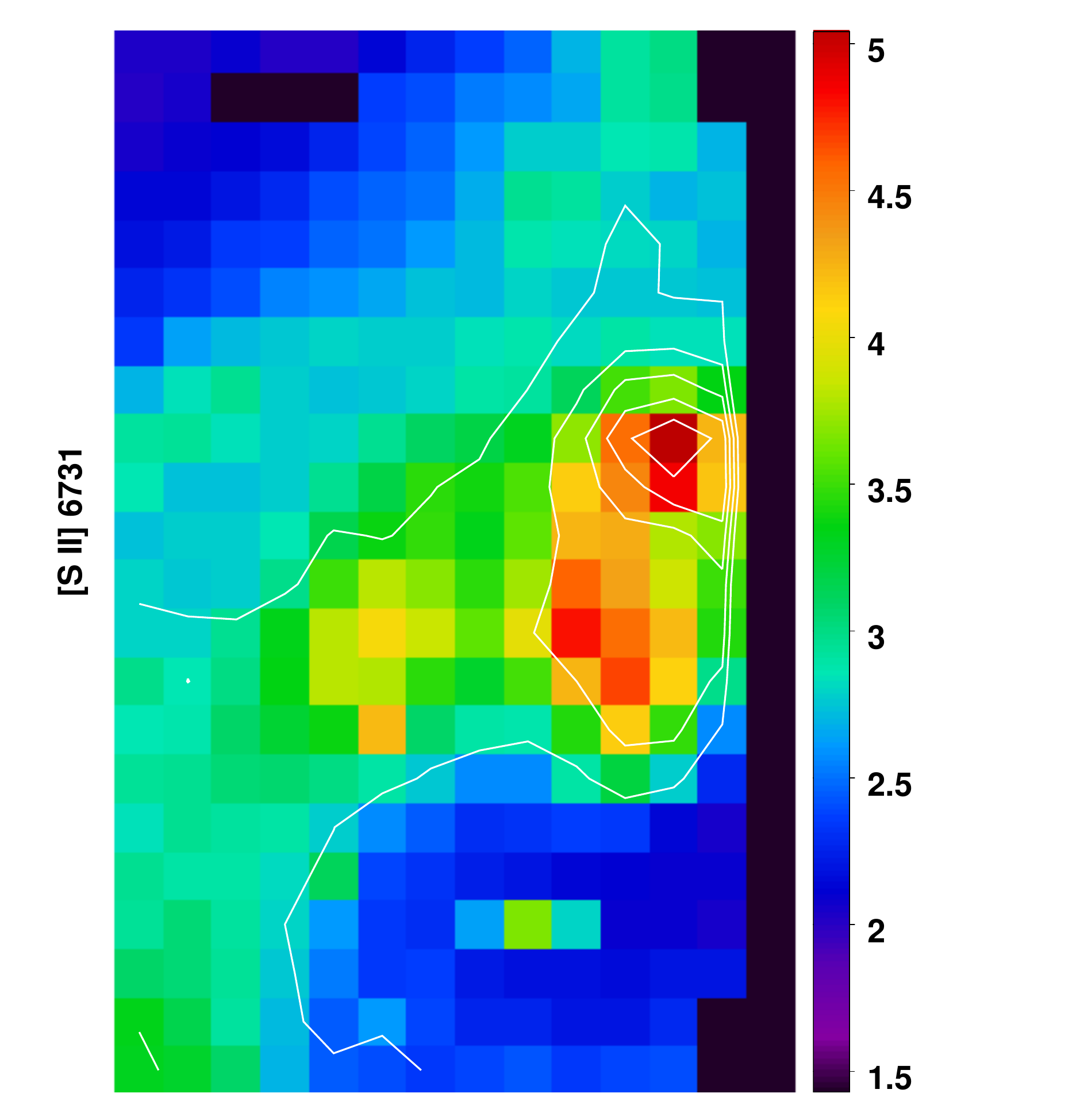}
    \includegraphics[]{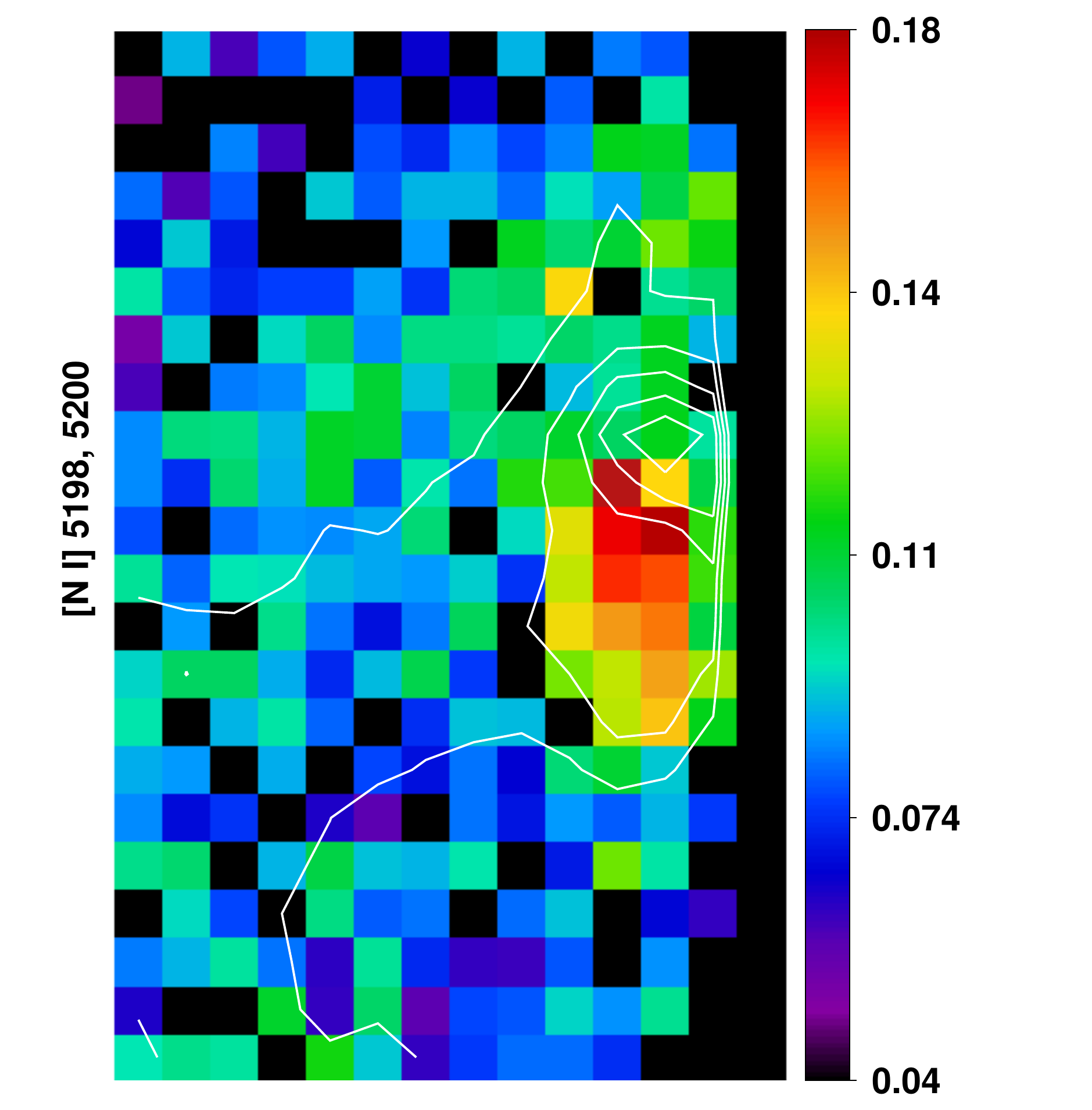}
        \includegraphics[]{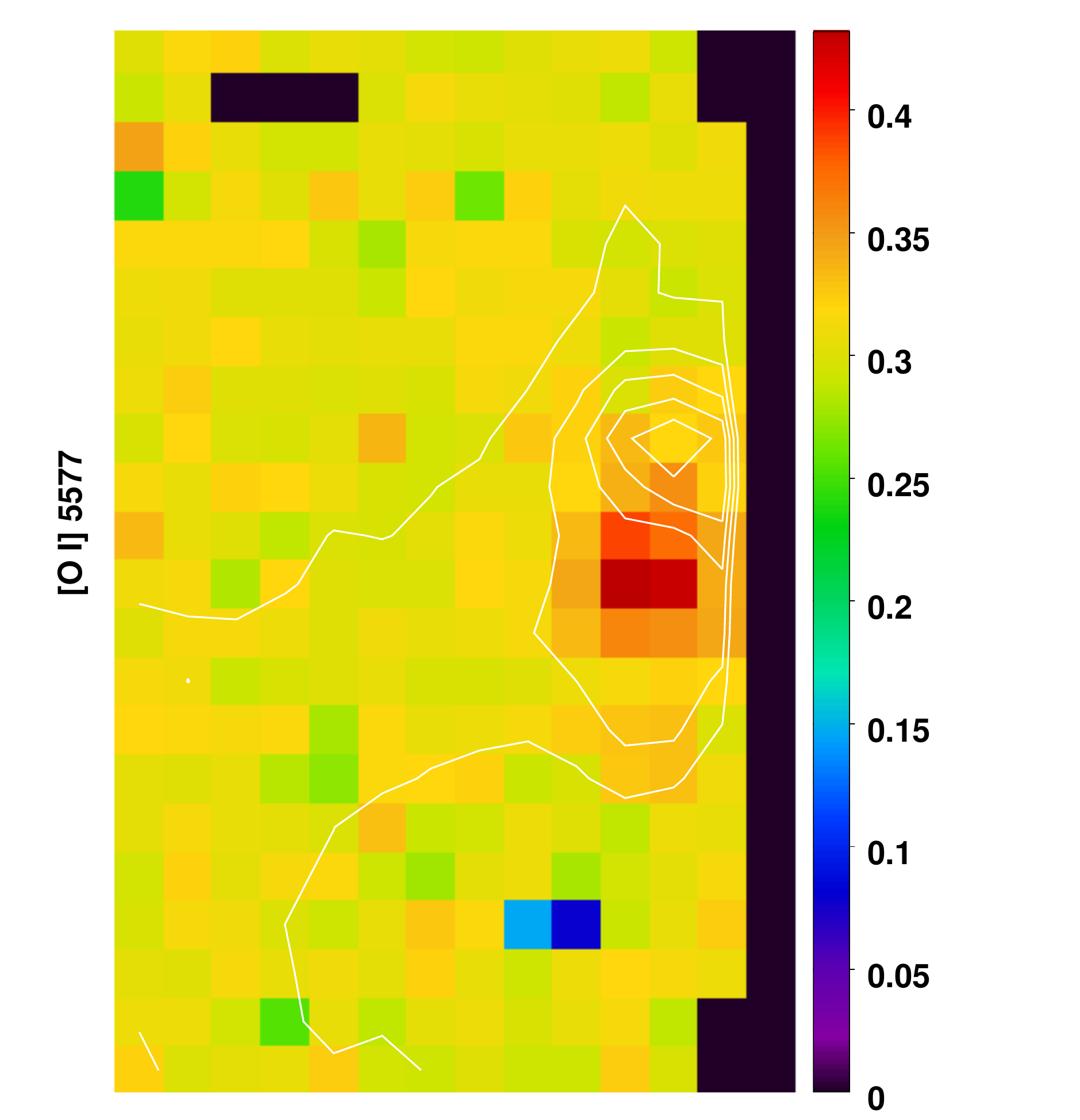}
            \includegraphics[]{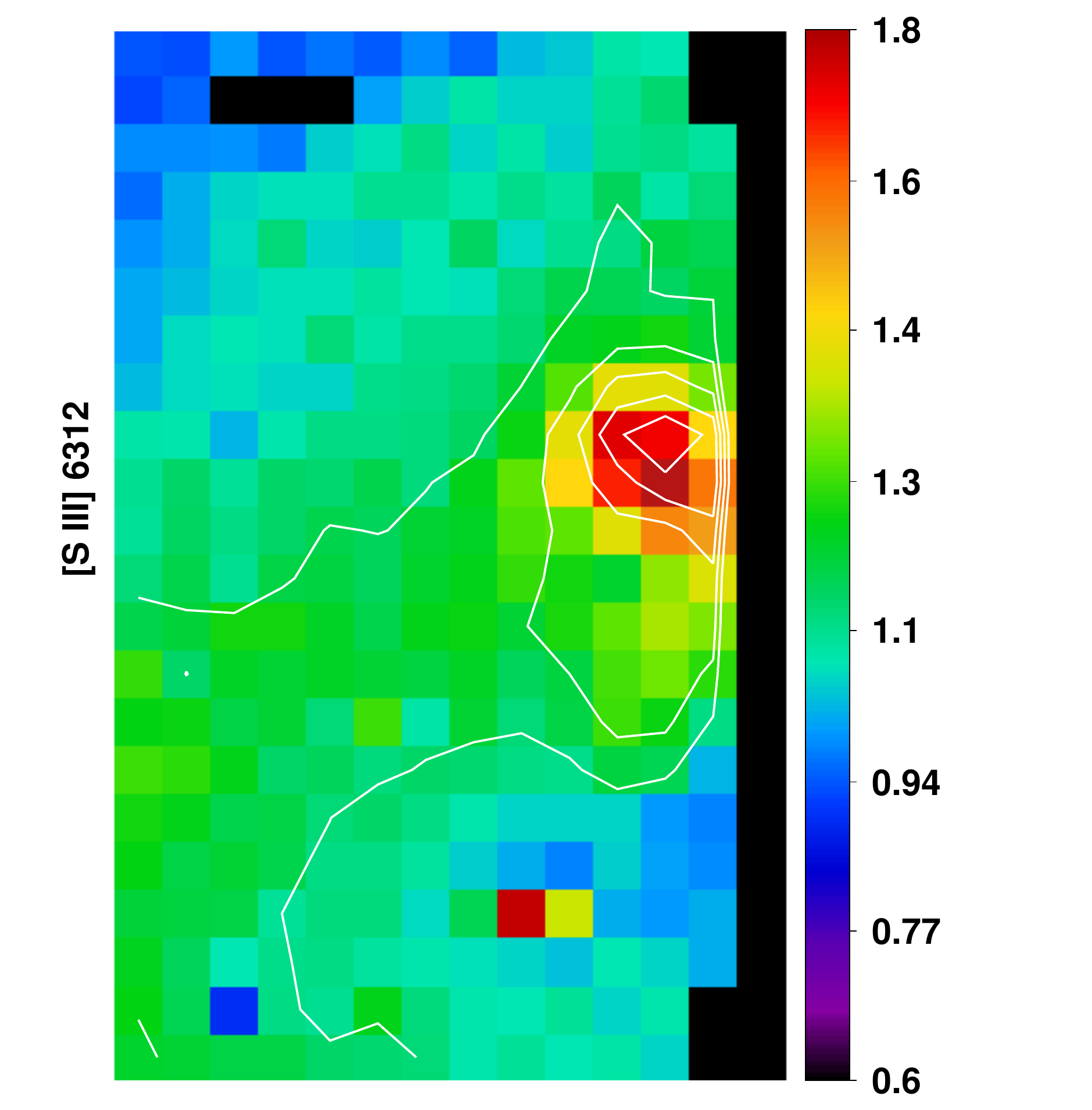}

\caption{Emission line maps (in units of 10$^{-15}$ erg\,s$^{-1}$\,cm$^{-2}$\,spaxel$^{-1}$, not corrected for reddening; the isocontours are the \ha\ line). The field of view is 6.8 $\times$ 4.3 arcsec$^2$ (22 by 14 spaxels). Dark masked outer columns are artefacts at the edges of the array introduced by the rebinning used to correct for differential atmospheric refraction and for the spatial shift between the LR1, 2, 3 and LR4, 5, 6 observations (see the text for details). Two spaxels in the fourth row are affected by a CCD defect and three blank spaxels in row 21 correspond to dead fibres.  North is up and east is to the left-hand side.}
\end{figure*}

\begin{figure*}

  \setkeys{Gin}{width=.32\linewidth}
  \setlength\tabcolsep{0pt}
  \centering
    \includegraphics[]{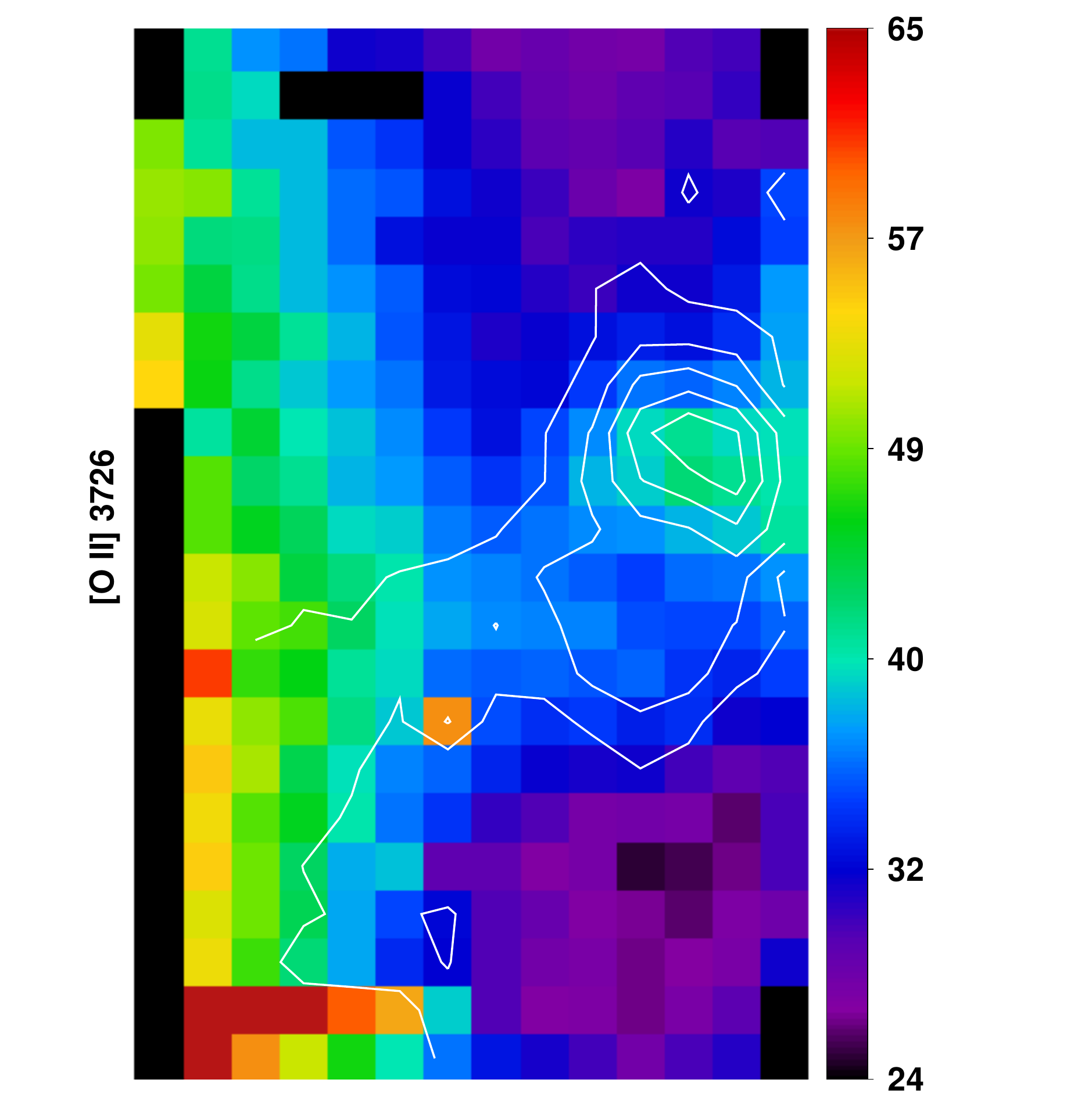}
    \includegraphics[]{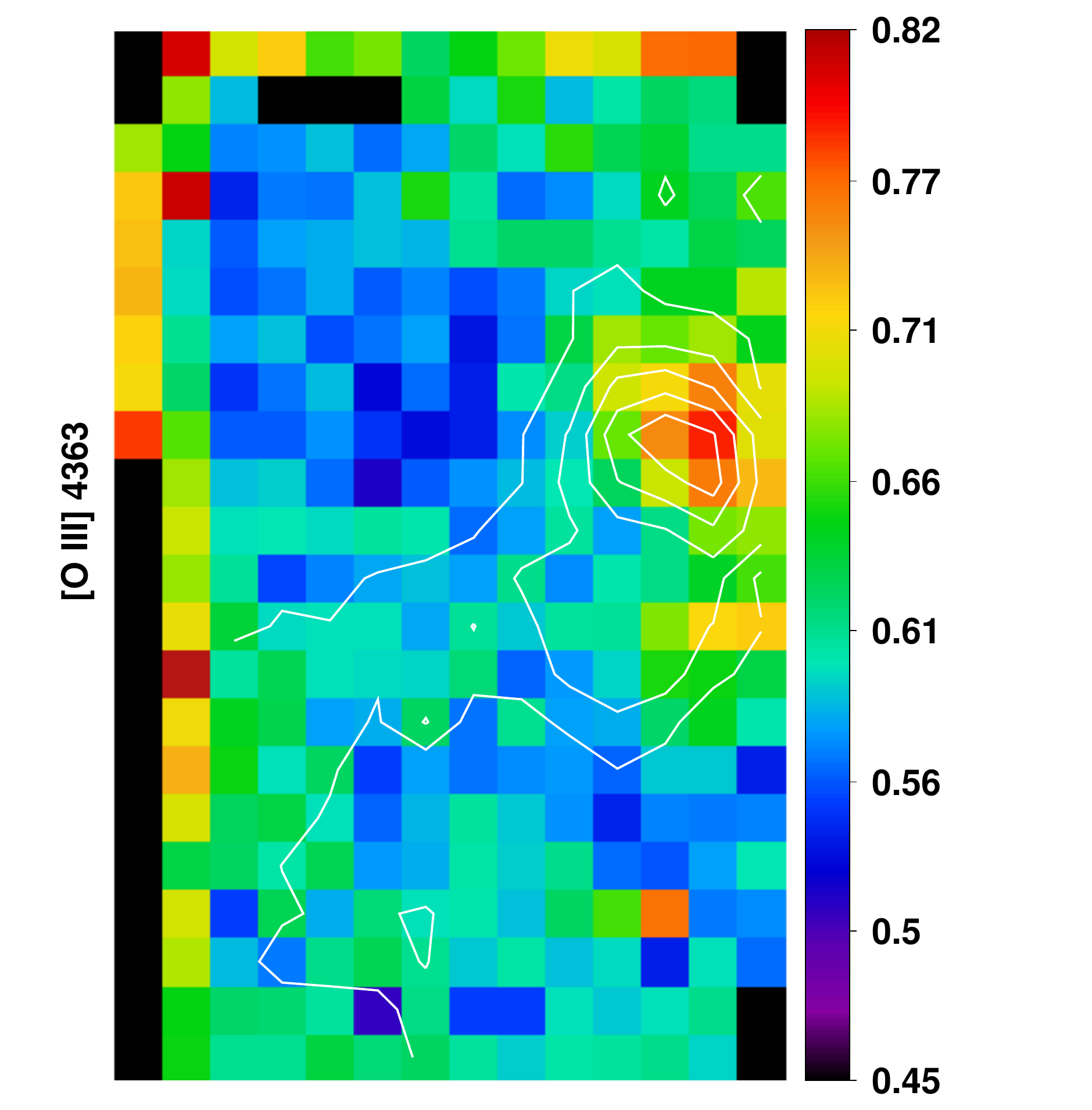}
    \includegraphics[]{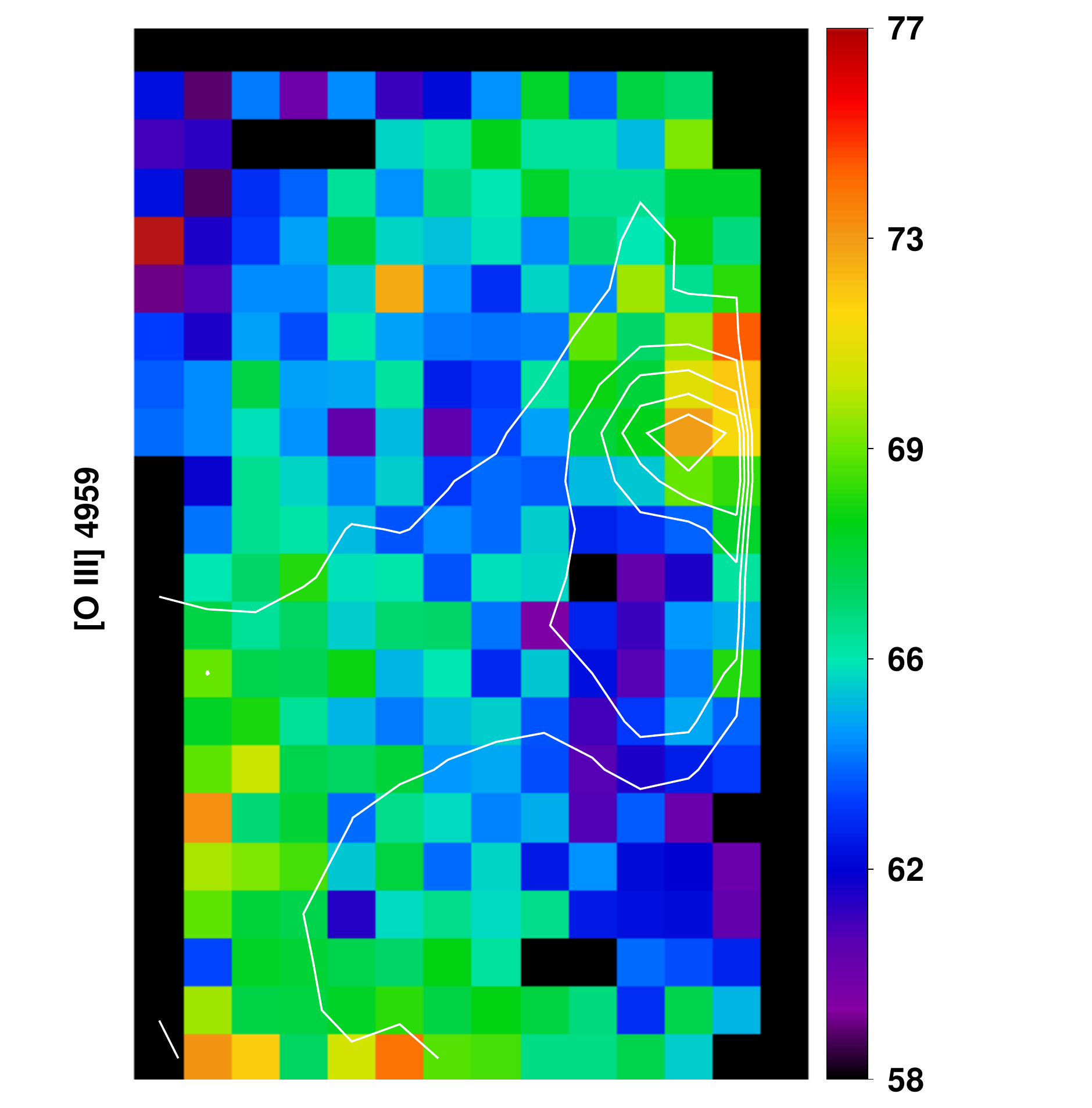}
    \includegraphics[]{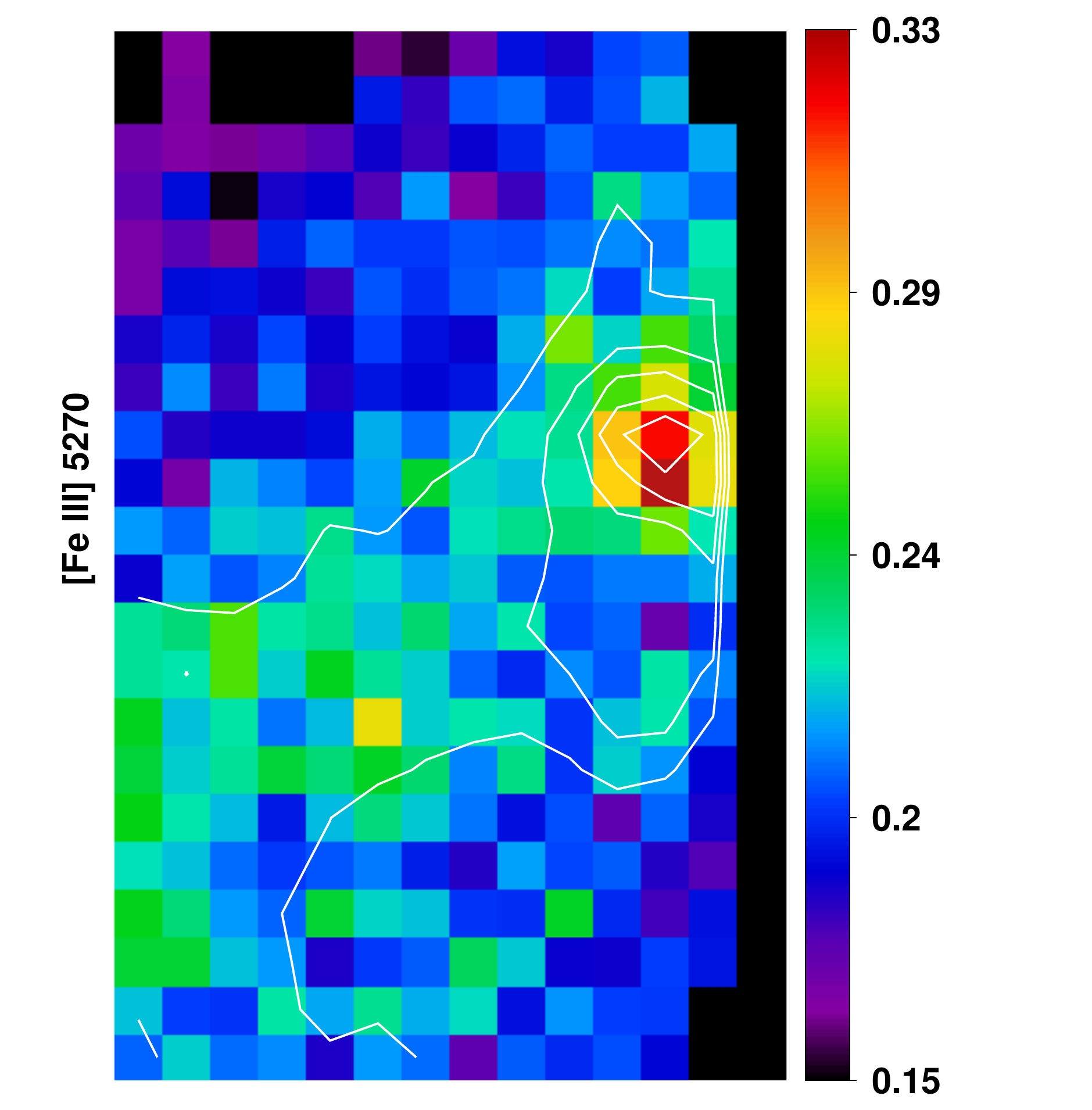}
    \includegraphics[]{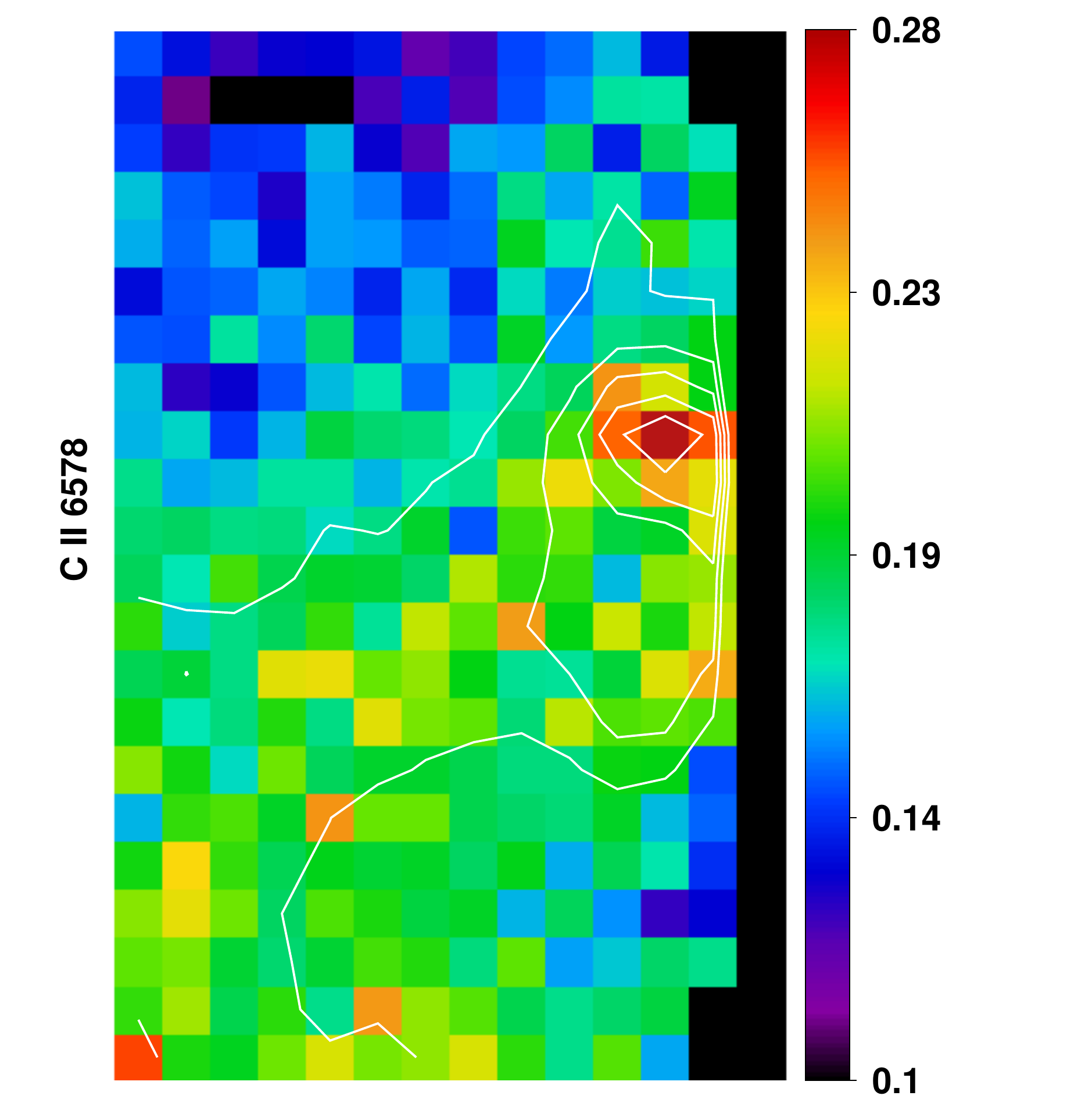}
    \includegraphics[]{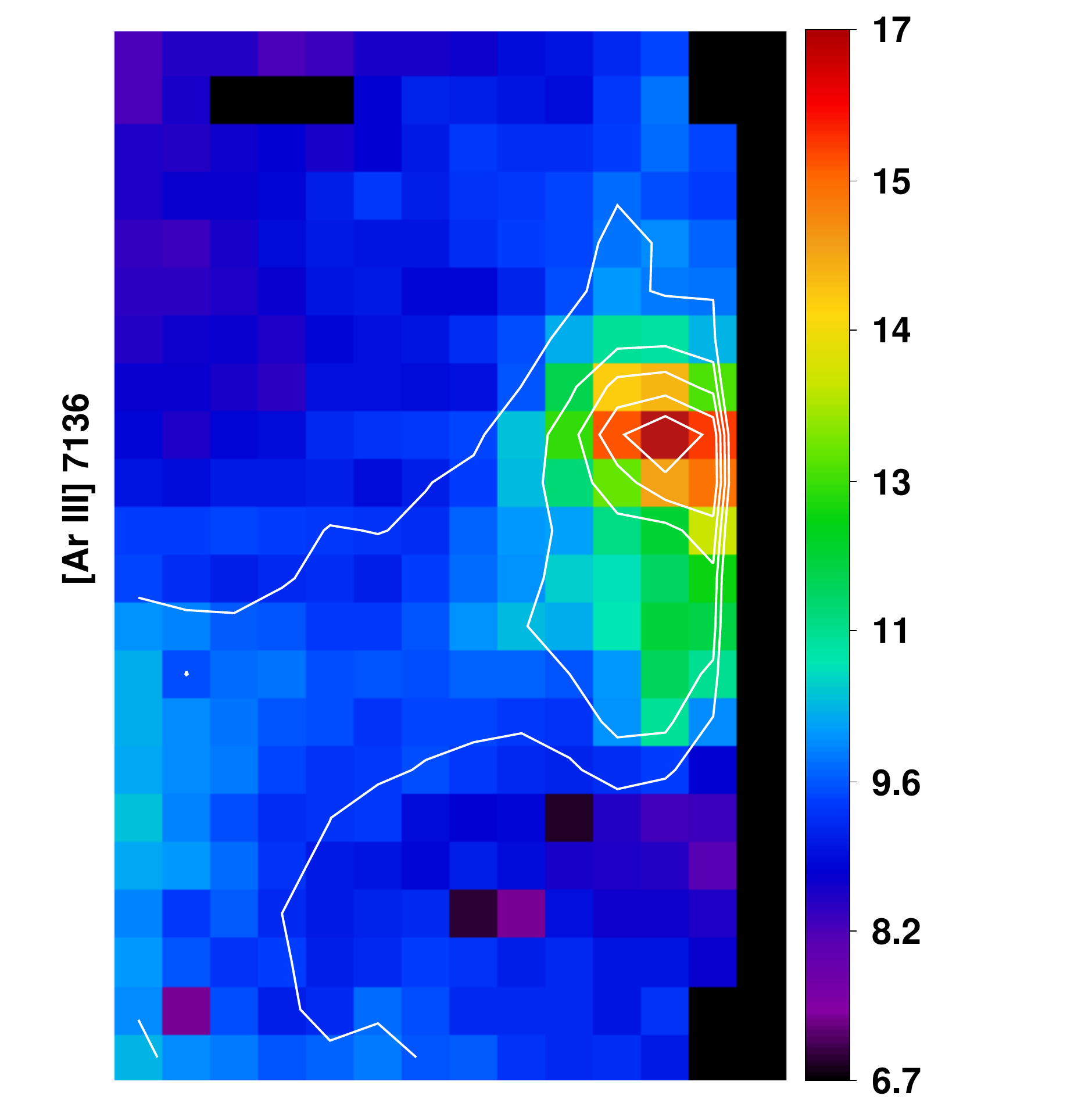}

\caption{As in the previous figure. The contours are \ha, except on the \foii\ \lam3726 and \foiii\ \lam4363 maps which show contours from \hd.}
\end{figure*}

\begin{figure*}


  \setkeys{Gin}{width=.32\linewidth}
  \setlength\tabcolsep{0pt}
  \centering
    \includegraphics[]{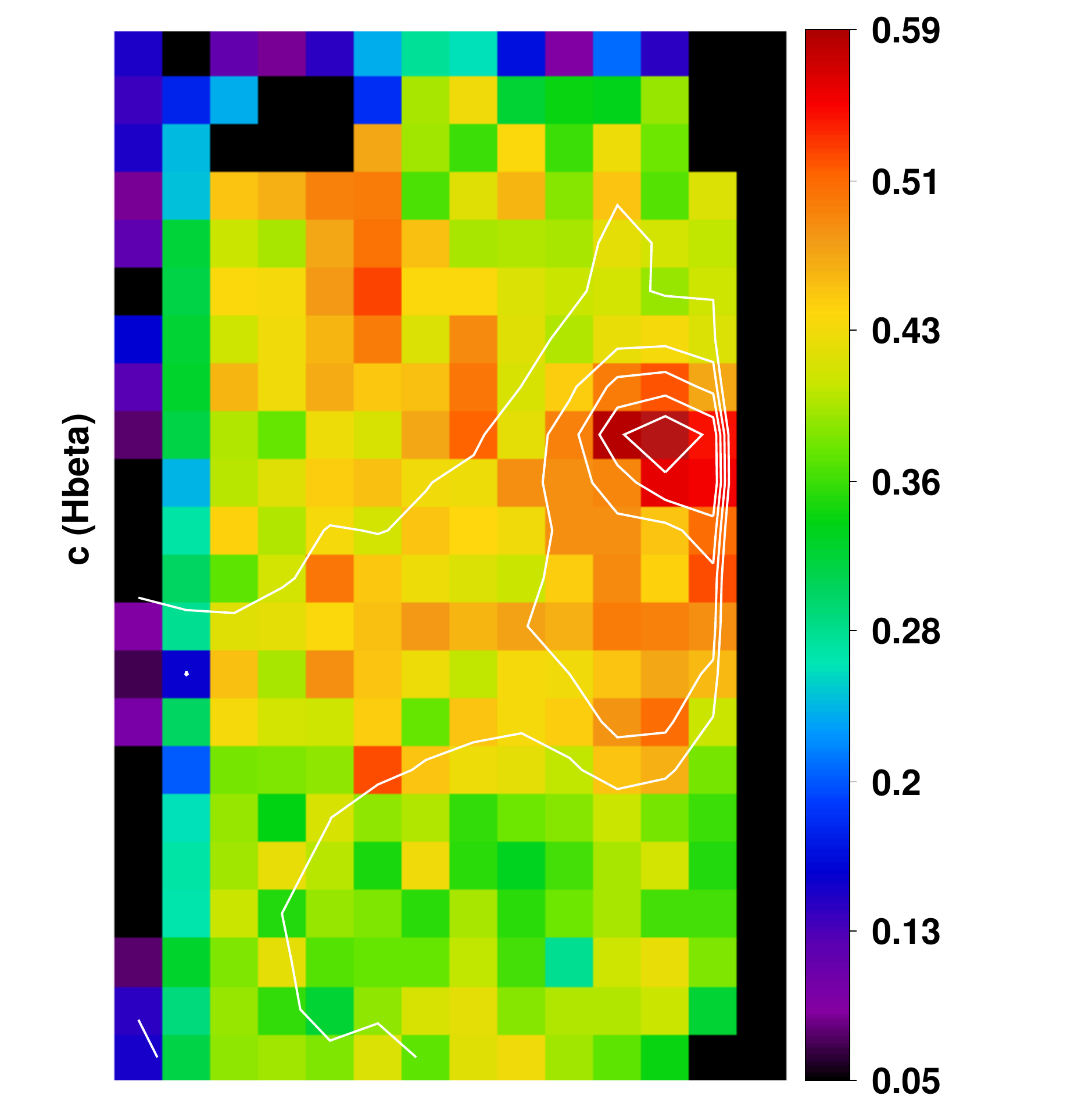}
    \includegraphics[]{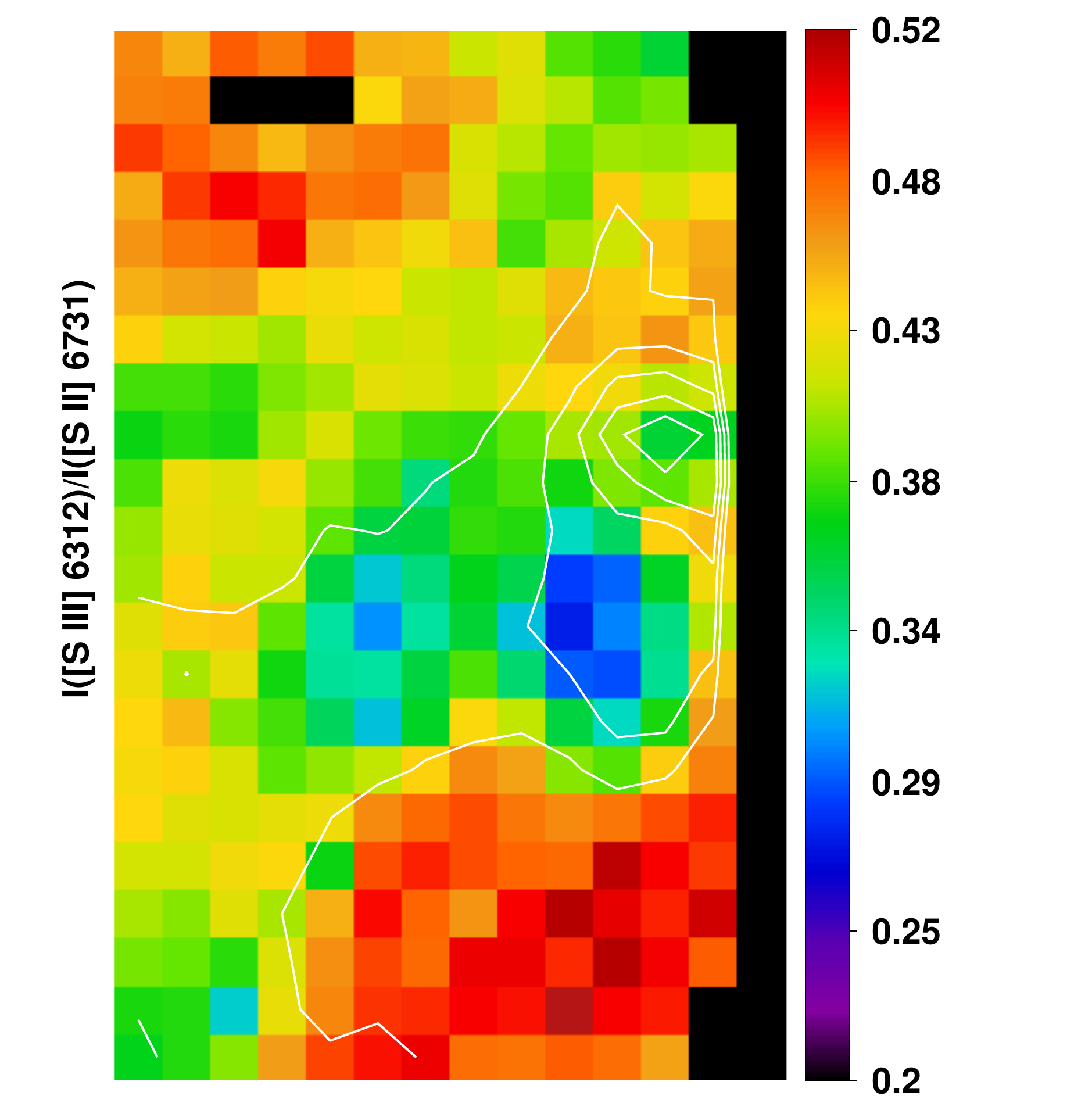}
    \includegraphics[]{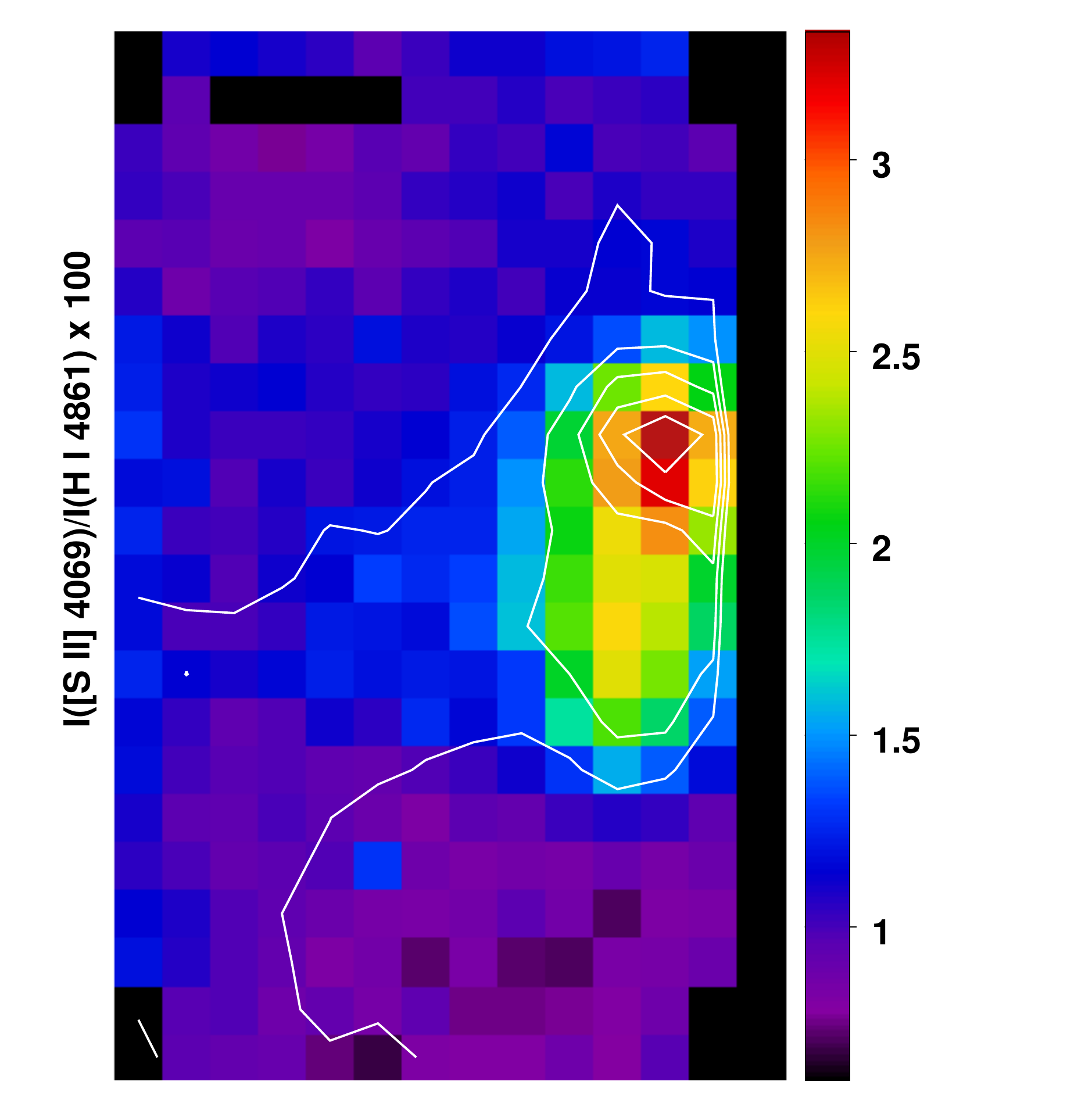}

\caption{The reddening parameter $c$(\hb) and a sample of dereddened maps: the \fsiii\ $\lambda$6312/\fsii\ $\lambda$6731 ratio and the \fsii\ $\lambda$4069/\hb\ ratio.}
\end{figure*}

\begin{figure*}

  \setkeys{Gin}{width=.32\linewidth}
  \setlength\tabcolsep{0pt}
  \centering
    \includegraphics[]{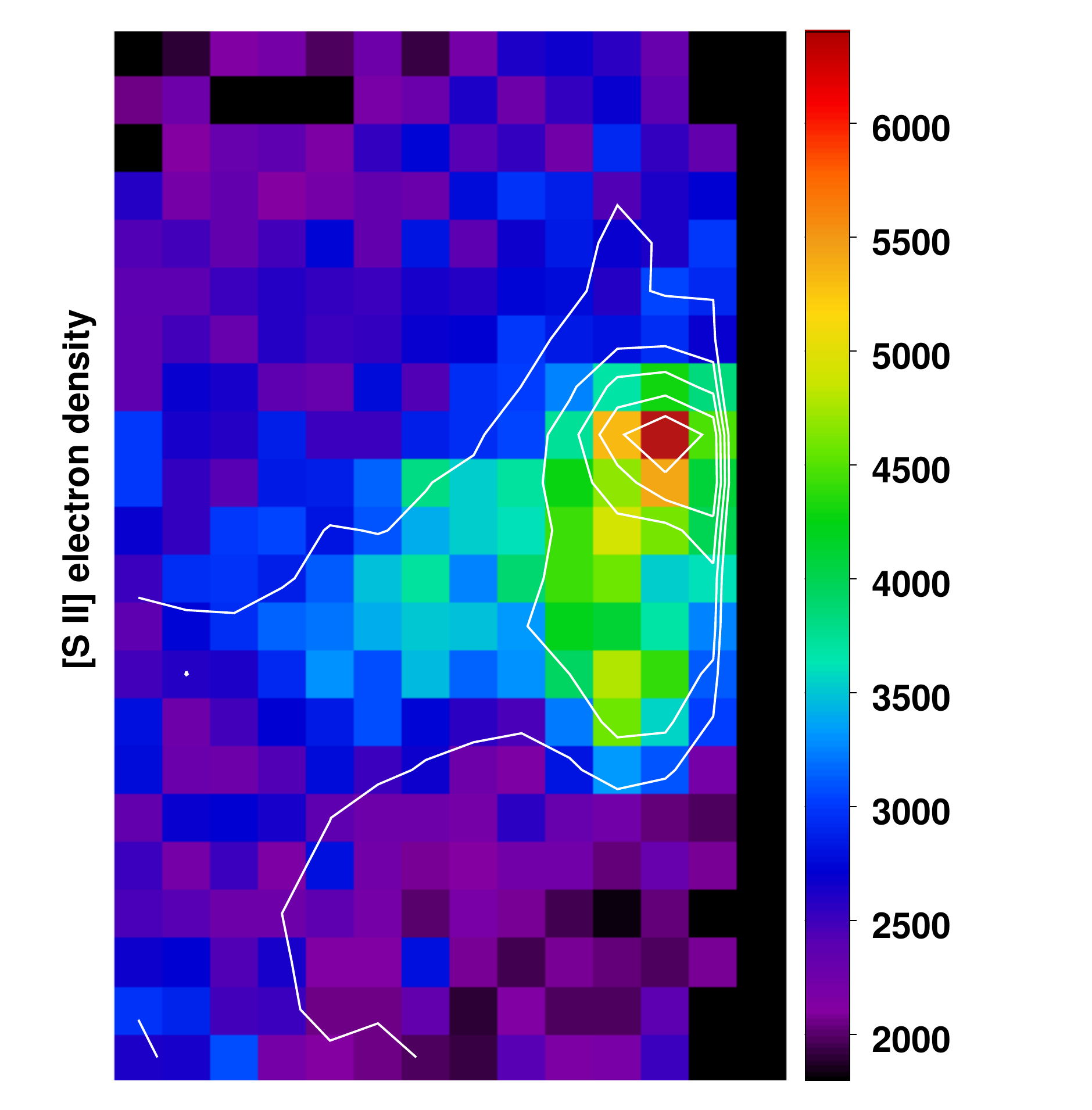}
    \includegraphics[]{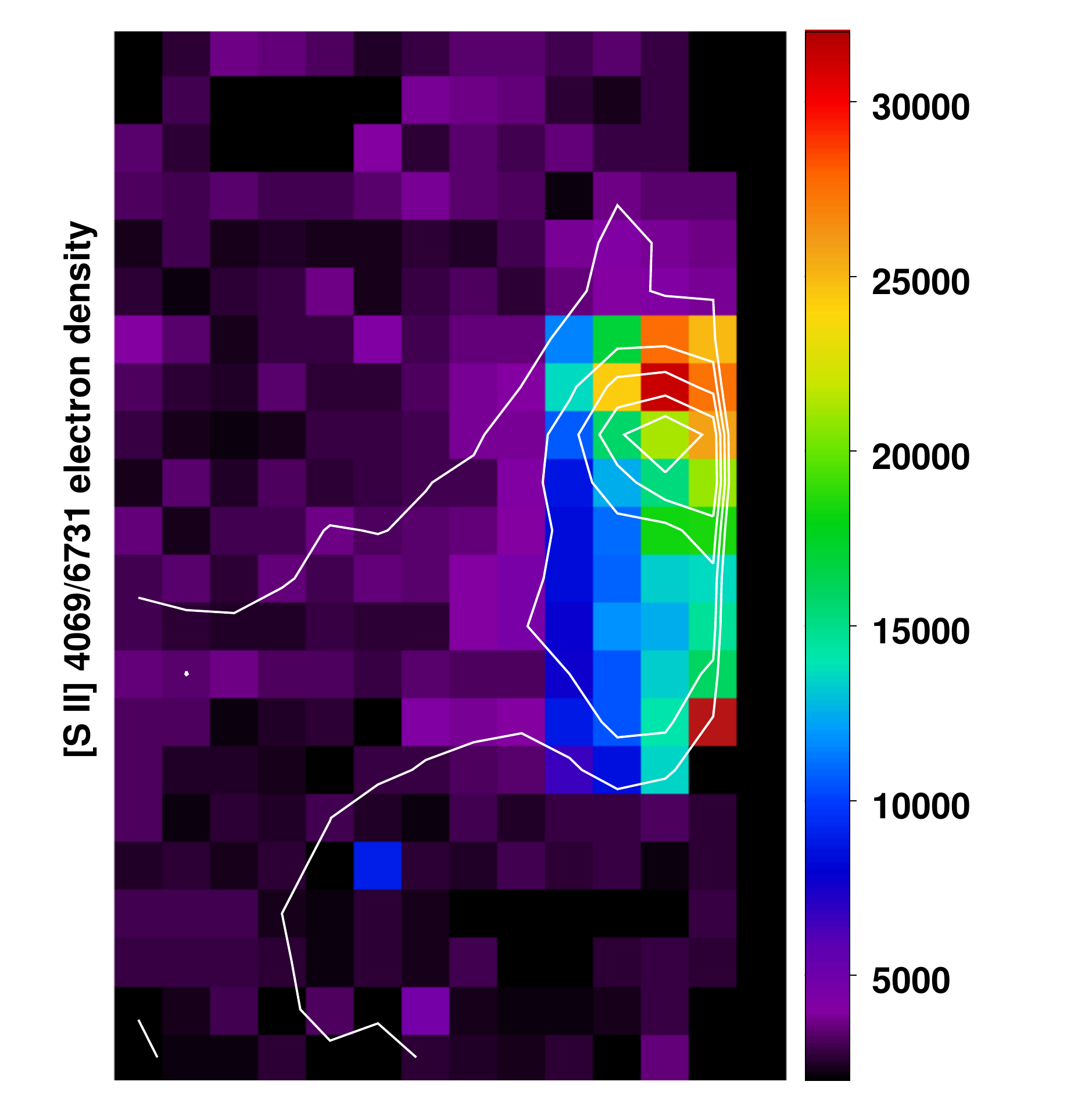}
    \includegraphics[]{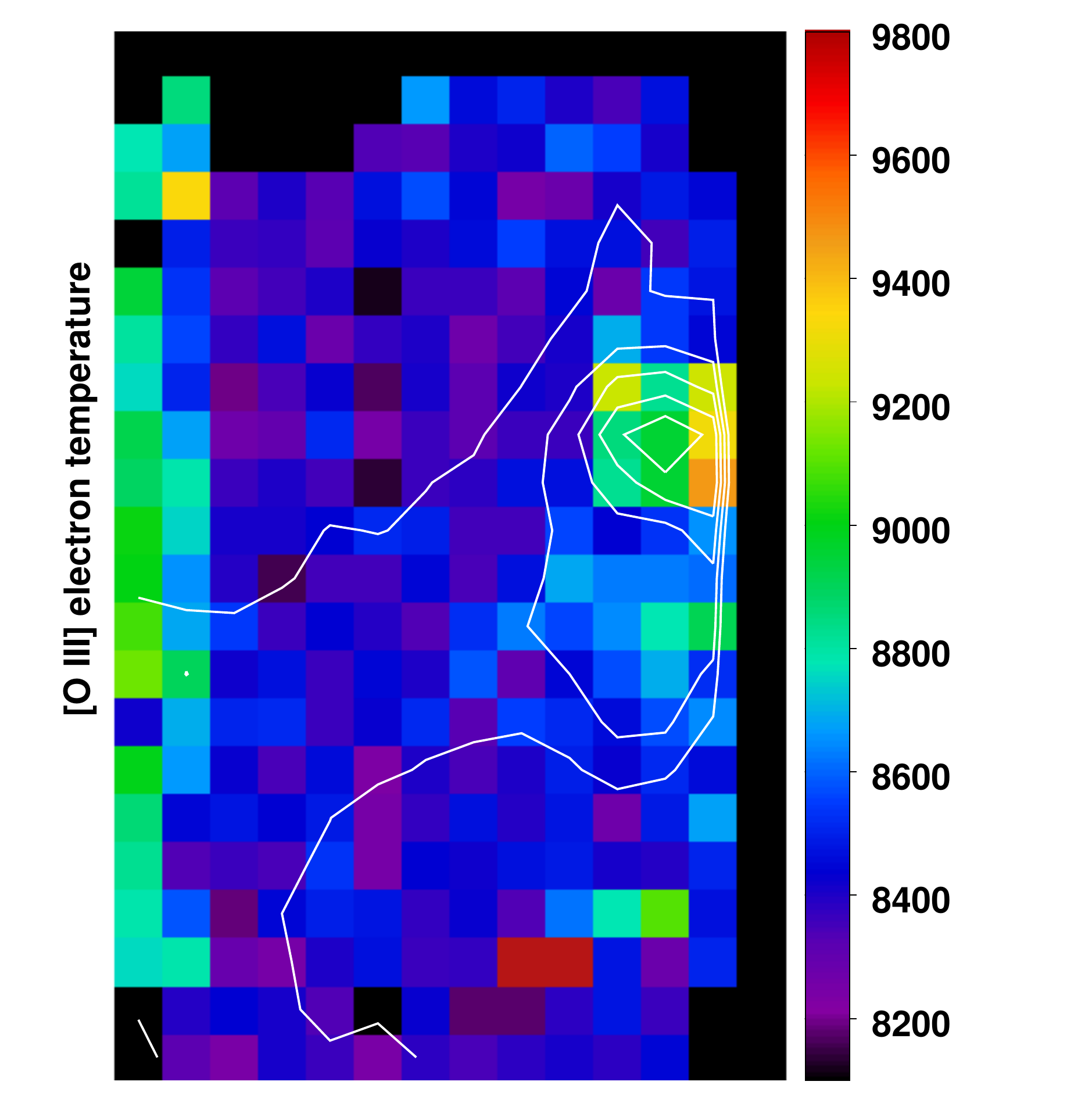}

\caption{Electron density from the \fsii\ \lam6716/\lam6731 and \lam4069/\lam6731 ratios (in \cmt), and electron temperature from the \foiii\ \lam4363/\lam4959 ratio (in Kelvin). The \lam4069/\lam6731 and \lam4363/\lam4959 diagnostic maps are based on background-subtracted fluxes over spaxels showing HST~10 emission.}
\end{figure*}

In Figs.\,~3 and 4 a collection of spectral maps of the HST~10 field is presented. These represent the observed line fluxes measured via Gaussian fitting as discussed above. The good seeing has resulted in a very good image quality, especially during the observations with the LR4, 5, 6 gratings. The proplyd shows an elongated morphology along the north-south direction resembling closely the higher resolution imagery of the \hst. A filamentary feature extending towards the southeast is evident in several lines, but is mostly prominent in \hi, \hei, \fnii\ $\lambda$6584, and particularly in \fsii\ $\lambda\lambda$6716, 6731 emission. This is probably the ionized skin of a background molecular filament that is protruding into the \hii\ region as discussed in Section 1.1. 

All lines, with the prominent exception of \foi\ $\lambda$5577 and \fni\ $\lambda\lambda$5198, 5200,\footnote{The mean flux from both \fni\ lines is shown in Fig.\,~3.} peak on the upper half (the `cusp') of the proplyd. The lines from neutral species, particularly \foi\ $\lambda$5577 (Fig.\,~1), peak at spaxels coincident with the position of the embedded circumstellar disc which has been detected in narrow band \ha\ absorption, and \foi\ $\lambda$6300 (Bally et al. 1998) and molecular hydrogen emission (Chen et al. 1998). These are the first detections of HST~10 in several other tracers, including the auroral \foiii\ \lam4363, \fsii\ \lam4069 and \fnii\ \lam5755 lines; \ffeiii\ \lam5270; \fariii\ \lam7136; and of the disc in \foi\ \lam5577. The \cii\ \lam4267 pure recombination line (RL) (Davey et al. 2000) is also detected from the cusp and is used in Section~4 as a \cpp\ abundance diagnostic.

The \ha\ and \foiii\ $\lambda$4959 maps show local minima at the position of the \foi\ $\lambda$5577 peak emission, consistent with the underlying disc being shielded from Lyman continuum photons
by the ionization front of the outer cusp.
The \fsii\ red doublet lines show peaks on both the cusp and the tail of HST~10. Most of the forbidden lines detected here are collissional lines (CLs) produced when low-lying ionic levels are thermally excited by electron impacts. Exceptions are the \foi\ lines which, near the disc, are emitted when the photo-dissociation of OH at the base of a neutral evaporated flow arising from the disc produces electronically excited atomic oxygen (St\"orzer \& Hollenbach 1998). On the other hand, at the proplyd's cusp, the emission of \foi\ \lam6300 is probably dominated by electronic collisional excitation.\footnote{The \foi\ line maps shown here contain a night-sky component which cannot be subtracted as the Argus sky allocated fibres could not be pointed to regions free from Orion nebula emission.} Also, the \fni\ doublet lines are emitted in atomic gas absorbing far-UV stellar radiation in a fluorescence process (Ferland et al. 2012). A similar fluorescence process may be affecting the \cii\ \lam6578 line (Fig.\,~4) whose utility as an abundance diagnostic is limited (e.g. Esteban et al. 2004).

\subsection{Reddening and ionization degree}

The interstellar reddening in the line of sight towards HST~10 was determined from a comparison of the observed \ha/\hb, \hd/\hb, \hi\ $\lambda$3970/\hb\ ratios with their predicted Case B recombination theory ratios from Storey \& Hummer (1995). We used the modification by Blagrave et al. (2007) of the reddening law of Cardelli, Clayton \& Mathis (1989) with a total to selective extinction ratio $R_V$ $=$ 5.5 applicable to M42. A map of the resulting logarithmic $c$(\hb) reddening parameter is shown in Fig.\,~5, where the individual results from the above line ratios were given weights of 3:1:1. The individual line ratios over the proplyd's cusp result in these mutually consistent $c$(\hb) values: 0.52 $\pm$ 0.10 (\ha/\hb), 0.49 $\pm$ 0.15 (\hd/\hb) and 0.60 $\pm$ 0.11 (\hi\ \lam3970/\hb); these are mean values over the 9 spaxels of our designated cusp area. Overall, the reddening appears to be higher over HST~10 and shows values of $\approx$0.6 at the position of the peak \ha\ emission. Our reddening determination is in good agreement with the value obtained from \hst\ image analysis by O'Dell (1998), and by O'Dell \& Yusef-Zadeh (2000) who calculated $c$(\hb) from the 20\,cm/\ha\ ratio. The latter study shows that there is a general trend of increasing reddening (0.45--0.60) from south to north across an area coincident with the Argus field of view, while our reddening map of Fig.\,~5 shows $\sim$0.4--0.55. The spatial variation is almost identical between the two techniques, whereas the absolute values of the extinction at \hb\
derived by O'Dell \& Yusef-Zadeh are 12 per cent higher. Adopting
such a higher extinction value would reduce by ~3 per cent the ratios
with respect to \hb\ of lines in the red portion of the spectrum, which
is smaller than the observational uncertainty in our measurements
(Table 2).


The enhanced $c$(\hb) at the proplyd position is not likely due to grains mixed in with the ionized gas.  This is because the dust optical depth through the ionized gas is proportional to the ionization
parameter, which for the proplyd is quite low. Our model in Section 4.2 indicates a $V$-band optical depth of only about 0.05 to the ionization front (see Fig.\,~10). Therefore most of the extra extinction compared to the Orion nebula must be due to dust between the ionization front and the surface of the embedded circumstellar disc, which should have $A_V$ of order unity. This is further supported by geometric arguments as the proplyd's axis of symmetry is tilted with respect to the line of sight by 150 degrees with the cusp facing away from the observer (Table~5). As a result, emission from the cusp suffers reddening as it passes through the intervening outflow arising from the disc.

The lower part of the Argus field containing Orion nebula emission shows lower $c$(\hb) values, by $\sim$0.3 dex.  Such moderate reddening is in overall agreement with other studies of the Orion \hii\ region which focused on Orion nebula emission (O'Dell \& Yusef-Zadeh 2000; Rubin et al. 2003; Esteban et al. 2004; Blagrave et al. 2007; Tsamis et al. 2011), proplyds (O'Dell 1998; Tsamis et al. 2011; Mesa-Delgado et al. 2012), and Herbig-Haro objects (Mesa-Delgado et al. 2009; N{\'u}{\~n}ez-D{\'{\i}}az et al. 2012).

The above $c$(\hb) map was used to convert the observed line maps to dereddened intensity maps relative to $I$(\hb), such that $I$($\lambda$)/$I$(\hb) = $F$($\lambda$)/$F$(\hb) $\times$ 10$^{c({\rm H\beta}) f(\lambda)}$, where $f$($\lambda$) is the Blagrave et al. (2007) reddening curve. The dereddened intensity ratio maps were used for the subsequent physical analysis of the HST~10 field. 

In Fig.\,~5 also the \fsiii/\fsii\ ratio is shown, which is a good indicator of the variation of the ionization degree across the field. HST~10 is a relatively low ionization structure overall. The ionization degree peaks in spaxels free from HST~10 emission in the southern part of the field. On HST~10, it shows distinct minima in the tail of the proplyd, southeast of the embedded disc, as well as across an extended area directly east of the proplyd. The former low ionization area roughly coincides with an area of polycyclic aromatic hydrocarbon emission imaged with adaptive optics at the Keck telescope (Kassis et al. 2007). The latter feature, which also displays relatively higher reddening than the nebula emission to the southwest, is the background \fsii- and \fnii-bright filament (Section 1.1).   

\subsection{Electron density and temperature}

\begin{figure}
  \setkeys{Gin}{width=1.15\linewidth}
  \setlength\tabcolsep{-100pt}
  \centering
  \begin{tabular}{l} 
        \includegraphics[]{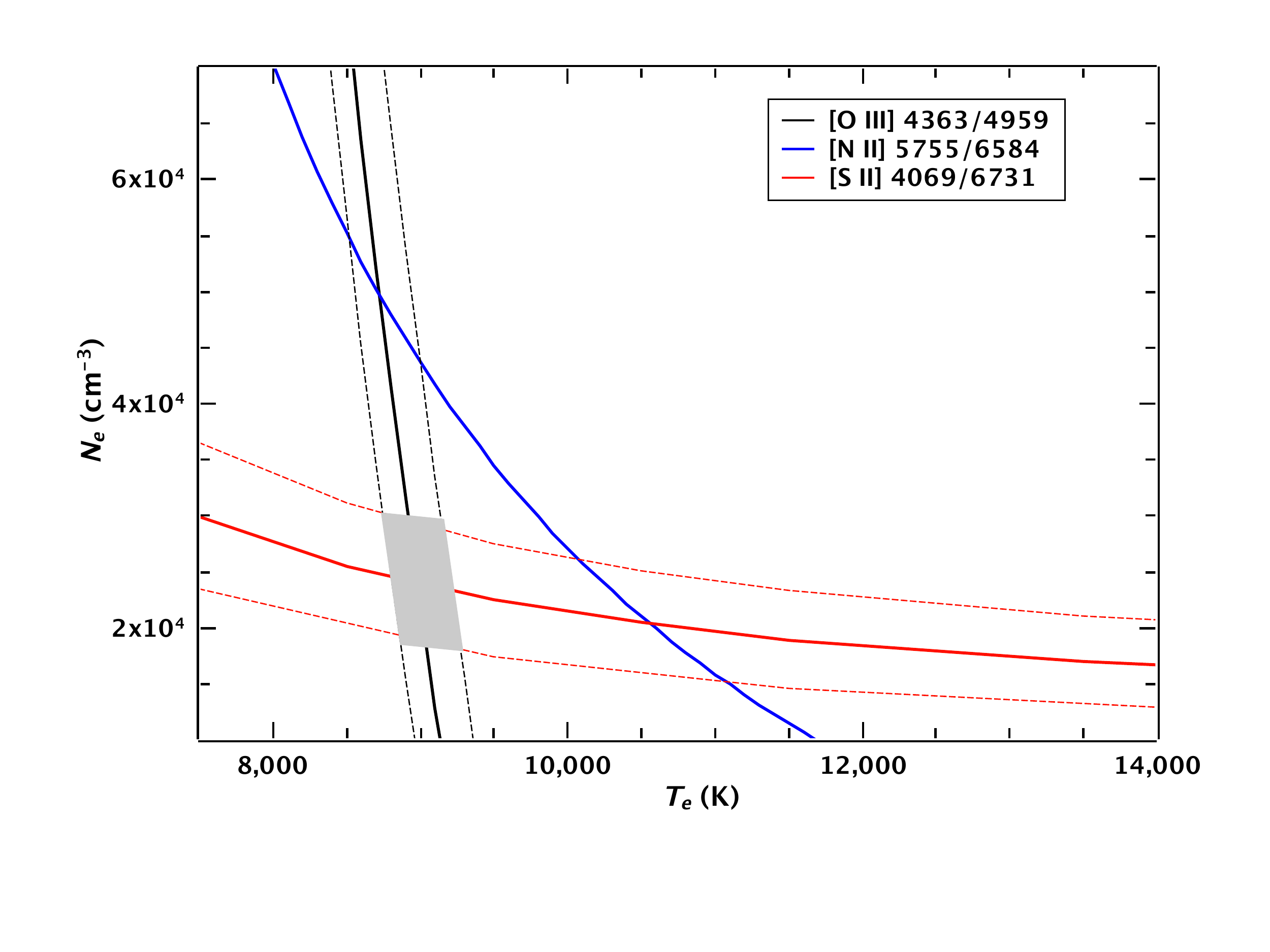}\\
  \end{tabular}
  \caption{Diagnostic diagram applicable to mean line ratio values (solid lines) over the cusp of HST~10 (9 spaxels): ($X$, $Y$) $=$ (13--15, 2--4) in Argus array coordinates, where $X$, $Y$ are respectively the long and short axes with a point of origin at the bottom right corner of the maps in Fig.\,~3. Dashed lines bracketing solid lines of the same colour denote the 1$\sigma$ range of the corresponding ratios. The equivalent range for the \fnii\ diagnostic is too large (see Table~2) and is not shown (the same holds for the \fsii\ \lam6731/\lam6716 ratio which is not included in this plot; see the text for details). The shaded area corresponds to the adopted conditions for the empirical analysis of the cusp in Section 4.1.}
\end{figure}

There is a variety of plasma diagnostics in the current dataset for the determination of the electron density, \eld, and temperature, \elt. The {\sc temden} task in {\sc iraf} v2.14 was employed. Maps of \eld\ are shown in Fig.\,~6, based on the \fsii\ \lam6731/\lam6716 doublet ratio and the \lam4069/\lam6731 ratio. In both cases the overall morphology of the proplyd is well reproduced by the observed density distribution. Based on the former ratio, the mean \eld(\fsii) is 4170 $\pm$ 730\,\cmt\ over an area within the 80 per cent \ha\ isophote. 
The \lam6731/\lam6716 ratio, however, approaches the high density limit dictated by the low critical densities (1200, 3300 \cmt) of the \fsii\ doublet upper levels, and its measured variation in the cusp of the proplyd indicates densities from $\sim$ 3000 \cmt\ to a poorly defined upper value.
The \lam6731/\lam6716 density map is therefore only shown for illustration as it is \emph{not accurate} over the proplyd area. We thus also employ the \fsii\ \lam4069/\lam6731 ratio which is sensitive to densities of up to $\sim$2 $\times$ 10$^6$ \cmt. The corresponding map, which has been computed from the background-subtracted ratio over an area of 32 spaxels covering HST~10, shows peak densities of 3.1 $\times$ 10$^4$ \cmt\ in the cusp of the proplyd.

For comparison, the proplyd density can also be estimated from the \ha\ surface brightness as follows. From the peak \ha\ emission, the dereddened brightness is $S$(\ha) $=$ 1.475 $\times$ 10$^{11}$ photon\,s$^{-1}$\,cm$^{-2}$\,sr$^{-1}$, whereas the background shows 8.53 $\times$ 10$^{10}$ photon\,s$^{-1}$\,cm$^{-2}$\,sr$^{-1}$. The emission measure ($EM$) is $EM$ $=$ \eld$^2$ $dL$ (where $dL$ is the photon path length), or 4$\pi$$S$(\ha)/$\alpha_{\rm eff}$(\ha). Substituting for the background-subtracted brightness and assuming a cusp radius or path length equal to two spaxels ($=$ 2 $\times$ 0.31 $\times$ 440 au), with $\alpha_{\rm eff}$(\ha) $=$ 10$^{-13}$ cm$^3$ s$^{-1}$ at 10\,000\,K, then \eld\ $=$ 4.45 $\times$ 10$^{4}$ \cmt. This is in fair agreement with the direct \fsii\ \lam4069/\lam6731 ratio measurement.

There is an area of increased density directly east of the proplyd that corresponds to a lower ionization area visible on the \fsiii/\fsii\ ratio image. The \eld\ of this filamentary feature is 3200 $\pm$ 300 \cmt\ over 20 spaxels based on the \lam6731/\lam6716 ratio and 3030 $\pm$ 340 based on the \lam4069/\lam6731 ratio. The agreement between the \eld\ determinations from the two ratios over spaxels free of HST~10 emission lends support to the validity of both diagnostics in the lower density regime of the local nebula. The peak density over this filament is up to a factor of $\sim$2 higher than over the background nebula in the southwest corner of the field. This is most likely a background filament projected near HST~10's position as mentioned in Section 1.1.

The \elt\ distribution is shown in the last panel of Fig.\,~6 based on the background-subtracted \foiii\ \lam4363/\lam4959 ratio over the HST~10 area. The temperature over the proplyd is 8740 $\pm$ 290\,K within the 80 per cent \ha\ isophote (28 spaxels). The \fsii\ \lam4069/\lam6731 density map has been adopted for the computation. The cusp of HST~10 shows \elt\ $=$ 9070 $\pm$ 240\,K, while the tail has 8620 $\pm$ 120\,K. The proplyd cusp is of higher temperature than the local M42 background that has \elt\ $=$ 8390 $\pm$ 100\,K. In Fig.\,~7 we show a diagnostic diagram of the \eld--\elt\ plane applicable to the cusp of HST~10. It is based on background-subtracted line ratios and demonstrates that the \fsii\ \lam4069/\lam6731 ratio is a fairly accurate density diagnostic, displaying a variation of 28 per cent about the mean at \elt\ $=$ 9070 $\pm$ 240. On the other hand, the \fnii\ ratio is substantially more sensitive to the temperature as well, showing a large variarion of 69 per cent about the mean in the same \elt\ range (Table~2), and cannot be relied upon. In the abundance analysis that follows the \fsii\ \lam4069/\lam6731 densities and  \foiii\ \lam4363/\lam4959 temperatures will be therefore adopted.

In Table~2 the mean line fluxes and physical conditions over the bright cusp of HST~10 as well as over spaxels free from proplyd emission are listed. In column 2 of Table 2 the values shown correspond to background-subtracted quantities following a straightforward extrapolation procedure, according to which spaxels close to the proplyd have been chosen as representative of the mean nebular background in front of and behind the source. The Orion nebula background shows appreciable structure down to scales of just a few spaxels ($<$ 1 arcsec). The scarcity of suitable background spaxels to the west of the proplyd might have somewhat affected the accuracy of our method. Inspection of {\it HST} images (Fig.\,~2, right panels) shows that to the west of HST~10, immediately outside the Argus field of view, the background appears as a mix of featureless nebulosity and the low ionization filament which passes behind the proplyd as discussed in Section 1.1. Features with similar characteristics are present within the Argus field covered by spaxels below and to the left/upper left of the proplyd and they have been included in our characterization of the background. \footnote{The cusp is adopted as the area enclosed within ($X$, $Y$) $=$ (13--15, 2--4) while the background is two regions below (5--6, 5--8) and (18--19, 9--12) above left of the proplyd; in Argus coordinates with point of origin the bottom right corner of the map (cf. Fig.\,~3).} Therefore, the location and number of spaxels used in the determination of the background in the proplyd's close vicinity is satisfactory for our purposes.

The quoted uncertainties in Table 2 are associated with the measured variation of a given quantity across the sampled spaxels (the rms deviation from the mean). The formal uncertainties, propagated using the reduction pipeline-generated error arrays, on the density or temperature sensitive line ratios that are important for the abundance determinations are much smaller (less than two per cent).

\begin{table*}
\begin{minipage}{80mm}
\caption{Dereddened line fluxes from the (background-subtracted) cusp of HST~10 and the local M42 background (in units
such that \hb\ $=$ 100).}
\begin{tabular}{lcc}
\noalign{\vskip3pt} \noalign{\hrule} \noalign{\vskip3pt}
					& HST~10			&M42 \\
Line              &$I$($\lambda$)  &$I$($\lambda$)              \\
\noalign{\vskip3pt} \noalign{\hrule} \noalign{\vskip3pt}
$c$(\hb)					&0.53 $\pm$ 0.04			&0.42 $\pm$ 0.06 \\
\foii\ \lam3726			    &89.0 $\pm$ 17.9			&73.3  $\pm$ 8.0			\\			
\fneiii\ $\lambda$3967     &2.54 $\pm$ 0.30           &5.21  $\pm$ 0.27                 \\
\hi\ \lam3970				&13.3 $\pm$ 0.38 			&15.8  $\pm$ 0.67 			\\
\fsii\ $\lambda$4069       &4.45 $\pm$ 0.47           &0.875 $\pm$ 0.056               \\
\hd\ $\lambda$4101         &22.1$\pm$ 0.55            &26.5 $\pm$0.72                   \\
\cii\ \lam4267			    &0.122 $\pm$ 0.048			&0.275$\pm$ 0.027  \\
\hg\ $\lambda$4340         &40.4 $\pm$ 4.6            &49.6 $\pm$0.67                 \\
\foiii\ $\lambda$4363      &0.796 $\pm$ 0.181         &1.16 $\pm$0.03                 \\
\hei\ \lam4471				&3.81 $\pm$ 0.72			&5.05 $\pm$ 0.09	\\
\oii\ \lam4649				& --						&0.130 $\pm$ 0.039	\\
\fariv\ \lam4711			&--							&0.059 $\pm$ 0.030  \\
\fariv\ \lam4740			&--							&0.066 $\pm$ 0.033	\\
\hb\ $\lambda$4861         &100.0 			          & 100.0        \\         
\foiii\ $\lambda$4959      &55.2 $\pm$ 9.4            &117.5 $\pm$3.6              \\
\foi\ $\lambda$5577        &0.193$\pm$ 0.033			& 0.495$\pm$ 0.116     \\                   
\ffeiii\ \lam5270			&0.297 $\pm$ 0.039			&0.330 $\pm$0.037   		 \\
\fcliii\ \lam5518			&0.172 $\pm$ 0.039 		    &0.400 $\pm$ 0.083	  \\	
\fcliii\ \lam5538			&0.267 $\pm$ 0.038 		    &0.480 $\pm$ 0.085		\\
\fnii\ $\lambda$5755       &2.32 $\pm$ 1.16     		&0.470 $\pm$0.049 		\\                 
\hei\ \lam5876				&10.3 $\pm$ 2.1				&13.9  $\pm$ 0.3	 \\
\foi\ $\lambda$6300        &2.82 $\pm$ 2.08           &0.535 $\pm$ 0.128	 \\                
\fsiii\ $\lambda$6312      &1.36 $\pm$ 0.39         	&1.74  $\pm$ 0.46 	\\             
\ha\ $\lambda$6563         &235.6 $\pm$ 2.2           &276.5 $\pm$ 7.9                 \\
\fnii\ $\lambda$6584       &72.1 $\pm$ 6.2            &31.7 $\pm$ 1.4 		\\           
\cii\ \lam6578			 	&0.176	$\pm$ 0.033			&0.350 $\pm$ 0.024 \\ 
\fsii\ \lam6716			    &1.82 $\pm$ 0.32			& 2.44$\pm$0.60 				\\
\fsii\ \lam6731			    &4.00 $\pm$ 0.67 			& 3.63$\pm$0.88		 \\
\fariii\ \lam7136		    &13.1 $\pm$ 1.38 			& 12.1$\pm$1.5 		\\
\noalign{\vskip3pt} 
\hb\ \lam4861$^a$				&148.7 $\pm$ 51.1    &142.0 $\pm$ 21.0 \\ 
\noalign{\vskip3pt} 
\elt(\foiii) (K)				&9070 $\pm$ 240 				&8390 $\pm$ 100 \\
\eld(\fsii) (\cmt)			&21560 $\pm$ 6227			& 2860 $\pm$  400 \\
\eld(\fnii) (\cmt)			&46785 $\pm$ 32281			& -- \\


\noalign{\vskip3pt} \noalign{\hrule}\noalign{\vskip3pt}
\end{tabular}
\begin{description}
\item[$^a$] In units of 10$^{-15}$ erg s$^{-1}$ cm$^{-2}$ spaxel$^{-1}$. 
The cusp area comprises 9 spaxels and the background area $\sim$20 spaxels. 
\end{description}
\end{minipage}
\end{table*}

\section{Chemical element abundances}

We employ two complementary approaches to determining the gas-phase
elemental abundances.  The first is the traditional semi-empirical method
(e.g. Seaton 1960), which in its simplest form assumes that each ion is emitted from a homogeneous
zone with a constant electron density and temperature, and employs
ionization correction factors (e.g. Peimbert \& Costero 1969; Kingsburgh \& Barlow 1994) to account for unobserved
ionization stages. The second is to construct self-consistent
physical models of the proplyd ionization front and photoevaporation
flow (e.g. Henney \& O'Dell 1999; Henney et al.\ 2002, 2005; Mesa Delgado et
al. 2012) that satisfy existing observational constraints from
high-resolution spectroscopy and \hst\ imaging, and in which the
abundances are adjusted so as to best reproduce the observed line
spectrum.  The second method has the great advantage that it can
account for the fact that the physical conditions may vary greatly,
even \emph{within} the zone of emission of a single ion.  It also
obviates the need for the use of ionization correction factors since
the full ionization structure of each element is calculated from first
principles, using the Cloudy microphysics code (Ferland at el. 1998).
On the other hand, it also has various disadvantages.  The models are
computationally costly to calculate, and there is no guarantee that a
model will be found that can simultaneously reproduce all of the
observations.  In the case of several models that can each reproduce
most, but not all, of the observed emission line spectrum, there is no
obvious criterion for deciding which is ``best''. 

\subsection{Empirical abundance determination}

\begin{figure*}




  \setkeys{Gin}{width=.3\linewidth}
  \setlength\tabcolsep{0pt}
  \centering
    \includegraphics[]{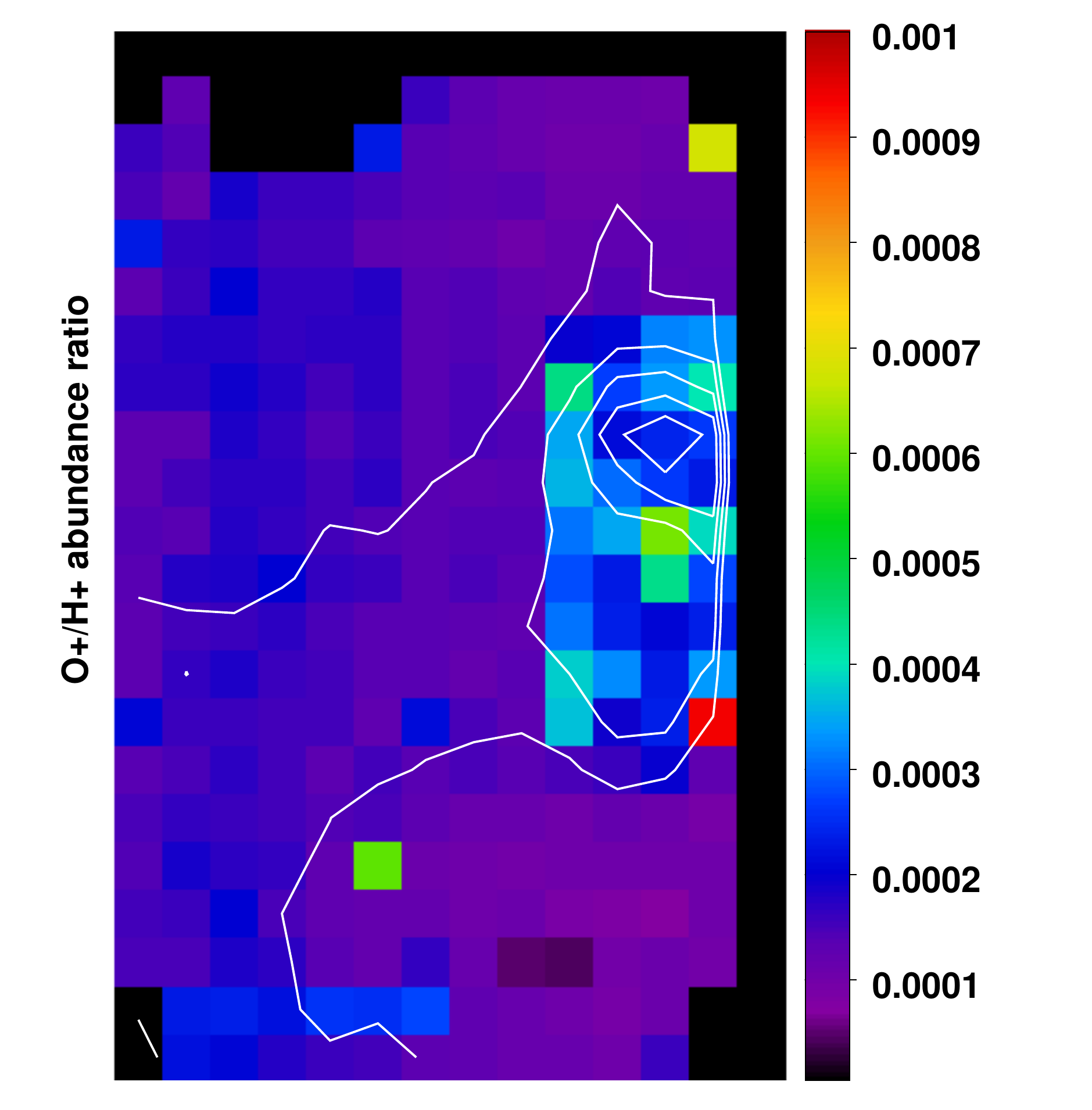}
    \includegraphics[]{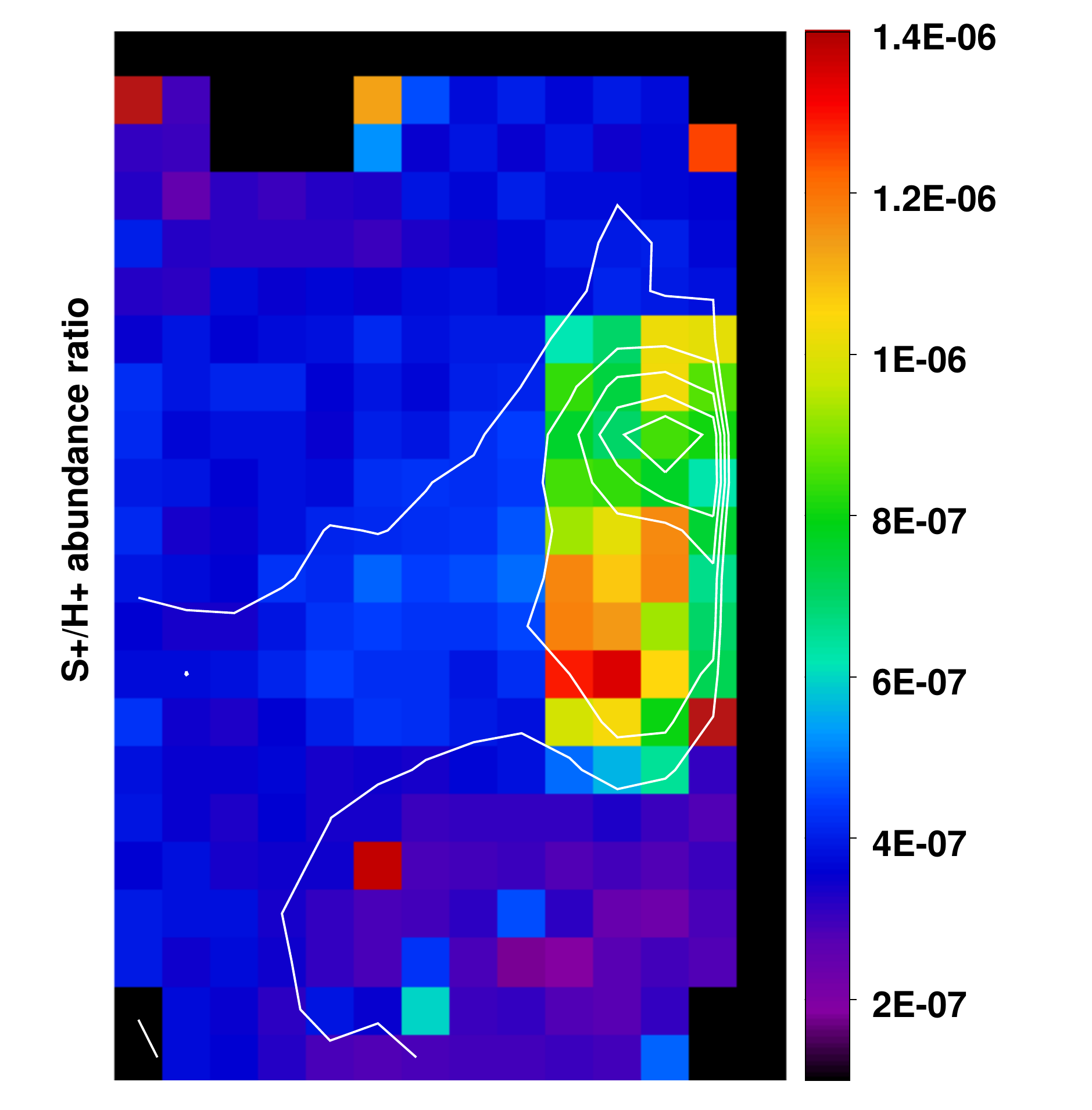}
    \includegraphics[]{map_o3tefc-eps-converted-to.pdf}
    \includegraphics[]{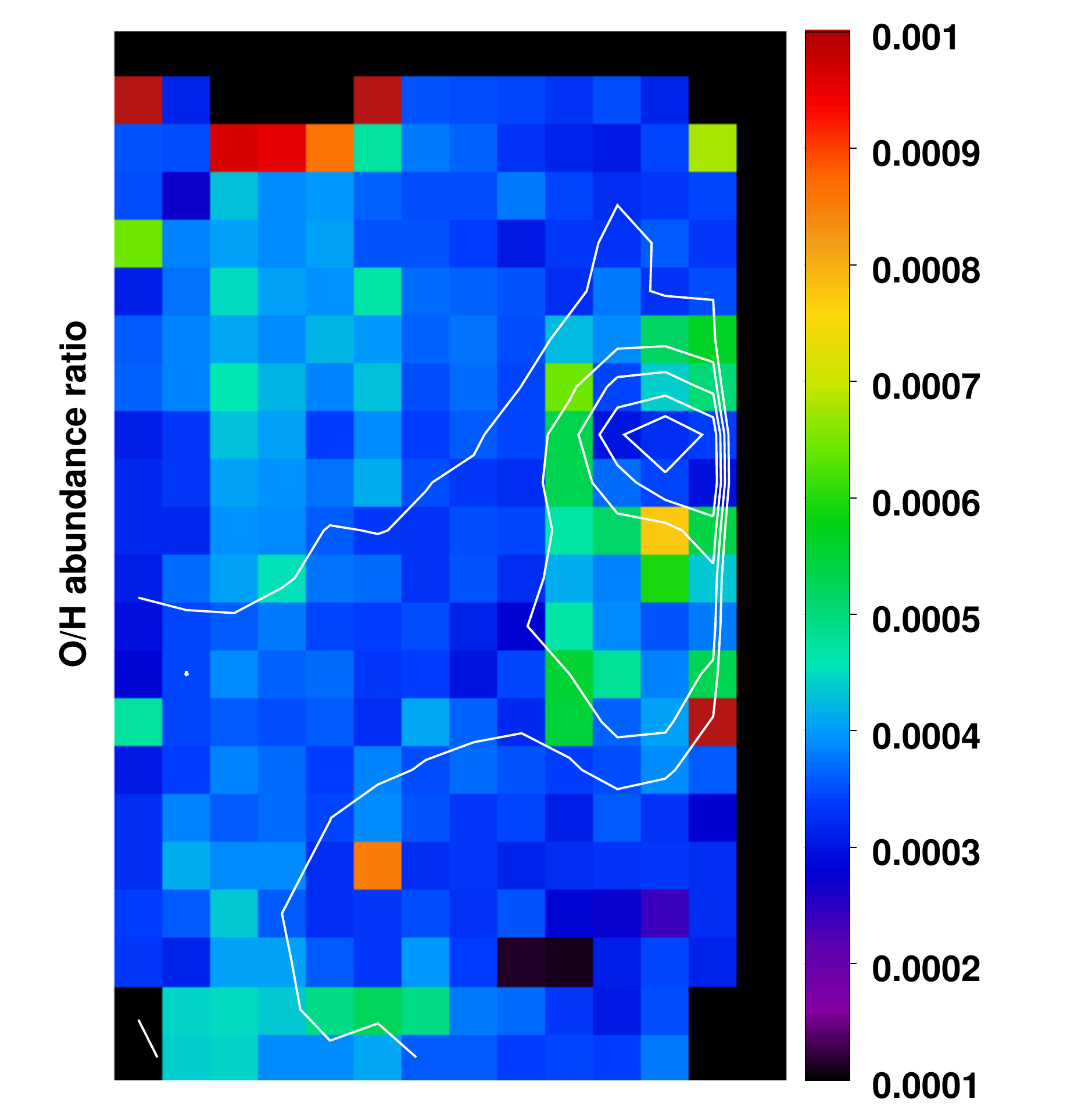}
    \includegraphics[]{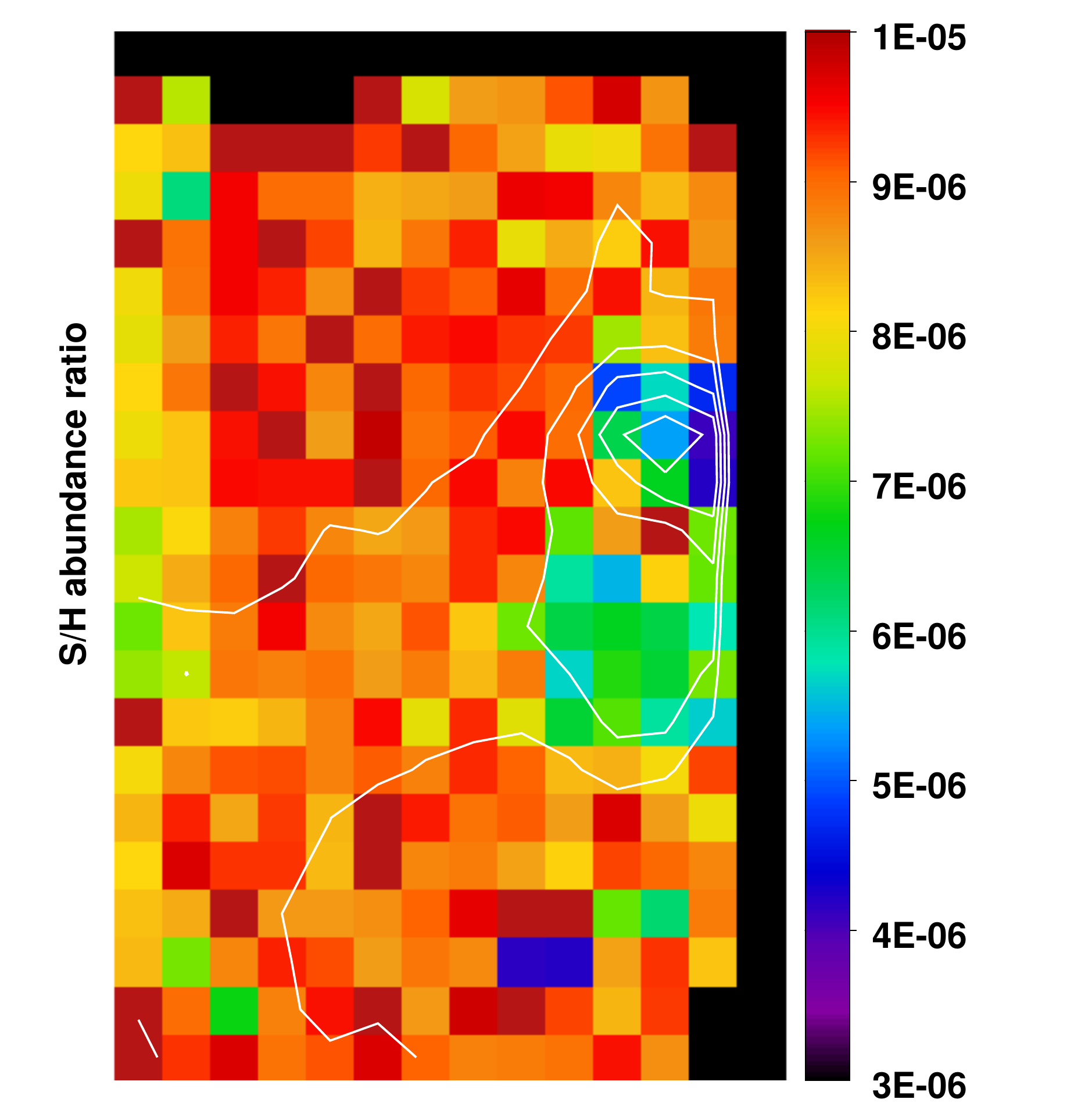}
    \includegraphics[]{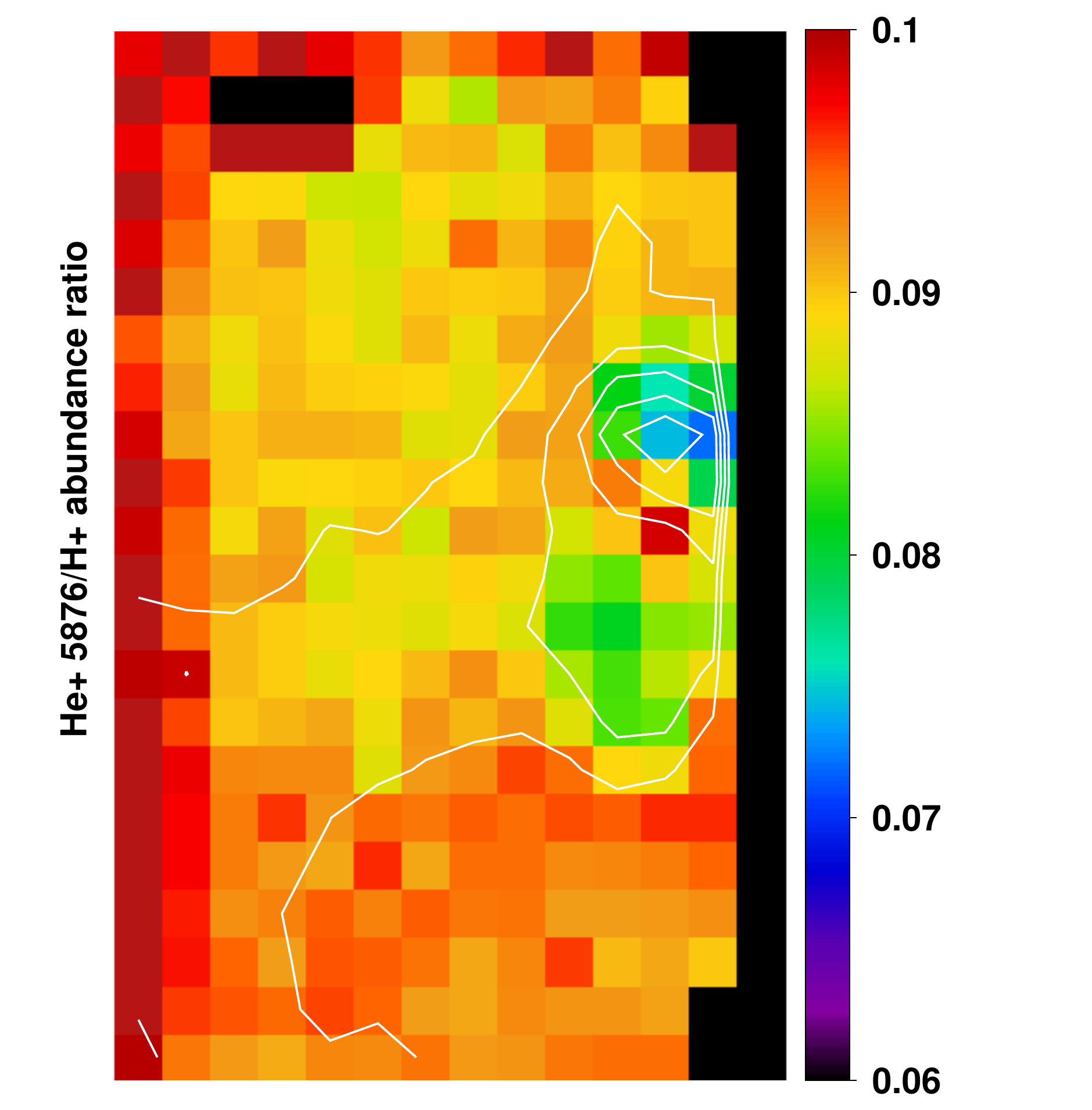}

\caption{Abundance maps for \op/\hp, O/H, \sulp/\hp, S/H and \hep/\hp\ (see the text for details).}
\end{figure*}

\begin{table*}
\begin{minipage}{90mm}
\caption{Gas-phase ionic and element abundances for HST~10 and the Orion nebula (in decimal units for He and units of log\,H $=$ 12 for the other species).}
\begin{tabular}{lcc}
\noalign{\vskip3pt} \noalign{\hrule} \noalign{\vskip3pt}
Ion            & M42				& HST~10 			  \\      
ratio				    & $t^2 = 0$			& $t^2 = 0$				\\
				    & this work			&this work			 \\
\noalign{\vskip3pt} \noalign{\hrule} \noalign{\vskip3pt}

\cpp/\hp\ $^a$			& 8.40 $\pm$ 0.04 		&8.06 $\pm$ 0.14			 \\
\np/\hp				 &7.00 $\pm$ 0.02 		&7.37 $\pm$ 0.03 					\\
\op/\hp              &8.13 $\pm$ 0.05     &8.45 $\pm$ 0.08             	\\ 
\opp/\hp             &8.38 $\pm$ 0.03     &7.98 $\pm$ 0.08             	\\ 
\opp/\hp\ $^a$       &8.54 $\pm$ 0.05 	     &--		                  	\\ 
\nepp/\hp            &7.58 $\pm$ 0.02     &7.16 $\pm$ 0.05             	\\
\sulp/\hp\           &5.51 $\pm$ 0.08      &5.90 $\pm$ 0.06             \\
\sulpp/\hp\          &6.95 $\pm$ 0.13      &6.68 $\pm$ 0.11            \\
\clpp/\hp			     &5.02 $\pm$ 0.07	&4.78 $\pm$ 0.06	   			 \\
\arpp/\hp			     &6.24 $\pm$ 0.05	&6.18 $\pm$ 0.05	   			 \\
\arppp/\hp\          &4.56 $\pm$ 0.18 	       &--				       		\\
\fepp/\hp\            &5.72 $\pm$ 0.05     & 5.59 $\pm$ 0.05      \\ 

\noalign{\vskip3pt}
\end{tabular}

\begin{tabular}{lcccc}
\noalign{\vskip3pt} \noalign{\hrule} \noalign{\vskip3pt}
Element            & M42				& HST~10 			& HST~10     	&M42 	\\      
ratio				    & $t^2 = 0$			&  $t^2 = 0$		& model			&$t^2 = 0.022$	\\	
					& this work			&this work			&this work		& E04$^b$  \\
					\noalign{\vskip3pt} \noalign{\hrule} \noalign{\vskip3pt}

He/H				 & 0.097 $\pm$ 0.003	& 0.095 $\pm$ 0.012  &0.095	&	0.097  \\
C/H					 &8.59 $\pm$ 0.04 		&8.66 $\pm$ 0.14		&8.48	&	8.42 $\pm$ 0.02	\\ 
N/H					 &7.46 $\pm$ 0.04 		&7.48 $\pm$ 0.05		&7.62	&	7.73 $\pm$	0.09\\
O/H					 &8.56 $\pm$ 0.03  	&8.63 $\pm$ 0.08		&8.63	&	8.65 $\pm$	0.03	\\
Ne/H				 &7.78 $\pm$ 0.04 		&7.75 $\pm$ 0.04		&8.16	&    8.05 $\pm$	0.07 \\
S/H                  &6.98 $\pm$ 0.10 	    &6.75 $\pm$ 0.09  &6.75   &   7.22 $\pm$	0.04	\\ 
Cl/H				 &5.07 $\pm$ 0.08 		&4.90 $\pm$ 0.08		&5.00	&	5.46 $\pm$	0.04 \\
Ar/H				 &6.27 $\pm$ 0.04     &6.25 $\pm$ 0.05		&6.30	&	6.62 $\pm$  0.05 \\ 
Fe/H				 &6.08 $\pm$ 0.05   	& 5.70 $\pm$ 0.05	    &5.78   &   5.99 $\pm$ 0.10 \\




\noalign{\vskip3pt} \noalign{\hrule}\noalign{\vskip3pt}
\end{tabular}
\begin{description}
\item[$^a$] Value derived from recombination lines.
\item[$^b$] From Esteban et al. (2004) for a nebula position 25$''$ south and 10$''$ west of $\theta^1$\,Ori~C ($\simeq$ 45$''$ northwest from HST~10).
\end{description}
\end{minipage}
\end{table*}





The density and temperature distributions derived in Section 3.3 were used in conjunction with the dereddened line ratio maps to compute ionic and elemental abundance maps relative to hydrogen. The {\sc ionic} task in {\sc iraf} v2.14 was implemented for this purpose. The \foiii\ \lam4363/\lam4959 temperature and \fsii\ \lam4069/\lam6731 density maps were adopted for the calculation. The \lam\lam6716$+$6731, 3726$+$3729, 3967, 6584, 7136, 4959, 6312 forbidden line ratios relative to \hb\ were respectively used to compute \sulp/\hp, \op/\hp, \nepp/\hp, \np/\hp, \arpp/\hp, \opp/\hp, and \sulpp/\hp\ ratios. Background-subtracted line intensities were adopted over the HST~10 area bounded by the 80 per cent \ha\ isophote.

Mean values of the ionic and total abundances for the HST~10 cusp area and for the local M42 background from spaxels clear of HST~10 emission are included in Table~3, where they are compared with the values from our best-fitting photoionization model described in Section~4.2. 


\subsubsection{Oxygen and sulfur}

The abundances for oxygen and sulfur should be well determined as they do not require the use of ionization correction factors. The amounts of O$^{3+}$  and S$^{3+}$ in Orion are estimated to be infinitesimally small. The results are shown in Fig.\,~8. The O/H abundance map is the sum of the \op/\hp\ and \opp/\hp\ ratios. The abundance of oxygen is dominated by that of \op\ which makes up $\approx$80 per cent of O/H in the proplyd's cusp. On the other hand, \sulpp\ makes up $\approx$80 per cent of S/H over the same area. Similar percentages apply in the tail area where again the dominant oxygen ion is \op\ and the dominant sulfur ion is \sulpp. The oxygen abundance is 12$+$log(O/H) $=$ 8.63 $\pm$ 0.08 and the abundance of sulfur is 6.75 $\pm$ 0.09 over HST~10. These are the mean values for the cusp area where the background-subtraction is considered to be more accurate than over the fainter tail. For the local M42 background we find 8.56 $\pm$ 0.03 and 6.98 $\pm$ 0.10 for oxygen and sulfur respectively. 

The oxygen abundance derived for the local Orion nebula is practically the same as the one measured in the vicinity of LV\,2 right within the Trapezium (Tsamis et al. 2011), and to the one measured by Esteban et al. (2004) at a position southwest of that (for a temperature fluctuation parameter, $t^2$ $=$ 0). Sulfur in HST~10 is 0.2 dex lower than in its Orion vicinity or just 0.1 dex lower than the Orion nebula benchmark value based on IR data (Rubin et al. 2011); this indicates that sulfur is not depleted onto dust grains to a degree substantially greater than in the Orion nebula.

The gas-phase abundance of oxygen in the proplyd is not significantly different from that in the Orion \hii\ region, and this is borne out from the model results too. Oxygen in HST~10 is equal within the uncertainties to the solar photospheric value of 8.69 $\pm$ 0.05 (Asplund et al. 2009), and to the mean value in B-type Orion OB1 stars of 8.74 $\pm$ 0.04 (Simon-Diaz 2010; Nieva \& Simon-Diaz 2011). Adding the small amount of oxygen ($\sim$8.08) estimated to be in grains would bring the total oxygen at $\sim$8.74, which is equal to the protosolar value (Asplund et al. 2009) or the value based on B-stars in the solar vicinity (Nieva \& Przybilla 2012).

\subsubsection{Helium}

The \hep/\hp\ abundance ratio, which represents a measure of He/H in low ionization objects such as HST~10 where \hepp\ is non-existent, was measured using the \lam4471 and \lam5876 lines, adopting effective recombination coefficients from Smits (1996) and correcting for the effects of collisional excitation using the formulae in Benjamin, Skillman \& Smits (1999). In both maps we measure a decreased helium abundance over the cusp of the proplyd by $\sim$13 per cent compared to the Orion background (Fig.\,~8). Unrealistically low densities and/or high temperatures over HST~10 need to be assumed before the He abundance begins to approximate that in the background nebula; therefore the discrepancy is not due to the adopted physical conditions.

In our model of HST~10's cusp discussed in Section 4.2 this behaviour can be explained as it is shown that in HST~10 the ionization of He lags considerably behind that of hydrogen. At the ionization front, defined where H is 50 per cent ionized, He is only about 10 per cent ionized. The He abundance maps therefore show that significant amounts of neutral helium are present in the \hp\ layer of the proplyd thus rendering the direct abundance determination of He inaccurate. Using the ionization correction factor (equation 2 in Tsamis et al. 2011) by Peimbert, Torres-Peimbert \& Ruiz (1992) to account for He$^0$ we estimate an upward correction of 14 per cent for He in the cusp (the correction for He$^0$ in the nebula is just 2 per cent).  The resulting value agrees with that obtained from our detailed model according to which He/H $=$ 0.095 in the cusp. This brings the HST~10 and Orion nebula helium abundances into very good agreement.


\subsubsection{Nitrogen and neon}

The abundances of nitrogen and neon were computed using the ionization correction factors (ICFs) by Kingsburgh \& Barlow (1994) and Peimbert  \& Costero (1969). For both elements we find no difference in their abundances between the proplyd and the nebula when comparing the ICF method determinations. However, our model of HST~10 (cf. Section~4) calls for a nitrogen abundance 0.14 dex higher than the one derived using the classic ICF, where N/O $=$ \np/\op. According to the model, the N abundance in HST~10 is 0.2 dex lower than in B-type Ori OB1 stars (Nieva \& Simon-Diaz 2011).

There is a substantial difference between the Ne abundance for HST~10 derived by the ICF method and the one obtained by the simulations. Our best fitting model (Section~4) has a neon abundance of 12 $+$ log(Ne/H) $=$ 8.16, i.e. 0.4 dex higher than the one derived using the classic ICF which calls for Ne/O $=$ \nepp/\opp. The ICF used to convert \nepp/\hp\ to Ne/H for HST~10 is already fairly large ($=$ 3.9) on account of the low ionization degree of the proplyd. A similar discrepancy was noted by Simon-Diaz \& Stasinska (2011) whose model of the Orion nebula position observed by Esteban et al. (2004) calls for a neon abundance 0.2 dex higher than the one derived using the classic ICF. On the other hand, in models of the much higher ionization degree 30 Doradus region, Tsamis \& P\'equignot (2005) obtained neon abundances that were only $\sim$20 per cent higher than those based on the ICF method. 

These results indicate that the ICF method will systematically underestimate Ne/H in the low ionization conditions prevailing in structures such as HST~10. Our model supports a neon abundance for HST~10 slightly higher than the photospheric solar value of 7.93 $\pm$ 0.10 by Asplund et al. (2009). If the Orion nebula benchmark abundance for neon of 8.01 $\pm$ 0.01 from the temperature and reddening insensitive IR \fneii\ and \fneiii\ Spitzer measurements is adopted (Rubin et al. 2011), then HST~10 is slightly neon-rich compared to the Orion nebula (by 0.15 dex). However, compared to B-type stars in Orion (Nieva \& Simon-Diaz 2011) the neon abundance from the model is equal within the uncertainties.

\subsubsection{Carbon}

\begin{figure}
  \setkeys{Gin}{width=1.3\linewidth}
  \setlength\tabcolsep{-70pt}
  \centering
  \begin{tabular}{l} 
    \includegraphics[]{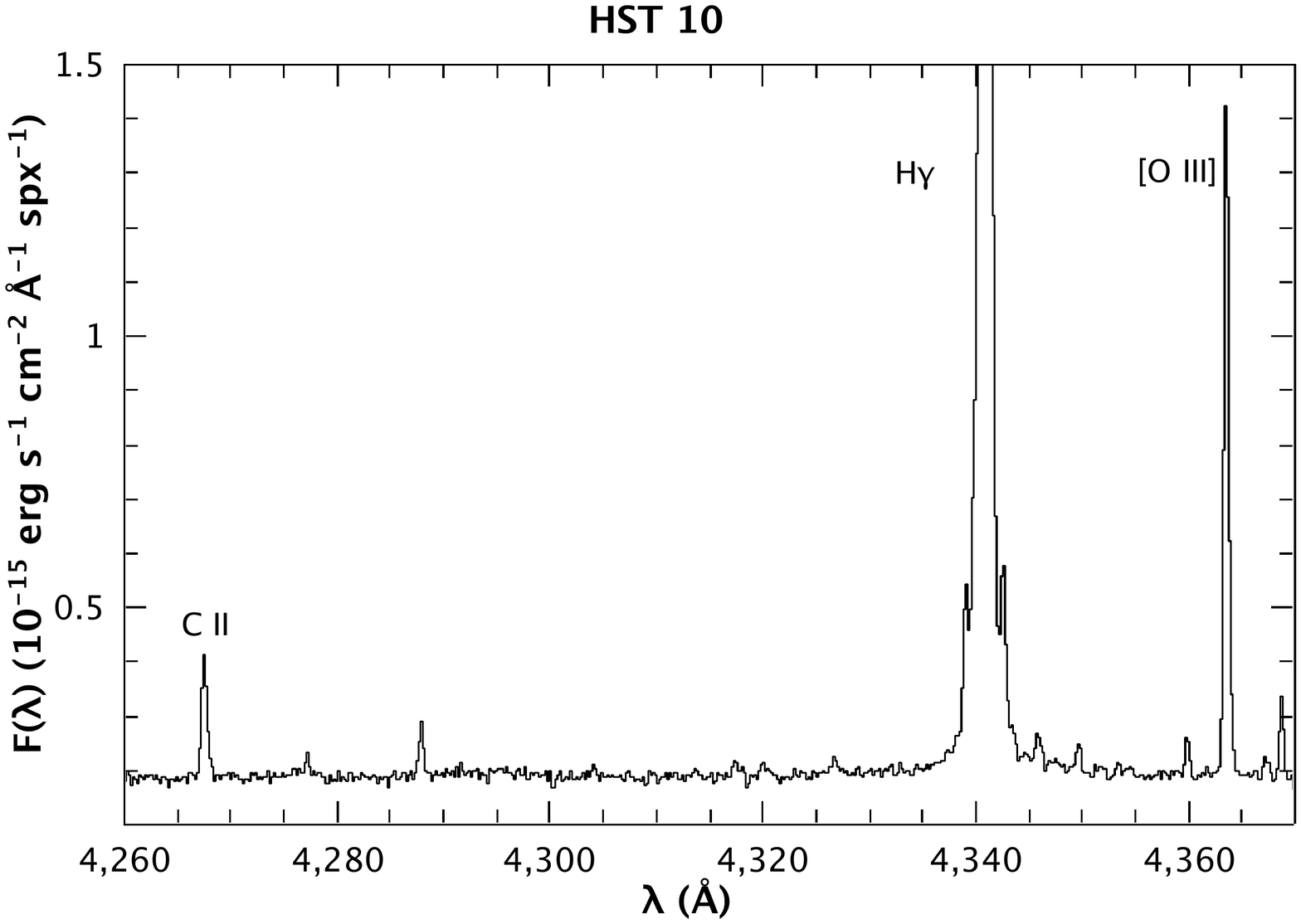}\\
  \end{tabular}
  \caption{Spectrum of HST~10 extracted over the cusp area (not corrected for reddening), showing the \cii\ \lam4267 recombination line and the \foiii\ \lam4363 collisional line. }
\end{figure}

The abundance of carbon rests on the determination based on the \cii\ \lam4267 recombination line (Fig.\,~9) adopting the effective recombination coefficients of Davey, Storey \& Kisielius (2000), which include both radiative and dielectronic
processes. The presence of \cp\ was corrected for adopting ICF(C) $=$ O/\opp\ (Kingsburgh \& Barlow 1994), necessarily computed here from collisionally excited lines (since oxygen RLs are not detected from HST~10). The \cii\ \lam6578 line, which was also detected, is dependent on optical depth effects and/or pumped by fluorescence and its diagnostic value is limited.

No difference is found in the abundance of carbon between the proplyd and the nebula, however the carbon abundance in our model of HST~10 is slightly lower than the one derived by the ICF method; this probably exposes the shortcomings of the latter when applied to HST~10. The carbon abundance from our preferred model is only $\sim$0.1 dex higher than that in B-type stars in Orion (Nieva \& Simon-Diaz 2011) or in a larger sample of B stars within 500\,pc from the Sun (Nieva \& Przybilla 2012).

\subsubsection{Chlorine, argon and iron}

The abundances of chlorine and argon were computed with the use of ICF equations 21 and 22 by Izotov et al. (2006), and are respectively almost equal in the proplyd and the nebula. The abundances of these species obtained by our detailed model of HST~10 do not differ substantially from those derived by the ICF method.

The abundance of iron was computed from the \ffeiii\ \lam5270 line (Fig.\,~4) using a 25-level atomic model for Fe~{\sc iii} (cf. Tsamis \& Walsh 2011). The resulting \fepp/\hp\ ratio was converted to Fe/H using the ICF by Rodr\'iguez \& Rubin (2005). The proplyd shows an iron abundance $\sim$0.3 dex lower than the Orion nebula. The abundance of gas-phase Fe therefore appears to be significantly different between the proplyd and the nebula. The abundance of Fe in the Orion \hii\ region is already depleted relative to solar by $\sim$1.5 dex (Esteban et al. 2004; Tsamis et al. 2011) due to iron being locked in dust grains. As a result, more iron appears to be trapped in condensates in HST~10 than is the case for the local Orion nebula. The observed overall depletion in HST~10 is however not as strong as the one encountered in LV\,2 ($-$2.5 dex relative to solar for LV\,2's rest velocity component); that proplyd shows a Fe depletion similar to those observed in the cold ISM (Tsamis et al. 2011; Tsamis \& Walsh 2011).


\subsubsection{The recombination/collisional-line abundance anomaly}

We note that the long-standing RL/CL abundance anomaly (e.g. Tsamis et al. 2003; Esteban et al. 2004; Garc\'ia-Rojas \& Esteban 2007; Mesa-Delgado \& Esteban 2010; Sim\'on-D\'iaz \& Stasi\'nska 2011; Tsamis et al. 2011; Mesa-Delgado et al. 2012) cannot be quantified in the case of HST~10. This is defined as the difference in ionic abundances derived from metallic recombination lines compared to those derived from collisionally excited lines emitted from the same ions (the so-called abundance discrepancy factor, ADF). In the Orion nebula the ADF for \opp\ is $\sim$0.1--0.3 dex (e.g. Mesa-Delgado et al. 2012 and references therein). On the other hand, the ADF is practically zero in the cusps of the LV\,2 (Tsamis et al. 2011) and 177--341  proplyds (Mesa-Delgado et al. 2012). The accurate removal of the contaminating Orion nebula emission and the adoption of realistic temperatures and densities for the proplyd doubly ionized carbon and oxygen regions are the most likely causes of these encouraging results. 

In the case of HST~10, with respect to carbon only \cii\ \lam4267 is observed while its CL counterpart \fciii\ \lam1908 is not available. Recombination lines of \opp, which are weaker than \cii\ \lam4267, are not detected in the background-subtracted spectra of HST~10 tallying with the overall low ionization degree of the proplyd. The intensity of the strongest predicted \oii\ \lam4649.13 V1 multiplet line is dominated by the Orion nebula emission. 

We derived the mean \opp/\hp\ ratio for $\sim$20 spaxels of the background nebula from the summed intensities ($I$ $=$ 0.3265 in units where \hb\ $=$ 100) of five \oii\ V1 RLs (4638.86, 4649.13, 4650.84, 4661.63, 4676.24\,\AA), adopting effective recombination coefficients from Storey (1994). Comparing with the abundance derived from the \foiii\ \lam4959 CL we obtain an ADF(\opp) $\simeq$ 0.16 dex for the Orion nebula within our field of view (Table~3).




\subsection{Photoevaporation models of HST~10}
\label{sec:phot-models-hst}

\begin{table}
  \centering
  \caption{Input parameters for example physical models of 182-413} 
  \begin{tabular}{@{\,}ll@{\,}}\hline
    Stellar spectrum:& 
    \(T_* = \SI{39000}{K}\)\\
    {Sim{\'o}n-D{\'{\i}}az} et al. (2006) & \(\log g = 4.1\)\\
    & \(L_* = \num{2.04e5}\,L_\odot\)\\
    Projected distance to \(\theta^1 C\): & 57 arcsec
    (O'Dell 1998)\\
    Inclination angle:& 30 deg (Henney \& O'Dell 1999)\\
    Derived physical distance to \(\theta^1 C\): & 7.51 $\times$ 10$^{17}$ cm\\
    Ionizing flux at proplyd:& 
    \(\Phi_{\mathrm{H}} = \SI{1.27e12}{cm^{-2} s^{-1}}\)
    \\
    Ionization front radius:& 
    \(r_0 = \SI{3.7e15}{cm}\) (Henney \& O'Dell 1999)
    \\
    Gas-phase abundances: & Model~A: Esteban et al. (2004)\\
    (\(12 + \log_{10} z/\mathrm{H}\)) & He 10.99, C 8.42, N 7.73, O 8.65\\
    & Ne 8.05, S 7.22, Cl 5.46, Ar 6.62, Fe 6.0 \\
    \noalign{\vskip3pt}
    & Model~B: {\it ad hoc} for HST~10\\
    & He 10.98, C 8.48, N 7.62, O 8.63\\
    & Ne 8.16, S 6.75, Cl 5.0, Ar 6.30, Fe 5.78
    \\
    Dust composition: & Standard Orion (Baldwin et al. 1991)\\
    \hline
  \end{tabular}
  \label{tab:model:pars}
\end{table}

\begin{figure*}
  \includegraphics[width=0.7\linewidth]{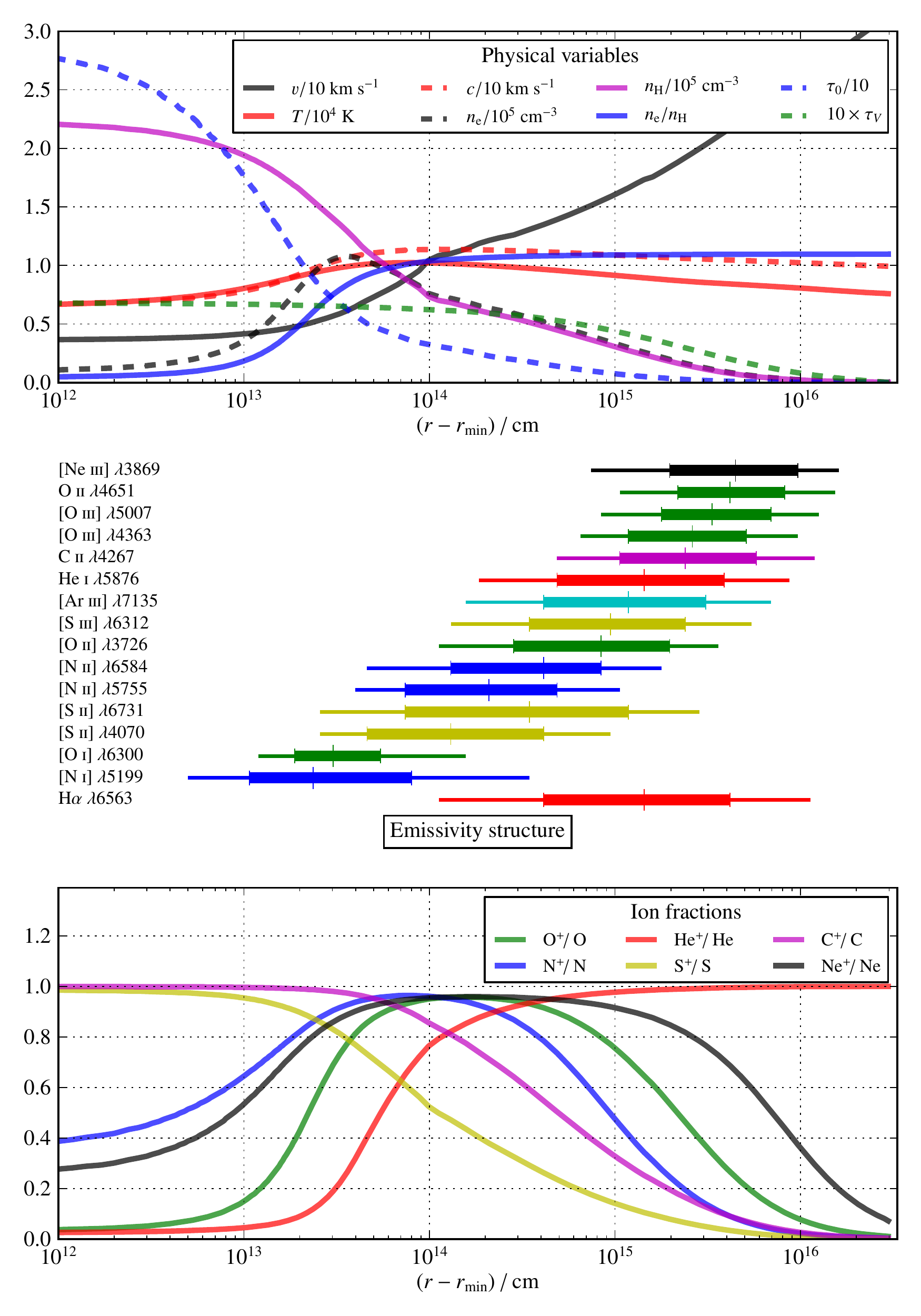}
  \caption{%
    Model structure  as a function of radius 
    along a line at an angle of \(\theta = 30^{\circ}\) to the symmetry axis 
    for our photoevaporation Model~B.
    The radius \(r\) is measured from the center of curvature of the ionization front, 
    shown on a logarithmic scale with respect to the deepest point in the model, 
    \(r_{\mathrm{min}} = 3.5 \times 10^{15}\ \mathrm{cm}\). The ionization front is at \(r\) $-$ \(r_{\mathrm{min}} \simeq 2 \times 10^{13}\ \mathrm{cm}\).
    \textit{Upper panel}: Physical variables. 
    Gas velocity, \(v\), in units of \(10\ \mathrm{km~s^{-1}}\) (solid black line).
    Gas temperature, \(T\), in units of \(10^4\ \mathrm{K}\) (solid red line).
    Isothermal sound speed, \(c\), in units of  \(10\ \mathrm{km~s^{-1}}\) (dashed red line).
    Electron density, \(n_{\mathrm{e}}\), in units of \(7 \times 10^4\ \mathrm{cm^{-3}}\) (dashed black line).
    Total hydrogen density, \(n_{\mathrm{H}}\), in units of  \(7 \times 10^4\ \mathrm{cm^{-3}}\) (solid purple line).
    Electron fraction, \(n_{\mathrm{e}}/n_{\mathrm{H}}\) (solid blue line).
    Lyman limit optical depth, \(\tau_0\) (divided by 10), measured from the ionizing source, 
    which is located off the graph at \(r =  6.32 \times 10^{17}\ \mathrm{cm}\) (dashed blue line).
    Visual dust extinction optical depth, \(\tau_V\) (times 10) (dashed green line). 
    \textit{Middle panel}: Emissivity structure for selected emission lines.  
    For each emission line, the vertical line indicates the median emission radius 
    (for which half the total line flux is emitted from gas at smaller radii and half from gas at larger radii),
    while the thick horizontal bar shows the range between the 25th and 75th quartiles of the emission,
    and the thin horizontal bar shows the range between the 10th and 90th quartiles of the emission.
    \textit{Lower panel}: Ionization structure.  
    The fraction of singly ionized ions for a selection of elements is shown by differently colored lines,
    as indicated in the key.
    Results for other angles and abundance sets are similar, 
    except with densities that scale roughly as \(\cos^{1/2}\theta\), 
    and temperatures that tend to decrease slightly with \(\theta\) and with increased metal abundances.
  }
  \label{fig:model-structure}
\end{figure*}

\begin{figure*}
  \setkeys{Gin}{width=0.7\linewidth}
  \setlength\tabcolsep{0pt}
  \centering
  \begin{tabular}{l} 
    (a)\\
    \includegraphics[]{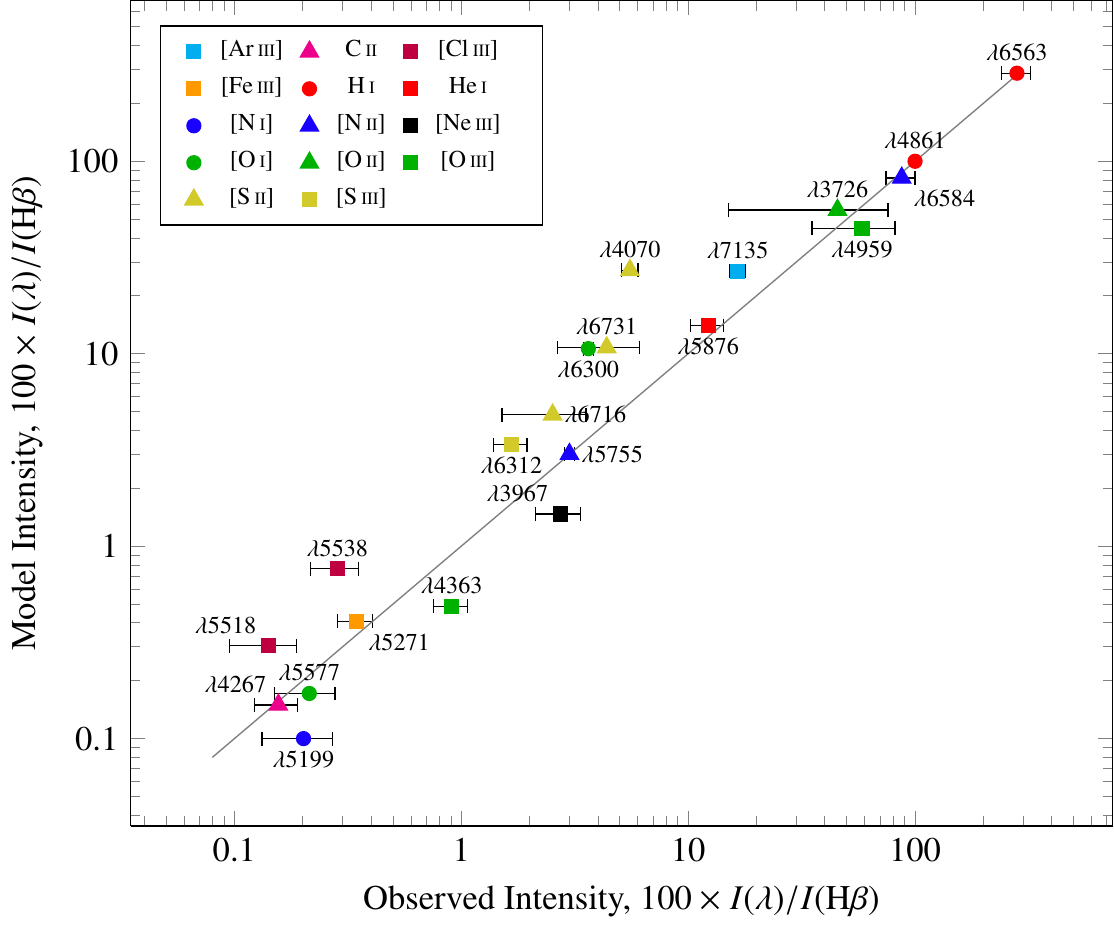}\\
    (b)\\
    \includegraphics[]{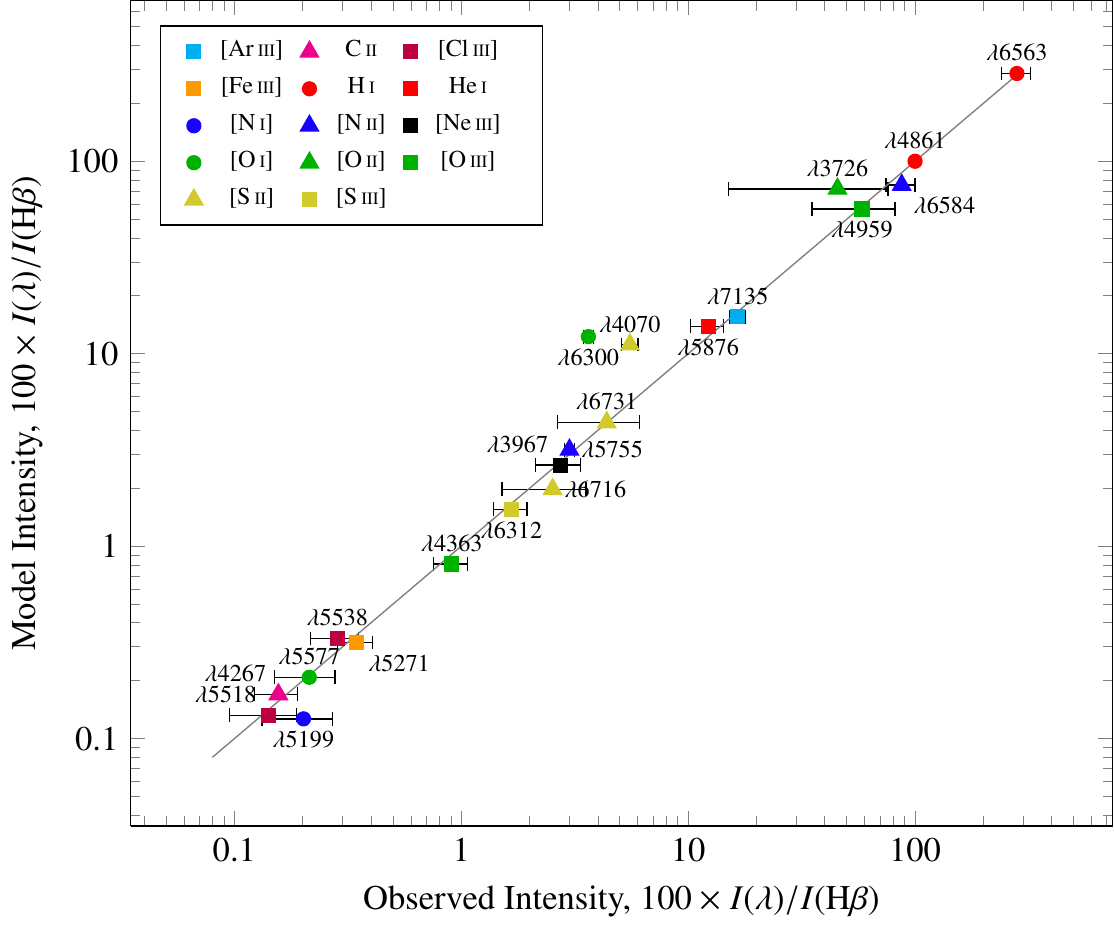}
  \end{tabular}
  \caption{Comparison between predicted and observed emission line fluxes 
    for photoevaporation models of HST~10 with 
    (a)~abundances as determined for the Orion Nebula by Esteban et al. (2004),
    (b)~abundances determined for HST~10 in this paper.  
    The agreement in panel~(b) is much better, but there are still notable discrepancies.
  }
  \label{fig:models}
\end{figure*}

We have calculated dynamic photoevaporation models of HST~10 with the photoionization code Cloudy v10.0 (Ferland et al. 1998), using the procedure outlined in Mesa-Delgado et al. (2012). 
The parameters for the models are shown in Table~\ref{tab:model:pars}. 
As compared with proplyd 177--341 (HST~1), which was modeled in Mesa-Delgado et al. (2012),
HST~10 receives a roughly 10 times smaller ionizing flux, $F$,
and is roughly twice as large. 
Since \(F \propto n^2 r\) for recombination-dominated photoevaporation flows (Henney 2001),
this implies that the densities in HST~10 should be \(\simeq 5\) times smaller
and the ionization parameter \(\simeq 2\) times smaller than in HST~1. 
Fig.\,~\ref{fig:model-structure} shows the variation of physical variables
along a representative radial cut through our best-fitting model described below.
The ionizing flux was adjusted until the model reproduced the correct
\hb\ flux for the proplyd cusp.  Assuming an ionizing flux of 9.0 $\times$ 10$^{48}$\,s$^{-1}$ for \oric\ and no intervening absorption, this implies a true
distance from the Trapezium of about twice the projected distance, or
an inclination of about 60 degrees from the plane of the sky.  This is
consistent with the appearance in the \hst\ images and with the
kinematics measured from Keck spectra.

Two different abundance sets were used in the models.  
The first (Model~A) is the Orion gas phase abundance set, 
as determined by Esteban et al. (2004) for a temperature fluctuation parameter, $t^2$ $>$ 0.
The second (Model~B) is a set of abundances starting from those determined in this paper using the semi-empirical method and assuming that $t^2$ $=$ 0 (Table~3). The emissivity structure of Model~B shown in the middle panel of Fig.\,~10 can be compared with the emission line maps presented in Section~3.1. The model predicts that the lower ionization lines, such as those of \foi, \fsii, and \fnii, should show  emission  that is offset towards the center of the proplyd with respect to the H\(\alpha\) emission by \(1\)--\(2 \times 10^{15}\)~cm (0.5--1 spaxel), which is very similar to what is observed (Fig.\,~3).   In contrast, the highest ionization lines, such as those of \foiii\ and \fneiii, should be offset away from the center of the proplyd by a similar amount, which is again consistent with the observed spatial distributions (Fig.\,~4).  The model also makes more detailed predictions about the emission distributions, such as the fact that for both \foiii\ and \fnii\ the nebular collisional lines should be weighted towards slightly larger radii than the auroral lines, a consequence of the temperature and density gradients within a given ionization zone.   There is some evidence for this from our emission maps, but the predicted magnitude of the effect (0.1--0.5 spaxel) is at the limit of what can be discerned with our spatial resolution.

The spectrum from the two models is shown in Fig.\,~11. 
It can be seen that the Esteban et al. (2004) abundances 
produce large discrepancies between the observed and predicted line fluxes 
(panel `a'). In particular, all sulfur, argon and chlorine lines are too strong in the model 
by a factor of three to 10, whereas, \foiii\ \lam4363 is too weak, indicating that the model temperature 
is too low in the highly ionized regions. On the other hand, the \fnii\ lines are well-reproduced by this model. 

The situation is much improved by using the Model~B abundances (Fig.\,~11b),
although some discrepancies remain.  
The sulfur line fluxes are now in good agreement with the observations,
with the notable exception of the \fsii\ \lam4069 auroral line,
which is still three times too strong in the model. 
The \foi\ \lam6300 shows a similar behaviour, being too strong by a factor of four. 
These two lines, together with the \fnii\ \lam5755 auroral line, 
show the strongest brightness contrast between the proplyd and the background nebula,
and hence are measured with a relatively small uncertainty,
making the disagreement highly significant. 
The \fnii{} lines do not agree with Model~B so well as they do with Model~A, 
but the agreement is still fair given the uncertainties. 
The remaining disagreement is with the \fni\ \lam5199 line,
but this is to be expected since the line arises through fluorescence in neutral gas
(Ferland et al. 2012), whereas the model extends only to a hydrogen ionization fraction of 0.1 
and therefore misses part of the zone emiting \fni.

The two most discrepant lines, \foi{}~\lam6300 and \fsii{}~\lam4069 have 
a considerable overlap in their zones of emission (Fig.\,~10).  
All the \foi{} emission, and roughly half of the \fsii{} \lam4069 arises 
from partially ionized gas at the ionization front itself.
If the temperature of this gas is overestimated in the models, 
then the discrepancy could be explained. 
With the standard Orion dust properties that we are using in all the models
(Baldwin et al. 1991), photoelectric emission from dust grains provides up to 15 per cent of the heating
in this zone. Since there is some evidence that dust is depleted in the proplyd flows
({Garc{\'{\i}}a-Arredondo} et al. 2001), it may be feasible to reduce this heating,
which is an avenue that will be explored in follow-up work. 

Notwithstanding these discrepancies, we consider that Model B for HST~10 has a much better defined O/H abundance than our model for HST~1. This is because the \foii\ \lam3726, 3729 lines were not measured from that proplyd due to their extreme weakness from collisional quenching in the prevailing high density conditions. With regard to chlorine and argon our preferred model for HST~10 does not support the high ($t^2$ $>$ 0) abundances for these elements in Orion obtained by Esteban et al. (2004). As a noble gas, Ar is not expected to be depleted on grains (even though it can form clathrates in ices; e.g. Mousis et al. 2010), and therefore the 0.3 dex difference with the Orion value by Esteban et al. is perhaps significant.

\subsection{Comparison with other proplyds}

\begin{figure}
  \centering
  \includegraphics[width=\linewidth]{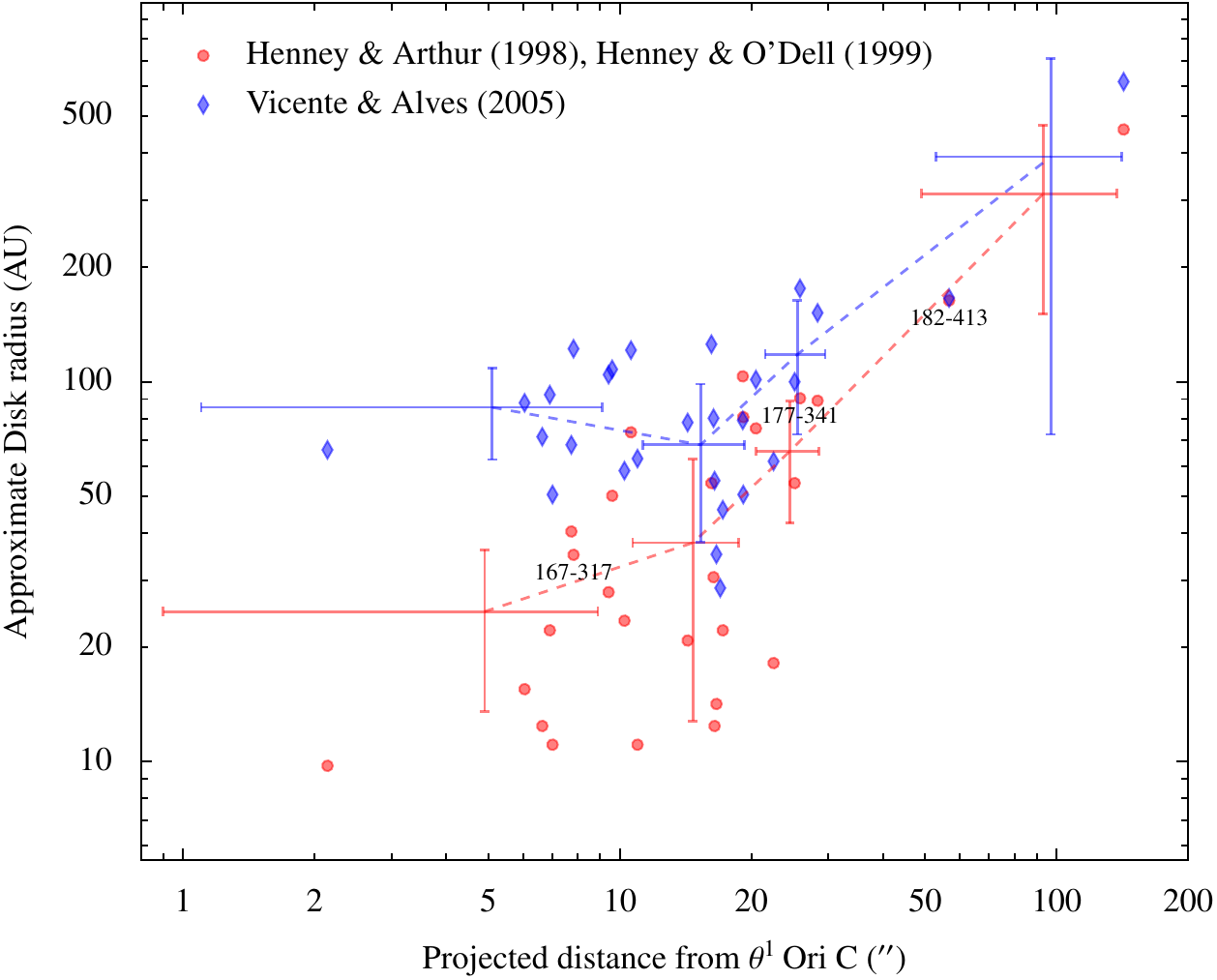}
  \caption{Variation with projected distance from the ionizing star of circumstellar disc sizes in the Orion proplyds.   Blue symbols show empirical determinations from Vicente \& Alves (2005), red symbols show results from fitting photoevaporation models (Henney \& Arthur 1998; Henney \& O'Dell 1999).  The three proplyds that have been subject to gas-phase abundance studies are marked: 167--317 (LV\,2), 177--341 (HST~1), and 182--413 (HST~10). In only a few cases, such as HST~10, is the molecular disc size directly measured, for the rest it is assumed to be half the size of the ionization front. Note that the Vicente \& Alves (2005) methodology significantly overestimates the true size of the ionization front (and by extension the enclosed disc) for small proplyds or those that are much brighter than the surrounding nebula, which tend to be those found close to the ionizing star.   Error bars show the mean and standard deviation measured in 4 broad spatial bins.}

  \label{fig:sizes}
\end{figure}

\begin{table*}\centering
  \caption{Comparison of physical properties between HST~10 and two other well-studied proplyds}
\label{tab:3props}
\setlength\tabcolsep{2\tabcolsep}
\begin{tabular}{llr rrr}\hline
 & Units & Note & LV\,2 & HST~1 & HST~10 \\ \hline
Coordinate-based designation & & 1 & 167--317 & 177--341 & 182--413 \\
\rule{0pt}{3ex}\textit{Relation to ionizing source}\\
Projected distance, \(D'\) & \(''\) & 2 & 7.83 & 25.84 & 56.7 \\
Inclination, \(i\) & \(^\circ\) & 3 & 50 & 70 & 150 \\
True distance, \(D\) & pc & 4 & 0.022 & 0.059 & 0.242 \\
\rule{0pt}{3ex}\textit{Ionized cusp} \\
Ionization front radius, \(r_0\) & AU & 5 & 53. & 136. & 247. \\
Peak electron density, \(n_0\) & \(10^{6}\ \mathrm{cm^{-3}}\) & 6 & \(2.0\) & \(0.4\) & \(0.1\) \\
Ionization parameter &  & 7 & 0.012 & 0.008 & 0.002 \\
Cusp mass-loss rate, \(\dot{M}\) & \(10^{-7}\ M_\odot\ \mathrm{yr}^{-1}\) & 8 & \(2.6\) & \(2.5\) & \(2.1\) \\
\rule{0pt}{3ex}\textit{Molecular disc} \\
Disc effective temperature, \(T_\mathrm{d}\) & K & 9 & 95 & 58 & 29 \\
Disc mass, \(M_\mathrm{d}\) & \(10^{-3}\ M_\odot\) & 10 & \(1.6\) & \(2.7\) & \(5.4\) \\
Disc radius, \(R_\mathrm{d}\) & AU & 11 & 34 & 89 & 160 \\
Evaporation age, \(t_\mathrm{evap}\) & \(10^4\ \mathrm{yr}\) & 12 & \(0.6\) & \(1.1\) & \(2.6\) \\
\hline

\end{tabular}
\par\smallskip
\begin{minipage}{0.7\linewidth}
  \def\NoteSep{\quad}
  \textit{Notes}:\NoteSep 
  (1) O'Dell \& Wenn 1994\NoteSep
  (2)~Angular separation from \thC{} (O'Dell 1998)\NoteSep
  (3)~Inclination of proplyd axis to line of sight estimated from kinematic studies of the velocity--ionization correlation in emission lines from the cusp (Henney \& O'Dell 1999; Henney et al. 2002).  Proplyds with \(i > 90^\circ\) have their head pointing away from the observer.\NoteSep
  (4)~\(D = D' / \sin i\).\NoteSep
  (5)~Estimated from fitting evaporation models to the \ha\ profiles of the cusps (Henney \& Arthur 1998).\NoteSep
  (6)~LV\,2 from \fciii\ density (Henney et al. 2002); HST~1 and HST~10 from model fitting (this paper and Mesa-Delgado et al. 2012).\NoteSep
  (7)~\(F / (n_0\,c)\).\NoteSep 
  (8)~Calculated by integrating model mass fluxes over the area of the cusp.\NoteSep
  (9)~Radiative equilibrium temperature, assuming that 25\% of the bolometric flux from \thC{} reaches the surface of the disc (see also Robberto et al. 2002).\NoteSep
  (10)~Estimated from observed fluxes at \(880~\mu\mathrm{m}\) (Mann \& Williams 2010) after subtracting the contribution from ionized free-free emission, assuming optically thin dust emission with opacity \(\kappa_\nu = 0.034~\mathrm{cm^2\ g^{-1}}\) and dust temperature equal to the effective temperatures derived above.\NoteSep
  (11)~Directly estimated from \textit{HST} images for HST~10.  For LV\,2 and HST~1, we assume \(r_\mathrm{d} = 0.65 r_0\), see Fig.\,~\ref{fig:sizes}.\NoteSep
  (12)~Nominal mass loss timescale: \(M_\mathrm{d} /\dot{M}\). 
\end{minipage}

\end{table*}

HST~10 is the third proplyd to be subject to a detailed abundance analysis, following earlier studies of LV\,2 (Tsamis et al. 2011; Tsamis \& Walsh 2011) and HST~1 (Mesa-Delgado et al. 2012).  
These proplyds cover a broad range in size and in separation from the Trapezium stars (see Fig.\,~\ref{fig:sizes}), and derived physical parameters for the three proplyds are summarised in Table~\ref{tab:model:pars}.  HST~10 shows a significantly lower ionization parameter than the two closer-in proplyds, which is reflected in its emission line spectrum that is relatively stronger in low ionization lines. Despite these differences, the estimated mass loss rate from the ionized cusp is very similar for all three proplyds, being of order \(2 \times 10^{-7}\ \mathrm{M_\odot}~\mathrm{yr^{-1}}\).  These values are somewhat lower than earlier estimates (e.g., Henney \& O'Dell 1999; Henney et al. 2002), which is partly because we are neglecting the contribution of mass loss though the proplyd tail.

Table~\ref{tab:3props} also shows estimates for the mass and radius of the embedded circumstellar accretion disc, which is the reservoir of mass in the proplyds.   All three proplyds show very similar sub-mm fluxes (Mann \& Williams 2010), of order 20~mJy once the contribution from ionized free-free emission has been subtracted.  However, conversion of this flux to a gas mass requires knowledge of the dust opacity per unit gas mass  and dust temperature, both of which have large uncertainties (Williams \& Cieza 2011).  The values in the table are calculated on the assumption that the dust temperature is the effective temperature of a disc in radiative equilibrium with the bolometric flux from the Trapezium stars,\footnote{Although the sub-mm emission is optically thin, the discs are likely to be optically thick at mid-infrared wavelengths where they emit the bulk of their radiation.   We assume that 50\% of the bolometric radiation from the Trapezium stars is absorbed in the ionized cusp of the proplyd, and that 50\% of the remainder is absorbed in the neutral photoevaporation flow, so that only 25\% reaches the surface of the disc.} and using the opacity recommended by Beckwith et al. (1990) of \(0.1 (\nu/1000~\mathrm{GHz})~\mathrm{cm^2~g^{-1}}\), giving \(0.034~\mathrm{cm^2~g^{-1}}\) at 880~\(\mu\)m.   It must be emphasised that the derived masses are highly uncertain, since the opacity could be up to 20 times smaller if substantial grain growth up to cm-sized bodies has occurred (D'Alessio et al. 2001).   Evidence for grain evolution has been found in the case of silhouette discs projected onto the Orion nebula (Miotello et al. 2012).   The masses given in the table are at least 6 times smaller than the `minimum mass solar nebula' (Weidenschilling 1977), which is the mass within \(30\)~AU required to account for the composition of the planets in the Solar System.  

The nominal photoevaporation ages, \(t_\mathrm{evap} = M_\mathrm{d}/\dot{M}\), are uncomfortably short compared with the estimated age of the photoionized nebula (\(\ge 10^5\)~yr; see discussion in section 8.2.2 of Henney \& O'Dell 1999), but would come into agreement if the masses were increased by a factor of 5--10, which cannot be ruled out (see previous paragraph). Interestingly, the evaporation age \(t_\mathrm{evap}\) \emph{increases} with increasing distance from the Trapezium. Given that \(t_\mathrm{evap}\) should be roughly equal to the elapsed time since the disc photoevaporation commenced (Johnstone et al. 1998), this is the opposite of what would be expected from a naive model of a roughly spherical \hii\ region, in which case proplyds at greater distances from the ionizing star would have entered the \hii\ region more recently.   However, evidence from both observations (O'Dell et al. 2009) and numerical simulations (Arthur et al. 2011) point to the continued survival of dense clumps of molecular gas well inside the apparent boundary of the \hii\ region, in which case it is reasonable that the proplyds closest to the Trapezium might have been shielded from ultraviolet radiation until relatively recently.

The abundance results for the three proplyds are rather disparate, so that it is very hard to see a consistent trend in the results.  For instance, although we find a roughly solar oxygen abundance for HST~10 from both empirical analysis and photoevaporation model fitting of \foii\ and \foiii\ CLs, the oxygen abundance was found to be \(\simeq 2 \times\) solar in LV\,2 (Tsamis et al. 2011) from a purely empirical analysis, with no discrepancy between collisional and recombination lines and with both \op\ and \opp\ stages observed (in \foii\ CLs for the former and \foiii\ CLs and \oii\ RLs for the latter).  In contrast, Mesa-Delgado et al. (2012) found oxygen to be about \(\simeq 0.4 \times\) solar in HST~1, from photoevaporation model fitting of \opp\ (\foiii\ CLs), whereas an empirical analysis of \oii\ RLs is consistent with the \foiii\ CLs. Given the wide range of characteristic ionization and density found in the three proplyds (Table~\ref{tab:3props}), it is possible that systematic errors in our abundance  analysis might be contributing to this wide spread.  In order to rule out any such effects, it would be very worthwhile to carry out a similar analysis on a sample of proplyds that all have \emph{similar} ionization parameters and densities.

\section{Summary and conclusions}

We have presented integral field spectroscopic observations of the HST~10 proplyd found in the Orion nebula obtained with the VLT and produced a collection of sub-arcsecond resolution monochromatic images of the proplyd and the surrounding nebula. The IFS observations facilitated the isolation of the proplyd's intrinsic emission from the emission of the Orion nebula. As a result, the reddening, ionization degree, electron temperature and density, as well as the ionic and elemental abundances for several species have been derived for the proplyd and for the local Orion neighbourhood. 

The abundances of oxygen and sulfur have been derived for HST~10 without resort to ionization correction factors and are not significantly different from those measured in the local Orion nebula. This conclusion is also borne out from a detailed photoevaporation model of the proplyd. The abundances of carbon, oxygen, and neon in HST~10 are practically the same as those in B-type stars in Orion. 

Based on our model, the abundance of neon in HST~10 is 0.2 dex higher than the photospheric solar value by Asplund et al. (2009) and 0.15 dex higher than the IR Spitzer value by Rubin et al. (2011). Compared to the model, the standard ICF method for neon significantly underestimates the element's abundance in HST~10 by 0.4 dex. This indicates that the classic ICF method applied to optical \fneiii\ emission lines can systematically underestimate Ne/H in low ionization Orion structures. 

Iron in HST~10 is depleted by $\sim$0.3 dex compared to the Orion nebula as a result of a higher percentage of Fe trapped in solids than in the ambient \hii\ region.

Based on our derived mass loss rates and observed disc fluxes at mm
wavelengths, the photoevaporation ages of the HST~10, LV\,2 and HST~1 proplyds are very short
(\(\simeq 10^4\)~years) if one assumes `standard' dust opacities
(Beckwith et al. 1990), which is in conflict with lower limits on the age of
the Orion nebula (\(> 10^5\)~years; Henney \& O'Dell 1999).  This
implies that the mm dust opacity must be lower, possibly due to grain
growth to cm sizes.   The photoevaporation ages are found to be lowest
for the proplyds closest to \thC{}, implying that these must have been
shielded inside dense molecular gas until relatively recently, which
is consistent with a highly irregular shape for the Orion nebula's ionization front,
as has been suggested by recent observational (O'Dell et al. 2009) and
theoretical (Arthur et al. 2011) studies.



In conclusion,  our study of HST~10 -- and those of LV\,2 and HST~1 based on IFS techniques -- shows that spatially resolved spectrophotometry accompanied by detailed simulations can be successfully applied to evaporating protoplanetary discs, helping to establish their chemical composition in Orion for the first time. These studies of the youngest component of this landmark region's ecosystem complement superbly the abundance analyses of ionized gas and stars in the same area. 

\section*{Acknowledgments}

We thank the VLT/FLAMES astronomers at ESO for supporting the service
mode observations (programme 078.C-0247(A); PI: Tsamis). We also thank the referee Dr. C. R. O'Dell for a constructive report. This research made use of the NASA ADS digital library portal operated by the Smithsonian Astrophysical Observatory (SAO); ESO-MIDAS on Mac/OSX provided by the European Southern Observatory; IRAF, distributed by the National Optical Astronomy Observatory, which is operated by the Association of Universities for Research in Astronomy; SAOImage DS9 developed by SAO.

WJH and NFF acknowledge financial support from DGAPA-UNAM through
project PAPIIT IN102012 and from a postdoctoral fellowship to
NFF. AMD acknowledges support from Comit\'e Mixto ESO-Chile and a Basal-CATA (PFB-06/2007) grant.

YGT acknowledges support from a Marie Curie intra-European Fellowship within
the 7$^{\rm th}$ European Community Framework Programme (project PIEF-GA-2009-236486), and thanks his collaborators at the Centro de Radioastronom\'ia y Astrof\'isica in Morelia for their warm hospitality during his visit.

\bibliographystyle{mn2e}

\begin{thebibliography}{}



\bibitem[{{Arthur} {et~al}\mbox{.}(2011){Arthur}, {Henney}, {Mellema}, {de
  Colle}, \& {V{\'a}zquez-Semadeni}}]{Arthur:2011}
{Arthur} S.~J., {Henney} W.~J., {Mellema} G., {de Colle} F.,
  {V{\'a}zquez-Semadeni} E., 2011, \mnras, 414, 1747


\bibitem[Asplund et al.(2009)]{} Asplund, M., Grevesse, N., Jacques Sauval, A., 
\& 
Scott, P.\ 2009, ARA\&A, 47,
481

\bibitem[Baldwin et al.(1991)]{1991ApJ...374..580B} Baldwin, J.~A., 
Ferland, G.~J., Martin, P.~G., et al.\ 1991, \apj, 374, 580 


\bibitem[{{Bally} {et~al}\mbox{.}(1998){Bally}, {Sutherland}, {Devine}, \&
  {Johnstone}}]{1998AJ....116..293B}
{Bally} J., {Sutherland} R.~S., {Devine} D., {Johnstone} D., 1998, \aj, 116,
  293

\bibitem[Bally et al.(2000)]{2000AJ....119.2919B} Bally, J., O'Dell, C.~R., 
\& McCaughrean, M.~J.\ 2000, \aj, 119, 2919 

\bibitem[Benjamin et al.(1999)]{1999ApJ...514..307B} Benjamin, R.~A., 
Skillman, E.~D., \& Smits, D.~P.\ 1999, \apj, 514, 307 

\bibitem[{{Beckwith} {et~al}\mbox{.}(1990){Beckwith}, {Sargent}, {Chini}, \&
  {Guesten}}]{Beckwith:1990}
{Beckwith} S.~V.~W., {Sargent} A.~I., {Chini} R.~S., {Guesten} R., 1990, \aj,
  99, 924



\bibitem[Blagrave et al.(2007)]{2007ApJ...655..299B} Blagrave, K.~P.~M.,
Martin, P.~G., Rubin, R.~H., Dufour, R.~J., Baldwin, J.~A., Hester, J.~J., \&
Walter, D.~K.\ 2007, \apj, 655, 299

\bibitem[Blecha \& Simond 2004]{} Blecha, A., \& Simond, G.\ 2004, GIRRAFE BLDR
Software Reference Manual 1.12 (http://girbldrs.sourceforge.net)

\bibitem[Cardelli et al.(1989)]{1989ApJ...345..245C} Cardelli, J.~A.,
Clayton, G.~C., \& Mathis, J.~S.\ 1989, \apj, 345, 245 (CCM)
D.~L.\ 
1994, \apj, 426, 170

\bibitem[Chen et al.(1998)]{1998ApJ...492L.173C} Chen, H., Bally, J., 
O'dell, C.~R., et al.\ 1998, \apjl, 492, L173 

\bibitem[{{D'Alessio} {et~al}\mbox{.}(2001){D'Alessio}, {Calvet}, \&
  {Hartmann}}]{DAlessio:2001}
{D'Alessio} P., {Calvet} N., {Hartmann} L., 2001, \apj, 553, 321


\bibitem[Davey et al.(2000)]{2000A&AS..142...85D} Davey, A.~R., Storey,
P.~J., \& Kisielius, R.\ 2000, \aap Supplement, 142, 85



\bibitem[Doi et al.(2004)]{2004AJ....127.3456D} Doi, T., O'Dell, C.~R., 
\& Hartigan, P.\ 2004, \aj, 127, 3456 


\bibitem[Esteban et al.(2004)]{2004MNRAS.355..229E} Esteban, C., Peimbert,
M., Garc{\'{\i}}a-Rojas, J., Ruiz, M.~T., Peimbert, A., \& Rodr{\'{\i}}guez,
M.\ 2004, \mnras, 355, 229

\bibitem[Esteban et al.(2009)]{2009ApJ...700..654E} Esteban, C., Bresolin,
F., Peimbert, M., Garc{\'{\i}}a-Rojas, J., Peimbert, A., \& Mesa-Delgado, A.\
2009, \apj, 700, 654

\bibitem[Ferland et al.(1998)]{1998PASP..110..761F} Ferland, G.~J., 
Korista, K.~T., Verner, D.~A., et al.\ 1998, \pasp, 110, 761 



\bibitem[Ferland et al.(2012)]{2012ApJ...757...79F} Ferland, G.~J., Henney, 
W.~J., O'Dell, C.~R., et al.\ 2012, \apj, 757, 79 


\bibitem[Garc{\'{\i}}a-Rojas 
\& Esteban(2007)]{2007ApJ...670..457G} Garc{\'{\i}}a-Rojas, J., \& Esteban, C.\ 2007, \apj, 670, 457 

\bibitem[Garc{\'{\i}}a-Arredondo et al.(2001)]{2001ApJ...561..830G} 
Garc{\'{\i}}a-Arredondo, F., Henney, W.~J., 
\& Arthur, S.~J.\ 2001, \apj, 561, 830 



\bibitem[Garc{\'{\i}}a-D{\'{\i}}az 
\& Henney(2007)]{2007AJ....133..952G} Garc{\'{\i}}a-D{\'{\i}}az, M.~T., \& Henney, W.~J.\ 2007, \aj, 133, 952 


\bibitem[Garc{\'{\i}}a-D{\'{\i}}az et al.(2008)]{2008RMxAA..44..181G}
Garc{\'{\i}}a-D{\'{\i}}az, M.~T., Henney, W.~J., L{\'o}pez, J.~A., \& Doi, T.\
2008, RevMexAA, 44, 181


\bibitem[Henney 
\& Arthur(1998)]{1998AJ....116..322H} Henney, W.~J., \& Arthur, S.~J.\ 1998, \aj, 116, 322 



\bibitem[Henney(2001)]{2001RMxAC..10...57H} Henney, W.~J.\ 2001, Revista 
Mexicana de Astronomia y Astrofisica Conference Series, 10, 57 

\bibitem[Henney \& O'Dell(1999)]{1999AJ....118.2350H} Henney, W.~J., \& O'Dell, 
C.~R.
\ 1999, \aj, 118, 2350

\bibitem[Henney et al.(2002)]{2002ApJ...566..315H} Henney, W.~J., O'Dell,
C.~R., Meaburn, J., Garrington, S.~T., \& Lopez, J.~A.\ 2002, \apj, 566, 315

\bibitem[Henney et al.(2005)]{2005ApJ...627..813H} Henney, W.~J., Arthur, 
S.~J., \& Garc{\'{\i}}a-D{\'{\i}}az, M.~T.\ 2005, \apj, 627, 813 



\bibitem[Izotov et 
al.(2006)]{2006A&A...448..955I} Izotov, Y.~I., Stasi{\'n}ska, G., Meynet, G., Guseva, N.~G., \& Thuan, T.~X.\ 2006, \aap, 448, 955 



\bibitem[Johnstone et al.(1998)]{1998ApJ...499..758J} Johnstone, D.,
Hollenbach, D., \& Bally, J.\ 1998, \apj, 499, 758

\bibitem[Kassis et al.(2007)]{2007AAS...210.7501K} Kassis, M., Shuping, 
R.~Y., Morris, M., Smith, N., 
\& Bally, J.\ 2007, Bulletin of the American Astronomical Society, 39, 181 


\bibitem[Kingsburgh 
\& Barlow(1994)]{1994MNRAS.271..257K} Kingsburgh, R.~L., \& Barlow, M.~J.\ 1994, \mnras, 271, 257 

\bibitem[Laques \& Vidal(1979)]{1979A&A....73...97L} Laques, P., \& Vidal, J.~L.
\ 
1979, \aap, 73, 97


\bibitem[{{Mann} \& {Williams}(2010)}]{Mann:2010}
{Mann} R.~K., {Williams} J.~P., 2010, \apj, 725, 430


\bibitem[Mesa-Delgado et al.(2009)]{2009MNRAS.395..855M} Mesa-Delgado, A.,
Esteban, C., Garc{\'{\i}}a-Rojas, J., Luridiana, V., Bautista, M.,
Rodr{\'{\i}}guez, M., L{\'o}pez-Mart{\'{\i}}n, L., \& Peimbert, M.\ 2009,
\mnras, 395, 855

\bibitem[Mesa-Delgado \& Esteban(2010)]{} Mesa-Delgado, A., \& Esteban, C.\ 
2010,
MNRAS, 405, 2651

\bibitem[Mesa-Delgado et al.(2012)]{2012MNRAS.426..614M} Mesa-Delgado, A., 
N{\'u}{\~n}ez-D{\'{\i}}az, M., Esteban, C., et al.\ 2012, \mnras, 426, 614 

\bibitem[{{Miotello} {et~al}\mbox{.}(2012){Miotello}, {Robberto}, {Potenza}, \&
  {Ricci}}]{Miotello:2012}
{Miotello} A., {Robberto} M., {Potenza} M.~A.~C., {Ricci} L., 2012, \apj, 757,
  78


\bibitem[Mousis et al.(2010)]{2010FaDi..147..509M} Mousis, O., Lunine, 
J.~I., Picaud, S., \& Cordier, D.\ 2010, Faraday Discussions, 147, 509 


\bibitem[Nieva 
\& Sim{\'o}n-D{\'{\i}}az(2011)]{2011A&A...532A...2N} Nieva, M.-F., \& Sim{\'o}n-D{\'{\i}}az, S.\ 2011, \aap, 532, A2 

\bibitem[Nieva 
\& Przybilla(2012)]{2012A&A...539A.143N} Nieva, M.-F., \& Przybilla, N.\ 2012, \aap, 539, A143 


\bibitem[N{\'u}{\~n}ez-D{\'{\i}}az et al.(2012)]{2012MNRAS.421.3399N} 
N{\'u}{\~n}ez-D{\'{\i}}az, M., Mesa-Delgado, A., Esteban, C., et al.\ 2012, 
\mnras, 421, 3399 


\bibitem[O'dell(1998)]{1998AJ....115..263O} O'Dell, C.~R.\ 1998, \aj, 115, 
263


\bibitem[{{O'Dell} \& {Wen}(1994)}]{1994ApJ...436..194O}
{O'Dell} C.~R., {Wen} Z., 1994, \apj, 436, 194


\bibitem[O'Dell(2000)]{2000AJ....119.2311O} O'Dell, C.~R.\ 2000, \aj, 119, 
2311 

\bibitem[O'dell(2001)]{2001ARA&A..39...99O} O'Dell, C.~R.\ 2001, ARA\&A, 39, 99


\bibitem[O'Dell et al.(2009)]{2009AJ....137..367O} O'Dell, C.~R., Henney, 
W.~J., Abel, N.~P., Ferland, G.~J., \& Arthur, S.~J.\ 2009, \aj, 137, 367 

\bibitem[O'Dell 
\& Yusef-Zadeh(2000)]{2000AJ....120..382O} O'Dell, C.~R., \& Yusef-Zadeh, F.\ 2000, \aj, 120, 382 

\bibitem[O'dell \& Wong(1996)]{1996AJ....111..846O} O'Dell, C.~R., \& Wong, K.\ 1996, \aj, 111, 846 

\bibitem[Pasquini et al.(2002)]{2002Msngr.110....1P} Pasquini, L., Avila, 
G., Blecha, A., et al.\ 2002, The Messenger, 110, 1 

\bibitem[Peimbert 
\& Costero(1969)]{1969BOTT....5....3P} Peimbert, M., \& Costero, R.\ 1969, Boletin de los Observatorios Tonantzintla y Tacubaya, 5, 3 


\bibitem[]{} Peimbert M., Torres-Peimbert S., Ruiz M. T.\ 1992, Revista Mexicana Astron.
Astrofisica, 24, 155


\bibitem[Reid et al.(2009)]{2009ApJ...700..137R} Reid, M.~J., et al.\ 2009, 
\apj, 
700, 137

\bibitem[{{Robberto} {et~al}\mbox{.}(2002){Robberto}, {Beckwith}, \&
  {Panagia}}]{Robberto:2002}
{Robberto} M., {Beckwith} S.~V.~W., {Panagia} N., 2002, \apj, 578, 897


\bibitem[Rodr{\'{\i}}guez 
\& Rubin(2005)]{2005ApJ...626..900R} Rodr{\'{\i}}guez, M., \& Rubin, R.~H.\ 2005, \apj, 626, 900 

\bibitem[Rubin(1989)]{1989ApJS...69..897R} Rubin, R.~H.\ 1989, \apjs, 69, 897

\bibitem[Rubin et al.(2003)]{2003MNRAS.340..362R} Rubin, R.~H., Martin, 
P.~G., Dufour, R.~J., et al.\ 2003, \mnras, 340, 362 

\bibitem[Rubin et al.(2011)]{} Rubin, R.~H., Simpson,
J.~P., O'Dell, C.~R., McNabb, I.~A., Colgan, S.~W.~J., Zhuge, S.~Y., Ferland,
G.~J., \& Hidalgo, S.~A.\ 2011, MNRAS, 410, 1320

\bibitem[Seaton(1960)]{1960RPPh...23..313S} Seaton, M.~J.\ 1960, Reports on 
Progress in Physics, 23, 313 


\bibitem[Sim{\'o}n-D{\'{\i}}az et al.(2006)]{2006A&A...448..351S} Sim{\'o}n-D{\'{\i}}az, S., Herrero, A., Esteban, C., \& Najarro, F.\ 2006, \aap, 448, 351

\bibitem[Sim{\'o}n-D{\'{\i}}az(2010)]{2010A&A...510A..22S} Sim{\'o}n-D{\'{\i}}az, S.\ 
2010, \aap, 510, A22

\bibitem[Sim{\'o}n-D{\'{\i}}az \& Stasi{\'n}ska(2011)]{2011A&A...526A..48S} Sim{\'o}n-D{\'{\i}}az, S., \& Stasi{\'n}ska, G.\ 2011, \aap, 526, A48

\bibitem[Smits(1996)]{1996MNRAS.278..683S} Smits, D.~P.\ 1996, \mnras, 278, 
683 

\bibitem[Storey(1994)]{1994A&A...282..999S} Storey, P.~J.\ 1994, \aap, 282, 999 

\bibitem[Storey \& Hummer(1995)]{1995MNRAS.272...41S} Storey, P.~J., \& Hummer, 
D.~G.
\ 1995, \mnras, 272, 41

\bibitem[Storzer \& Hollenbach(1998)]{1998ApJ...502L..71S} Storzer, H., \& Hollenbach, D.\ 1998, \apjl, 502, L71 

\bibitem[Tsamis et al.(2003)]{2003MNRAS.338..687T} Tsamis, Y.~G., Barlow, M.~J., 
Liu, X.-W., Danziger, I.~J., \& Storey, P.~J.\ 2003, \mnras, 338, 687

\bibitem[Tsamis \& P{\'e}quignot(2005)]{2005MNRAS.364..687T} Tsamis, Y.~G., \& P{\'e}quignot, D.\ 2005, \mnras, 364, 687 


\bibitem[Tsamis et al.(2011)]{2011MNRAS.412.1367T} Tsamis, Y.~G., Walsh,
J.~R., V{\'{\i}}lchez, J.~M., \& P{\'e}quignot, D.\ 2011, \mnras, 412, 1367

\bibitem[Tsamis \& Walsh(2011)]{2011MNRAS.417.2072T} Tsamis, Y.~G., \& Walsh, J.~R.\ 2011, \mnras, 417, 2072 


\bibitem[Vasconcelos et al.(2005)]{2005AJ....130.1707V} Vasconcelos, M.~J., 
Cerqueira, A.~H., Plana, H., Raga, A.~C., \& Morisset, C.\ 2005, \aj, 130, 1707

\bibitem[{{Vicente} \& {Alves}(2005)}]{Vicente:2005}
{Vicente} S.~M., {Alves} J., 2005, \aap, 441, 195

\bibitem[Walsh \& Roy(1990)]{1990ESOC...34...95W} Walsh, J.~R., \& Roy, J.~R.\ 1990, European Southern Observatory Conference and Workshop Proceedings, 34, 95 


\bibitem[{{Weidenschilling}(1977)}]{Weidenschilling:1977a}
{Weidenschilling} S.~J., 1977, \apss, 51, 153

\bibitem[{{Williams} \& {Cieza}(2011)}]{Williams:2011b}
{Williams} J.~P., {Cieza} L.~A., 2011, \araa, 49, 67






\end{thebibliography}

\label{lastpage}

\end{document}